\documentclass[usenatbib,useAMS]{mnras}

\usepackage{natbib}
\usepackage{amsmath}
\usepackage{graphicx}
\usepackage{color}
\usepackage{mathrsfs}
\usepackage{epsfig}
\usepackage{comment}
\usepackage{ulem}

\usepackage{graphicx}
\usepackage[hang,small,bf]{caption}
\usepackage[subrefformat=parens]{subcaption}
\usepackage{amssymb}
\usepackage[T1]{fontenc}

\newcommand{\msun}{{\rm M}_\odot}
\newcommand{\mchirp}{{\rm M_{\rm chirp}}}
\newcommand{\rsun}{{\rm R}_\odot}
\newcommand{\zsun}{Z_\odot}

\newcommand{\gpc}{{\rm Gpc}}

\newcommand{\yr}{\rm yr}

\newcommand{\beq}{\begin{equation}}
\newcommand{\eeq}{\end{equation}}

\defcitealias{K14}{K14}
\defcitealias{K16}{K16}
\defcitealias{Visbal_2015}{VHB15}

\title[Chirp Mass and Spin of Pop III BBHs]
{Chirp Mass and Spin of Binary Black Holes from First Star Remnants}

\author[T. Kinugawa et al.]
{Tomoya Kinugawa$^{(1)}$\thanks{E-mail: kinugawa@icrr.u-tokyo.ac.jp},  Takashi Nakamura$^{(2)}$, and Hiroyuki Nakano$^{(3)}$\\
\\
$^{1}$Institute for Cosmic Ray Research, The University of
  Tokyo, Kashiwa, Chiba 277-8582, Japan\\
$^{2}$Department of Physics, Graduate School of Science, Kyoto University,
Kyoto 606-8502, Japan\\
$^{3}$Faculty of Law, Ryukoku University, Kyoto 612-8577, Japan}

\begin{document}

\date{\today}
\maketitle
%\label{firstpage}

\begin{abstract}
We performed Population III (Pop III) binary evolution by using population synthesis simulations for seven different models. We found that Pop III binaries tend to be binary black holes (BBHs) with chirp mass $M_{\rm chirp} \sim 30~\msun$ and they can merge at present day due to long merger time.
The merger rate densities of Pop III BBHs at $z=0$ ranges 3.34--21.2 $\rm /yr/Gpc^3$ which is consistent with the aLIGO/aVIRGO result of 9.7--101 $\rm /yr/Gpc^3$.
These Pop III binaries might contribute to some part of the massive BBH gravitational wave (GW) sources detected by aLIGO/aVIRGO.
 We also calculated the redshift dependence of Pop III BBH mergers.
 We found that Pop III low spin BBHs tend to merge at low redshift, while Pop III high spin BBHs do at high redshift, which can be confirmed by
future GW detectors such as ET, CE, and DECIGO.
These detectors can also check the redshift dependence of BBH merger rate and spin distribution. {Our results show that except for one model, the mean effective spin $\left\langle \chi_{\rm eff} \right\rangle$ at $z=0$ ranges
$0.02$--$0.3$ while at $z=10$ it does $0.16$--$0.64$.}
Therefore, massive stellar-mass BBH detection by GWs will be a key for the stellar evolution study in the early universe. 
\end{abstract}

\section{Introduction}
% LIGO detections

\begin{figure}
\begin{center}
\includegraphics[width=\hsize]{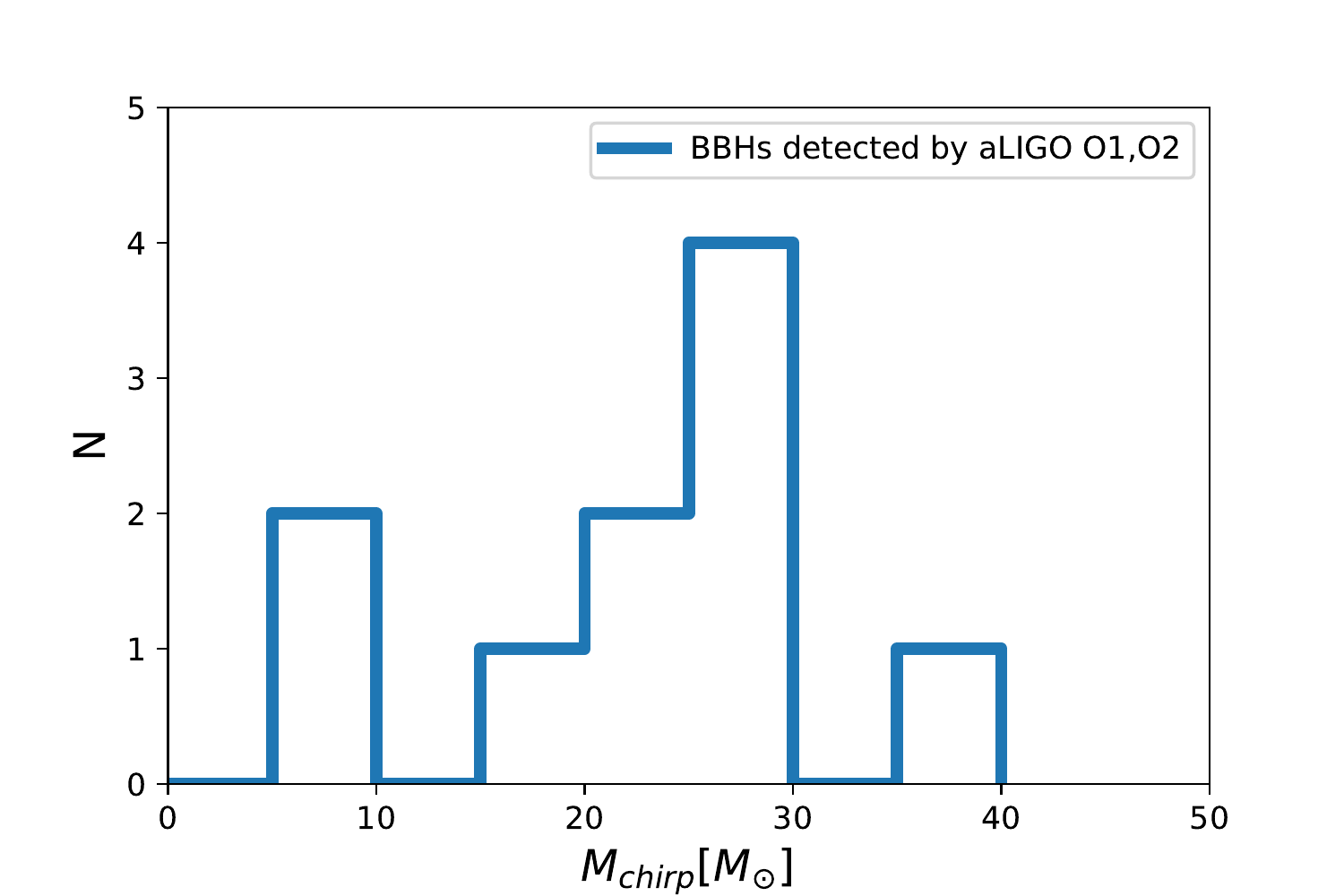}
\end{center}
\caption{Chirp mass distribution of ten BBHs presented in GWTC-1 \citep{Abbott_2019}.
The horizontal axis is the chirp mass. The vertical axis is the number of observed BBHs.}  
\label{aLIGOMc}
\end{figure}

Advanced LIGO (aLIGO) and Advanced Virgo (aVIRGO) have detected gravitational waves (GWs) from binary black hole (BBH) mergers \citep{Abbott_2019}.
The observed binary mass and the chirp mass ($\mchirp$) of aLIGO/aVIRGO O1 and O2 runs are summarized in Table~\ref{gw_events}.
Figure \ref{aLIGOMc} shows the chirp mass distribution of BBHs detected by aLIGO/aVIRGO O1 and O2 runs.
%inferred as
%$(35.6^{+4.7}_{-3.1}~\msun, 30.6^{+3.0}_{-4.4}~\msun)$, 
%$(23.2^{+14.9}_{-5.5}~\msun, 13.6^{+4.1}_{-4.8}~\msun)$
%$(13.7^{+8.8}_{-3.2}~\msun, 7.7^{+2.2}_{-2.5}~\msun)$,
%$(30.8^{+7.3}_{-5.6}~\msun, 20.0^{+4.9}_{-4.6}~\msun)$, 
%$(11.0^{+5.5}_{-1.7}~\msun, 7.6^{+1.4}_{-2.2}~\msun)$,
%$(50.2^{+16.2}_{-10.2}~\msun, 34.0^{+9.1}_{-10.1}~\msun)$,
%$(35.0^{+8.3}_{-5.9}~\msun, 23.8^{+5.1}_{-5.2}~\msun)$,
%$(30.6^{+5.6}_{-3.0}~\msun, 25.2^{+2.8}_{-4.0}~\msun)$,
%$(35.4^{+7.5}_{-4.7}~\msun, 26.7^{+4.3}_{-5.2}~\msun)$,
%$(39.5^{+11.2}_{-6.7}~\msun, 29.0^{+6.7}_{-7.8}~\msun)$
%for GW150914, GW151012, GW151226, GW170104, GW170608, 
%GW170729, GW170809, GW170814, GW170818, and GW170823, 
%respectively.
Seven out of ten BBHs have massive stellar-mass black holes with $\mchirp \sim30~\msun$ (see Fig. \ref{aLIGOMc}). 
Candidates for the astrophysical origins of such massive compact BBHs have been proposed: 
isolated massive stellar binaries \cite[e.g.][]{Dominik_2012,Dominik_2013,Kinugawa2014,Kinugawa2016,Belczynski_2016_a,Miyamoto2017,Belczynski2020}, dynamical formation in dense stellar clusters \citep[e.g.][]{Portegies2000,O2006,Tanikawa2013,Rod2015,Kumamoto2019}, 
 rapidly rotating massive stars \citep[e.g.][]{Mandel_2016}, and the compact binary formation in disk region \citep[e.g.][]{Tagawa2019} and so on.
%Each model can produce  merging rate of BBHs consistent with aLIGO/aVIRGO observations of $R_{\rm BBH}\sim$ 9.7--101 $\gpc^{-3}~\yr^{-1}$ \citep{Abbott_2019}.

% PopIII scenario

One possible candidate providing aLIGO/aVIRGO sources is massive field binaries of Population III (Pop III) stars \cite[e.g.][]{Kinugawa2014,Kinugawa2016},
which are the first generation stars in the Universe.
Pop III stars are expected to be born as massive as $\sim$  10--100 $\msun$ \citep[e.g][]{Hosokawa_2011,Hirano_2014,Susa_2014} and would initiate massive binary BHs heavier than those from Pop I, II stars.
Although they form at the early epoch of the Universe (at $z\ga 10$), some fraction of Pop III BBHs would merge due to GW emission at present taking a long timescale of a Hubble time so that these GWs can be detected within the detection horizon of aLIGO/aVIRGO ($z\la 1$).
\cite{Kinugawa2014} first predicted that Pop III binaries tend to become massive stellar BBHs with $\mchirp \sim 30~\msun$ and they can merge at  present, which was before aLIGO detection of  the first GW from BBH in 2015 since the paper was published in 2014.
Furthermore, the merger rate of Pop III BBHs $\sim 30~\gpc^{-3}~\yr^{-1}$ \citep{Kinugawa2014,Kinugawa2016} is consistent with aLIGO/aVIRGO results $R_{\rm BBH}\sim$ 9.7--101 $\gpc^{-3}~\yr^{-1}$ \citep{Abbott_2019}.

\begin{table}
	\caption{GW events from BBH mergers in GWTC-1 \citep{Abbott_2019}
	which is the gravitational-wave transient catalog of the first and second Observing runs of aLIGO/aVIRGO. The event names, individual masses and chirp mass of BBHs are presented here.}
	\label{gw_events}
	\centering
	\begin{tabular}{c| c c c c}
	\hline
	Event name & BH mass 1 & BH mass 2 & Chirp mass \\
	\hline
	GW150914 & $35.6^{+4.7}_{-3.1}~\msun$ & $30.6^{+3.0}_{-4.4}~\msun$ 
	& $28.6^{+1.7}_{-1.5}~\msun$ \\
    GW151012 & $23.2^{+14.9}_{-5.5}~\msun$ & $13.6^{+4.1}_{-4.8}~\msun$
    & $15.2^{+2.1}_{-1.2}~\msun$ \\
    GW151226 & $13.7^{+8.8}_{-3.2}~\msun$ & $7.7^{+2.2}_{-2.5}~\msun$
    & $8.9^{+0.3}_{-0.3}~\msun$ \\
    GW170104 & $30.8^{+7.3}_{-5.6}~\msun$ & $20.0^{+4.9}_{-4.6}~\msun$
    & $21.4^{+2.2}_{-1.8}~\msun$ \\
    GW170608 & $11.0^{+5.5}_{-1.7}~\msun$ & $7.6^{+1.4}_{-2.2}~\msun$
    & $7.9^{+0.2}_{-0.2}~\msun$ \\
    GW170729 & $50.2^{+16.2}_{-10.2}~\msun$ & $34.0^{+9.1}_{-10.1}~\msun$
    & $35.4^{+6.5}_{-4.8}~\msun$ \\
    GW170809 & $35.0^{+8.3}_{-5.9}~\msun$ & $23.8^{+5.1}_{-5.2}~\msun$
    & $24.9^{+2.1}_{-1.7}~\msun$ \\
    GW170814 & $30.6^{+5.6}_{-3.0}~\msun$ & $25.2^{+2.8}_{-4.0}~\msun$
    & $24.1^{+1.4}_{-1.1}~\msun$ \\
    GW170818 & $35.4^{+7.5}_{-4.7}~\msun$ & $26.7^{+4.3}_{-5.2}~\msun$
    & $26.5^{+2.1}_{-1.7}~\msun$ \\
    GW170823 & $39.5^{+11.2}_{-6.7}~\msun$ & $29.0^{+6.7}_{-7.8}~\msun$
    & $29.2^{+4.6}_{-3.6}~\msun$ \\
	\hline
	\end{tabular}
\end{table}

However, there are some objections for the Pop III BBH scenario after the GW detection.
\cite{Hartwig_2016} suggested that the merger rate of Pop III BBHs is lower than our model in \cite{Kinugawa2014}, using a low metal binary evolution model ($Z=10^{-1}\zsun$) \citep{demink2015} and a latest Pop III star formation rate (SFR) with the constraint from \cite{Visbal_2015}.
\cite{Belczynski_2017} calculated low metal binary evolution using a modified low metal stellar evolution ($Z=5\times10^{-3}\zsun$), and suggested that most Pop III BBHs merged at the early universe and the merger rate at present day is much smaller than the aLIGO/aVIRGO result. 

In these objections, the authors did not use the Pop III evolution model but the low metal Pop II one, although the Pop III stellar evolution is mighty different from the evolution of Pop II stars.
All Pop I, II stars generally evolve via a red giant phase, but Pop III stars whose masses are less than $50~\msun$ end the evolution in a blue giant phase.
Binary interaction such as the mass transfer stability, and the tidal evolution strongly depends on whether the star is a red giant or a blue giant.
Blue giant stars have a radiative envelope which tends to be stable for a mass transfer and have low efficiency for the tidal interaction.
On the other hand, red giant stars have a convective envelope which tends to be unstable for a mass transfer and have high efficiency for the tidal interaction. These facts show that evolution of low metal Pop II stars used in \cite{Belczynski_2017} is largely different from that of Pop III ones in \cite{Kinugawa2014}.

Therefore, in this paper we calculate the Pop III binary evolution using the latest Pop III SFR and binary prescriptions,
%parameters \tn{What are the binary parameters? Data of BBH from GW detections?}, 
and clarify the Pop III BBH merger rate in details by comparing seven different models.
We also estimate the spin distribution and the redshift evolution of Pop III BBH mergers for future observations by ET \citep{ET}.
%, CE \citep{Reitze:2019iox} and DECIGO \citep{Seto:2001qf}.
% spin alignment

% organization

This paper is organized as follows.
In Sec.~2, we present our Pop III binary evolution code.
A brief review is given in Sec. 2.1,
and the initial conditions and calculated models are
summarized in Sec. 2.2. 
Details of the binary evolution are discussed
in Sec. 2.3 to Sec. 2.8, which can be skipped  to proceed directly to Sec. 3.
In Sec. 3, we show the results  including the dependence on
the evolution channels,
the chirp mass and merger time distributions of BBHs,
the merger rate, spin distribution and the detection rate
for the second and third-generation GW detectors.
Finally, Sec. 4 is devoted to summary and discussion.
Some complementary figures are shown in Appendix.

%\section{Gravitational-wave detections}

\section{Method}
In order to perform Pop III binary evolutions, we use Pop III binary evolution code \citep{Kinugawa2014,Kinugawa2016,Kinugawa2017} which is an upgraded version from BSE code \footnote{http://astronomy.swin.edu.au/~jhurley/} \citep*{Hurley_2002} to Pop III stars case.
We describe basic equations to calculate the binary evolution for Pop III binary evolution in this section.

\subsection{Review of Pop III stellar evolution and star formation}
We use the Pop III stellar evolution code \citep{Kinugawa2014,Kinugawa2016,Kinugawa2017} which is based on Pop III stellar evolution simulations \citep{Marigo_2001}.
There are two main differences in properties and evolution of Pop III stars from those of Pop I, II stars.

First, the Pop III initial mass is much more massive than that of Pop I, II.
In the case of Pop I, II stars, the initial mass distribution is described well by the Salpeter one and the typical mass is $\lesssim1~\msun$.
On the other hand, the typical mass of Pop III stars is $\sim$ 10--100 $\msun$ due to less coolant during the formation \cite[e.g.][]{Hosokawa_2011,Hosokawa_2012,Stacy_2011,Hirano_2014,Susa_2014}.
The metal content has transition from Pop III stars to Pop I, II ones at $Z/\zsun\sim10^{-3.5}$--$10^{-6}$ \cite[e.g.][]{Schneider2002,Schneider2003,Omukai2005}.

Second, the late phase evolution of Pop III stars is completely different from that of Pop I, II stars.
BBH progenitors of Pop I, II has a red supergiant (RSG) phase with a convective envelope. Such stars tend to experience  unstable mass transfer and have a common envelope Phase \citep{Paczynski_1976,Iben_1993,  Taam_2000,  Ivanova_2013}. In fact, most of Pop I, II BBH progenitors evolve via the common envelope phase \citep[e.g.][]{Belczynski_2002,Dominik_2012, Dominik_2013, Belczynski_2016_a, Belczynski2020}. On the other hand, Pop III stars with masses less than $50~\msun$ end at blue supergiant (BSG) phase with radiative envelope \cite[e.g.][]{Marigo_2001,Ekstrom_2008}.
They tend to experience a stable mass transfer, details of which will be  described in Sec. \ref{sec:MTstability}. In such stable mass transfer phase the star generally loses less stellar mass than the mass loss in the common envelope phase.
Thus, Pop III BBHs tend to become more massive than Pop I, II BBHs \citep{Kinugawa2014,Inayoshi2017}.
This transition of evolutional path occurs around $Z/\zsun\sim10^{-5}$ \citep{Tanikawa2019}.

\subsection{Initial conditions and calculation models}
\label{sec:IandC}
The result of binary evolution depends on initial distribution functions and binary evolution parameters such as the initial mass function (IMF), the mass ratio distribution, the separation distribution, the eccentricity distribution, common envelope parameters $\alpha\lambda$, and the accretion fraction $\beta$ of transferred stellar mass.
In order to check the parameter dependence, we calculate seven models.

Tables \ref{tab.init} and \ref{tab.bin} describe the initial conditions and binary evolution parameters.
First we assume the binary fraction $f_b=0.5$ which is the same as the binary fraction of nearby stars \cite[e.g.][]{Sana2012,Sana2013}.
We perform Monte Carlo simulations using $10^6$ zero age main sequence binaries of zero metal Pop III stars for each model. 
In ``M100'' model the maximum initial mass is changed from our fiducial model in Table 2 (150~$\msun$) to 100~$\msun$.
In ``$\beta$=0.5'' model, $\beta=0.5$ is adopted, while in our fiducial model  $\beta=1$, where $\beta$ means the accretion fraction for secondary star during the mass transfer (Sec. \ref{stable MT}).
In ``$\alpha\lambda$=0.1'' model  $\alpha\lambda=0.1$ is adopted, while in our fiducial model $\alpha\lambda=1$, 
where $\alpha$ is the efficiency factor of the energy conversion which depends on the interaction between the giant's envelope and the companion,
%(= How much fraction of orbital energy used is used in order to eject the giant's envelope ?),
$\lambda$ is the parameter of the binding energy of giant's envelope.
In ``K14'' model  the same initial condition, the binary parameter, the mass transfer rate and the tidal interaction treatment in \cite{Kinugawa2014} are used to compare the results of the fiducial model with the previous one, {although some parts of fitting fomulae are updated \citep{Kinugawa2017}}.
In  ``FS1'' and ``FS2'' models,  the initial conditions, the values of $\beta$ and $\alpha\lambda$   are the same as those of \cite{Belczynski_2017} but the Pop III evolution model and numerical code are our ones in order to {compare the results  with those of} \cite{Belczynski_2017}.
%\tk{In our paper, we consider the pair instability supernova and the no remnant, if the CO core mass is more than $60~\msun$ ($M_{\rm initial}\sim140~M_{\odot}$) \citep{Heger2002}. }

In the following, {for those who are familiar with our stellar evolution code
or want to immediately read the results,
the remaining part of this section (Sec. 2) can be skipped to  proceed directly to Sec. 3.}

\begin{table*}
	\caption{Initial conditions for evolution of Pop~III binaries for seven models. As the initial conditions, we specify the initial mass function (IMF), the initial mass
	ratio, separation and eccentricity distributions of binaries.}
	\label{tab.init}
		\centering
		\begin{tabular}{c| c c c c}
		\hline
		Model & IMF& Initial mass ratio & Initial separation & Initial eccentricity\\
		\hline
		Fiducial& flat                          & flat                                & logflat  & power-law (index:1) \\ 
	
		$\beta$=0.5& $10\msun<M_1<150\msun$          & $10\msun/M_1<q<1$                   & $\log a_{\rm min}<\log (a/\rsun)<6$                  & $0<e<1$ \\
		$\alpha\lambda$=0.1 & &  &   &\\
		\hline
		M100& flat                          & flat                                & logflat  & power-law (index:1) \\ 
		K14		& $10\msun<M_1<100\msun$          & $10\msun/M_1<q<1$                   & $\log a_{\rm min}<\log (a/\rsun)<6$                  & $0<e<1$ \\
		\hline
		& Gaussian                           & Gaussian                  & logflat                               & Gaussian \\ 
		FS1& $\sigma=52.2\msun$, $M_0=128\msun$ & $\sigma=0.29$, $q_0=0.92$ & $72\%$ in range1: $2000$--$2\times10^5\rsun$    & $\sigma=0.25$, $e_0=0.8$ \\  
		&  $9.6\msun<M_1<138\msun$           &  $0.03<q<0.99$     & $28\%$ outside range1: $20$--$2\times10^8\rsun$ &$0.10<e<1.0$ \\
		
		\hline 
		& power-law + Gaussian               & power-law + Gaussian                & Gaussian                              & linear \\ 
		& $50\%$ in range1: $3$--$70\msun$   & range1: $0.002$-$0.3$ ($M>70\msun$) & $\sigma=71.6\rsun$, $a_0=90.1\rsun$   & slope $=0.08$ \\  
		FS2& index:$-0.55$                     & index:$-0.35$                      & range: $1.1$--$1075\rsun$             & range: $0.04$--$0.99$ \\
		& $50\%$ in range2: $70$--$181\msun$ & range2: $0.1$--$1.0$ ($M<70\msun$)  &                                       & \\
		& $\sigma=11.0\msun$, $M_0=144\msun$ & $\sigma=0.14$, $M_0=0.78$       
		    &                                       & \\
		\hline
		
	\end{tabular}	
\end{table*}

\begin{table*}
	\caption{Binary evolution parameters for evolution of Pop~III binaries for seven models. As the evolution parameters, we have the mass transfer rate of the donor (see Sec.~\ref{stable MT}), the accretion fraction $\beta$ of transferred stellar mass (see Secs.~\ref{sec:MTstability} and~\ref{stable MT}), the common envelope parameters $\alpha\lambda$ (see Sec.~\ref{sec:CE}), and the tidal coefficient factor $E$ (see Sec.~\ref{tidal}).}
	\label{tab.bin}
		\centering
		\begin{tabular}{c| c c c c}
		\hline
		Model & Mass transfer rate & Accretion fraction $\beta$& Common envelope parameter $\alpha\lambda$ & Tidal coefficient factor $E$\\
		\hline
		Fiducial&  Eq. (\ref{eq:MTrate1})                          & 1                                &1  & Eq. (\ref{eq:E2new}) \\ 
	
		M100&        &                    &  &  \\
		\hline
		$\beta$=0.5& Eq. (\ref{eq:MTrate1})                      & 0.5                              & 1  & Eq. (\ref{eq:E2new}) \\ 
		\hline
		$\alpha\lambda$=0.1 & Eq. (\ref{eq:MTrate1})  & 1&0.1 &Eq. (\ref{eq:E2new})\\
		\hline
		K14		&  Eq. (\ref{eq:m1dotf})           & 1                   & 1                & Eq. (\ref{eq:E2,1}) \\
		\hline
		FS1&  Eq. (\ref{eq:MTrate1})  & 0.5 & 0.1   & Eq. (\ref{eq:E2new}) \\  
		FS2&                    &                     &        &  \\
		\hline
		
	\end{tabular}
	
\end{table*}

\subsection{Stability of mass transfer}
\label{sec:MTstability}

If the Roche lobe around a star in the binary system is filled, the material of the star is transferred 
to the companion star through the first Lagrangian point.
This process is called as Roche lobe overflow (RLOF).
The radius of the donor star's Roche lobe (Roche lobe radius) is approximately expressed as \citep{Eggleton_1983}:
\begin{equation}
R_{\rm L,1}\simeq \frac{0.49q_1^{2/3}}{0.6q_1^{2/3}+\ln(1+q_1^{1/3})}a \,,
\end{equation}
where $a$ is the orbital separation and $q_1=M_1/M_2$ is the mass ratio where $M_1$, and $M_2$ are the mass of the donor and that of the accretor, respectively.

The mass transfer rate is determined by the values of the Roche lobe radius and 
the stellar radius when it loses the mass \citep{Paczynski_1976}.
{The Roche lobe radius changes as a function of the mass lost through mass transfer, 
\begin{align}\label{eq.stMT}
\zeta_{\rm L}&=\frac{d{\rm{log}}R_{\rm L,1}}{d{\rm{log}}M_1}\notag\\
                  &=\frac{(0.33+0.13q_1)[1+(1-\beta) q_1)]+(1-\beta)(q_1^2-1)-\beta q_1}{1+q_1} \,,
\end{align}   
where $\beta$ is the fraction of the gas that accretes to the accepting star
\citep{Eggleton_1983, Eggleton2011}}.
For $\zeta_{\rm L}<\zeta_{\rm ad}$ ($\equiv (d\ln R_1/d\ln M_1)$ within the dynamical time scale), the stellar radius $R_1$ of the donor shrinks and 
becomes smaller than the Roche lobe radius just after the mass of the donor star is transferred. 
On the other hand, for $\zeta_{\rm L}>\zeta_{\rm ad}$, the stellar radius becomes much larger than the Roche lobe radius so that the mass transfer would proceed unstably and 
the two stars in the binary system would merge or become a common envelope phase.
The value of $\zeta_{\rm ad}$ depends on the stellar envelope of the donor star. If the donor star is a red giant star which
has a convective envelope, $\zeta_{\rm ad}$ is given by
\begin{equation}
    \zeta_{\rm ad}=-1+\frac{2}{3}\frac{M_1}{M_{\rm env,1}} \,,
\end{equation}
where $M_1$ and $M_{\rm env,1}$ are the mass and the envelope mass of the red giant. 
When the donor star is in the other evolution stages, $\zeta_{\rm ad}$ is equal to 2.59, 6.85, 1.95 and 5.79 for the main sequence, the blue giant star with the radiative envelope \citep{Hjellming_1989}, the
naked-He main sequence and the naked-He giant star \citep{Ivanova_2002,Belczynski_2008}, respectively.

\subsection{Stable Roche lobe overflow}\label{stable MT}
%We adopt the binary module in \texttt{MESA} to treat the MT process.
When the star fulfils the Roche lobe ($R_1>R_{\rm L,1}$) and the stable RLOF ($\zeta_{\rm ad}>\zeta_{\rm L}$) occurs, 
we use two fomulae to calculate the mass transfer rate.
In our fiducial model, the mass transfer rate of the donor is calculated as
\begin{align}\label{eq:MTrate1}
\dot{M}_1 = -\frac{f(\mu) M_1}{\sqrt{R_1^3/GM_1}}\left(\frac{\Delta R_1}{R_1}\right)^{n+1.5}d_{n} \,,
\end{align}
where 
\begin{equation}
f(\mu)=\frac{4\mu\sqrt{\mu}\sqrt{1-\mu}}{(\sqrt{\mu}+\sqrt{1-\mu})^4}\left(\frac{a}{R_1}\right)^3 \,,
\end{equation}
$\mu=M_1/(M_1+M_2)$, $\Delta R_1=R_1-R_{\rm L,1}$, 
$n$ and $d_n$ are the polytropic index and the normalization factor depending on $n$, respectively 
\citep[e.g.][]{Paczynski_1972,Savonije_1978,Edwards_Pringle_1987,Inayoshi2017}.
Note that $n=3/2$ and $n=3$ are assumed 
in Eq. \eqref{eq:MTrate1} for a red giant phase and for the other phases, respectively.
For each case, $d_{3/2}=0.2203$ and $d_3=0.0364$, respectively.

In our previous study (K14 model), we calculate the mass transfer rate using the fitting formula of \cite{Hurley_2002}
given by
\begin{equation}
\dot{M_1}=F(M_1)\left[{\rm{ln}} \left(\frac{R_{1}}{R_{\rm L,1}}\right)\right]^3~\rm{M_{\odot}~yr^{-1}} \,,
\label{eq:m1dotf}
\end{equation}
where
\begin{equation}
F(M_1)=3\times10^{-6}\left\{{\rm{min}}\left[\left(10\frac{M_1}{10~\msun}\right),5.0\right]\right\}^2 \,.
\end{equation}
Since the stellar radius of the donor changes by the thermal timescale or more slowly,
the maximum value of the mass transfer rate is given by
\begin{equation}
\dot{M}_{1,\rm{max}}=\frac{M_1}{\tau_{\rm{KH,1}}} \,,
\label{eq:mkhmax}
\end{equation}
where $\tau_{\rm{KH,1}}$ is the thermal timescale of the donor.

%The first and second terms in eq. \ref{MTrate} correspond to the mass transfer rate for isothermal atmosphere and adiabatic atmosphere, respectively.
The accretion rate of the accretor is given by
\begin{equation}
    \dot{M}_2=-\beta\dot{M}_1 \,.
\end{equation}
We assume the conservative mass transfer, i.e., $\dot{M}_{\rm total}=\dot{M}_1+\dot{M}_2=0$ ($\beta=1$) in our fiducial model.
If the accretor is a compact object such as a neutron star or a black hole, we consider
the maximum of the accretion rate is limited by the Eddington mass accretion rate given by
\begin{align}
\dot{M}_{\rm{Edd}}&=-\frac{4\pi c R_2}{\kappa_{\rm T}}\\\nonumber
&=2.08\times10^{-3}(1+X)^{-1}
\left(\frac{R_2}{\rsun}\right)~\rm{M_{\odot}~yr^{-1}} \,,
\end{align}
where $R_2$ is the stellar radius of the accretor, $\kappa_{\rm T} =0.2(1+X)\ \rm cm^2\ g^{-1}$ is the Thomson scattering opacity and $X(=0.76)$ is the hydrogen mass fraction. 

We calculate the spin evolution of a binary system during the mass transfer. The angular momentum is carried from the donor to the accretor.
We estimate the angular momentum transferred in this process with a thin shell approximation:
\begin{equation}
\frac{dJ_{\rm sp,1}}{dt}= \frac{2}{3}\dot{M}_1R_1^2\Omega_{\rm spin,1} \,,
\end{equation}
where $\Omega_{\rm spin,1}$ is the spin angular velocity of the donor.
For the accretor's spin, we consider the two cases of accretion whether the material accretes via accretion disk or not.
First, if there is no accretion disk, i.e., the secondary radius is larger than the critical radius ($r_{\rm cri} = 0.07225a(q_2(1+q_2))^{1/4}$, where $q_2=M_2/M_1)$  \citep{LubowShu1974,UlrichBurger1976, Hurley_2002},
we assume that the angular momentum of the transferred material evaluated by using the critical radius {is added}  directly to the accretor's spin. 
Thus, the angular momentum transferred to the accretor is calculated as
\begin{equation}
\frac{dJ_{\rm sp,2}}{dt}=\dot{M}_2\sqrt{GM_2r_{\rm cri}} \,.
\end{equation}
Secondly, if the transferred material accretes through a disk, 
the accretor's spin angular momentum is altered assuming that the transferred material falls onto the accretor surface with Keplerian velocity.
Then the angular momentum transferred via the accretion disk is calculated as
\begin{equation}
\frac{dJ_{\rm sp,2}}{dt}=\dot{M}_2\sqrt{GM_2R_2} \,.
\end{equation} 
Note that the total angular momentum $J_{\rm orbit}+J_{\rm sp,1}+J_{\rm sp,2}$ is conserved in this calculation.

\subsection{Common envelope phase}\label{sec:CE}

When the mass transfer is unstable ($\zeta_{\rm ad}<\zeta_{\rm L}$) or when the companion plunges into a giant's envelope, a common envelope phase occurs. 
In order to calculate the separation just after the common envelope phase, we use the $\alpha\lambda$ formalism for the common envelope phase \citep{Webbink1984}.
When a binary of a giant star and a non giant star enters into a common envelope phase, they satisfy the condition given by 
\begin{equation}
\alpha\left(\frac{GM_{\rm{c,1}}M_2}{2a_{\rm{f}}}-\frac{GM_1M_2}{2a_{\rm{i}}}\right)=\frac{GM_{\rm{1}}M_{\rm{env,1}}}{\lambda R_1} \,,
\label{eq:ce1}
\end{equation} 
where $a_{\rm i}$, $a_{\rm f}$, $R_1$, $M_1$, $M_{\rm c,1}$, $M_{\rm env,1}=M_1-M_{\rm c,1}$, and $M_2$ are the separation just before the common envelope phase, the separation just after the common envelope phase, the radius, the mass, the core mass and the envelope mass of the donor giant, and the mass of the companion, respectively.
The value of $\alpha$ is the efficiency parameter to express {how much  orbital energy is needed to  eject the envelope.}  $\lambda$ is the parameter to express {the amount of } the binding energy of the envelope.

When the companion is also a giant, Eq.~\eqref{eq:ce1} is changed to
\begin{align}
	\alpha\left(\frac{GM_{\rm{c,1}}M_{c,2}}{2a_{\rm{f}}}-\frac{GM_1M_2}{2a_{\rm{i}}}\right)=&\frac{GM_{\rm{1}}M_{\rm{env,1}}}{\lambda R_1}\notag\\
	&+\frac{GM_{\rm{2}}M_{\rm{env,2}}}{\lambda R_2} \,,
	\label{eq:ce2}
\end{align}
where $M_{\rm c,2}$, $M_{\rm env,2}=M_2-M_{\rm c,2}$, and  $R_2$ are the core mass, the envelope mass, and the radius of the companion giant, respectively \citep{Dewi2006}.
The common envelope parameters $\alpha$ and $\lambda$ are not well understood yet \citep{Ivanova_2013}.
In this paper, we use the typical common envelope parameter values adopted in the previous binary population synthesis studies ($\alpha\lambda=1$ or $0.1$) \citep[e.g.][]{Belczynski2007,Kinugawa2014,Belczynski_2017}.
In our fiducial mode, we use $\alpha\lambda=1$.

\subsection{Tidal interaction}
\label{tidal}
Tidal interaction plays an important role in the evolution of the orbit and the spins. 
There are two mechanisms for the dissipation of the tidal kinetic energy.
One mechanism is the convective damping on the equilibrium tide for the stars with an outer convection envelope such as red giants.
The other mechanism is the radiative damping on the dynamical tide for
the stars with an outer radiative zone \citep{Zahn1977}. 
The time evolution of the separation, the eccentricity, and the spin are calculated as follows \citep{Zahn1977, Hut1981}.
{
\begin{align}
\frac{da}{dt}=&-6\frac{k}{T}q(1+q)\left(\frac{R_i}{a}\right)^8 \frac{a}{(1-e^2)^{\frac{15}{2}}}\notag\\
&\times\left[f_1(e^2)-(1-e^2)^{\frac{3}{2}}f_2(e^2)\frac{\Omega_{{\rm spin},i}}{\Omega_{\rm orb}}\right]\label{eq:dadt} \,, \\
\frac{de}{dt}=&-27\frac{k}{T}q(1+q)\left(\frac{R_i}{a}\right)^8\frac{e}{(1-e^2)^{\frac{13}{2}}}\notag\\
&\times\left[f_3(e^2)-\frac{11}{18}(1-e^2)^{\frac{3}{2}}f_4(e^2)\frac{\Omega_{{\rm spin},i}}{\Omega_{\rm orb}}\right] \,, \label{eq:dedt}\\
\frac{d\Omega_{{\rm spin},i}}{dt}=&3\frac{k}{T}\frac{q^2}{r_g^2}\left(\frac{R_i}{a}\right)^6\frac{\Omega_{\rm orb}}{(1-e^2)^6}\notag\\
&\times\left[f_2(e^2)-(1-e^2)^{\frac{3}{2}}f_5(e^2)\frac{\Omega_{{\rm spin},i}}{\Omega_{\rm orb}}\right] \,, \label{eq:dodt}
\end{align}
}
where
\begin{align}
f_1(e^2)&=1+\frac{31}{2}e^2+\frac{255}{8}e^4+\frac{185}{16}e^6+\frac{25}{64}e^8 \,, \label{eq:f1} \\
f_2(e^2)&=1+\frac{15}{2}e^2+\frac{45}{8}e^4+\frac{5}{16}e^6 \,, \label{eq:f2}\\
f_3(e^2)&=1+\frac{15}{4}e^2+\frac{15}{8}e^4+\frac{5}{64}e^6 \,, \label{eq:f3}\\
f_4(e^2)&=1+\frac{3}{2}e^2+\frac{1}{8}e^4 \,, \label{eq:f4}\\
f_5(e^2)&=1+3e^2+\frac{3}{8}e^4 \,. \label{eq:f5}
\end{align}
Here $k/T,~q,~R_i,~\Omega_{{\rm spin},i},~\Omega_{\rm orb},$ and $r_g$ are a coupling parameter depending on the tidal interaction mechanism, the mass ratio of the companion to the star, the radius of the star, the spin angular velocity of the star, the orbital angular velocity, and the dimensionless gyration radius of the star, respectively.

\subsubsection{convective damping in equilibrium tide}
In the case that the stellar envelope is convective, the energy dissipation by the convective motions causes the time lag of the tidal deformation.
It yields the misalignment between the direction of the maximum
tidal deformation and the direction to the companion. This misalignment generates the torque so that the angular momentum is transferred from the spin angular momentum to the orbital one or vice versa.
The coupling parameter for the equilibrium tide is given by
\begin{equation}
\frac{k}{T}=\frac{2}{21}\frac{
f_{\rm{con}}}{\tau_{\rm{con}}}\frac{M_{{\rm env},i}}{M_{\rm{i}}} \,,
\end{equation}
where $M_{{\rm env},i}$ is
the stellar envelope mass, the factor $f_{\rm{con}}$ is the correction factor of the tidal torque, and $\tau_{\rm{con}}$
is the eddy turnover timescale \cite[e.g.][]{Rasio1996,Hurley_2002}.
They are calculated as
\begin{align}
\tau_{\rm{con}}&=\left[\frac{M_{{\rm env},i}R_{{\rm env},i}
\left(R_i-\frac{1}{2}R_{{\rm env},i}\right)}{3L_i}\right]^{1/3} \,,
\\
f_{\rm{con}}&={\rm{min}}\left[1,\left(\frac{\pi|\Omega_{\rm{orb}}-\Omega_{{\rm spin},i}|^{-1}}{\tau_{\rm{con}}}\right)^2\right] \,.
\end{align}
where $L_i$, and $R_{{\rm env},i}$ are
the stellar luminosity and the stellar envelope radius, respectively.

\subsubsection{radiative damping in dynamical tide}
In the case of radiative envelope, a tidal mechanism is the radiative damping of the dynamical tide \citep{Zahn1975}.
$k/T$ given by \cite{Zahn1977, Hurley_2002} is 
\begin{align}
\frac{k}{T}=&4.3118\times10^{-8}\left(\frac{M_i}{\msun}\right)\left(\frac{R_i}{{\rm R}_{\odot}}\right)^2\notag\\
&\times\left(\frac{a}{1\,{\rm{AU}}}\right)^{-5}(1+q)^{5/6}E~\rm{yr^{-1}} \,,
%\textcolor{red}{\rm ***~No~!~***}
\end{align}
where $E$ is the tidal coefficient factor.

In the BSE code, $E$ is described by 
\begin{equation}\label{eq:E2,1}
E=1.101\times10^{-6}\left(\frac{M_1}{10\,\msun}\right)^{2.84} \,.
\end{equation}
However, recently $E$ % $E_2$
is better fitted by using the dependence of $E$ on the ratio  between the convective core radius and the stellar radius $R_{\rm con}/R$ \citep{Yoon2010,Qin2018}.
%It formulation has been widely used in recent detailed studies of rotation in massive stars \citep{de Mink et al. 2009; Song et al. 2013, 2018, Qin2018}.
In our fiducial model, we use $E$ calculated by \cite{Qin2018} as
\begin{equation}\label{eq:E2new}
    E = \begin{cases}
    10^{-0.42}\left(\frac{R_{\rm con}}{R}\right)^{7.5} & {\rm for\ H-rich\ stars} \,, \\
    \\
    10^{-0.93}\left(\frac{R_{\rm con}}{R}\right)^{6.7} & {\rm for\ He-rich\ stars} \,.
    \end{cases}
\end{equation}
We roughly fitted $R_{\rm con}=1\rsun$ and $ 0.5\rsun$ for main sequence stars and the naked herium stars, respectively, using the extremely metal poor ($Z=10^{-8}Z_{\odot}$) star evolution \citep{Tanikawa2019}.

\subsection{Gravitational wave emission}
\label{sec:GW}

The compact binary loses the angular momentum and the binding energy by the GW radiation. 
Based on \cite{Peter_Mathews_1963} and
\cite{1964PhRv..136.1224P}, the equations of the separation and the eccentricity are given by %\cite{Peter_Mathews_1963,1964PhRv..136.1224P},
\begin{equation}
\frac{\dot{a}}{a}=-\frac{64G^3M_1M_2M_{\rm{total}}}{5c^5a^4}\frac{1+\frac{73}{24}e^2+\frac{37}{96}e^4}{(1-e^2)^{7/2}} \,,
\label{eq:semimajor}
\end{equation}
and
\begin{equation}
\frac{\dot{e}}{e}=-\frac{304G^3M_1M_2M_{\rm{total}}}{15c^5a^4}\frac{1+\frac{121}{304}e^2}{(1-e^2)^{5/2}}\label{eq:edot} \,,
\end{equation} 
where $M_{\rm total}=M_1+M_2$.

\subsection{Pop III star formation rate}
In order to calculate the merger rate of Pop III BBH, we need to know when Pop III stars were born and how many Pop III stars were born.
At present, we have not been able to estimate the Pop III star formation rate from observations, yet.
However the Pop III SFR has been estimated by the cosmological simulation.
In our previous study, we use \cite{DeSouza_2011} semi-analytical approach estimate of SFR in which Pop III stars are formed in dark matter halos at their collapse. 

\begin{figure}
\begin{center}
\includegraphics[width=\hsize]{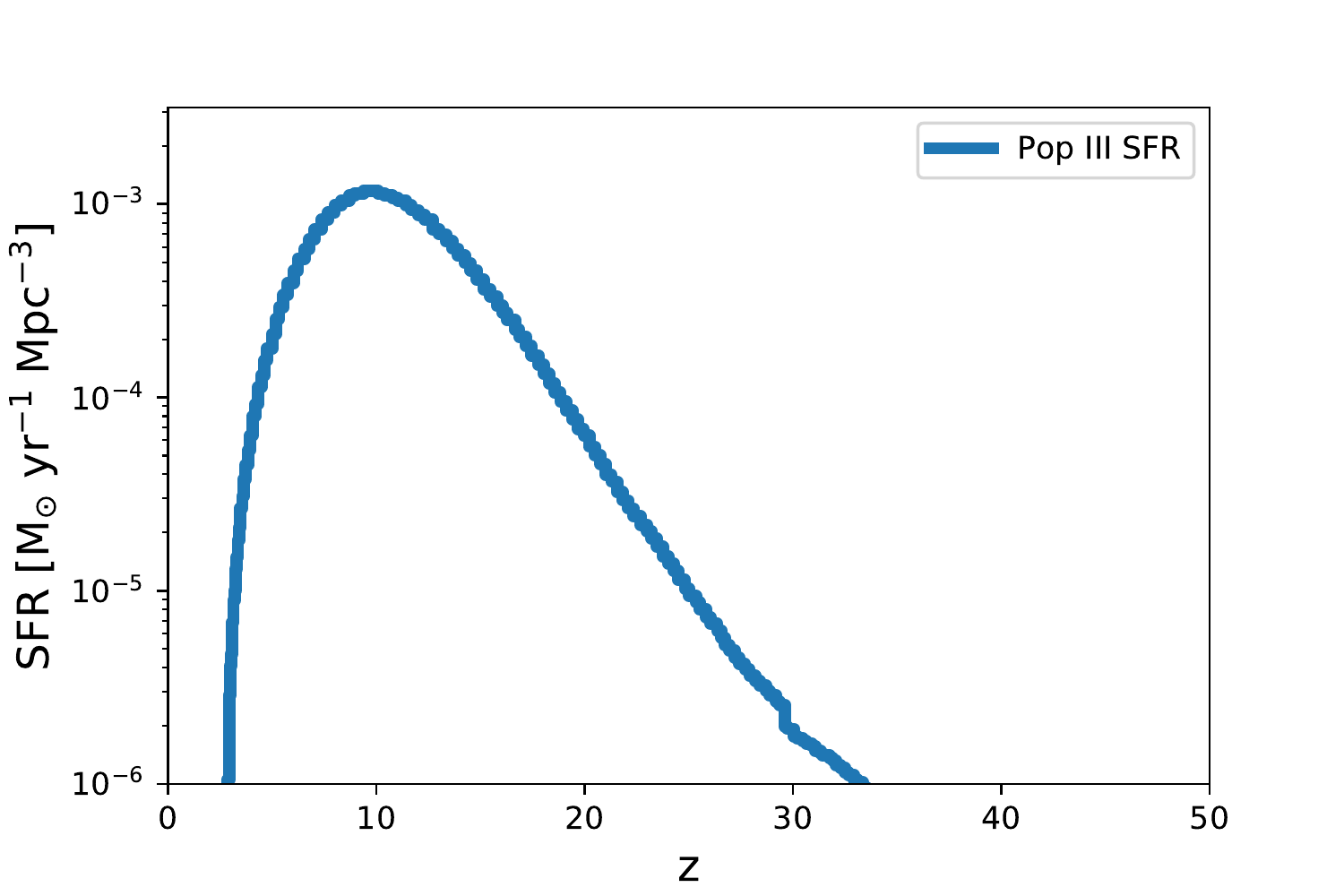}
\end{center}
\caption{Pop III star formation rate. The horizontal axis shows the redshift.}  
\label{sfr}
\end{figure}

Recently, the Thomson scattering optical depth measured by Planck decreases less than that of WMAP \citep{Planck_2016a}.
\cite{Visbal_2015,Inayoshi_2016} have taken into account the constraint of the star formation from the Thomson scattering optical depth $\tau_{\rm e}=0.066+1\sigma$ where $\sigma=0.016$, which is measured by \cite{Planck_2016a}. This change of $\tau_{\rm e}$ yields the new constraint on models of the Pop III SFR, although it strongly depends on some parameters such as the escape fraction of photon, the IMF, and so on.  
\cite{Inayoshi_2016} have shown that the constraint of the total Pop III star formation is $\rho_{*,III}\lesssim6\times10^5~\msun\rm~Mpc^{-3}$ for $\tau_{\rm e}=0.066+1\sigma$,  $f_{\rm esc}=0.1$, and the flat IMF ($10~\msun<M<100~\msun$) {where $f_{\rm esc}$ is the escape fraction of ionizing photons from mini halos}.
The SFR in \citet{DeSouza_2011} is 3 times the limit of \cite{Inayoshi_2016} so that we adopt the modified \citet{DeSouza_2011} SFR decreasing by a factor of 3.
{Figure \ref{sfr} shows the Pop III star formation rate used in this paper.}

\section{Results}

\begin{table*}
	\caption{Formation channels of merging Pop~III BBHs.
	``NoCE'' means that BBHs evolve only via mass transfers, but not via a common envelope phase. 
	``1CE$_P$'', ``1CE$_S$'', and ``1CE$_D$'' mean that BBHs experienced one common envelope phase where
	Subscripts ``P'', ``S'', and ``D'' mean that the common envelope phase is caused by the primary giant, the secondary giant, or double giant stars, respectively.
	``2CE'' means that BBHs experienced two common envelope phases. }
	\label{tab.chan}
		\centering
		\begin{tabular}{c| c c c c c}
		\hline
		model & NoCE & 1CE$_P$ & 1CE$_S$ & 1CE$_D$ & 2CE\\
		\hline
		Fiducial&    8747      (15.77\%)&  25802      (46.52\%) &    563       (1.02\%) & 16009      (28.86\%) &   4346       (7.84\%)

 \\ \hline
	
		M100&    10480      (21.23\%)&  18520      (37.52\%)&    870       (1.76\%) & 12631      (25.59\%) &   6859      (13.90\%)

 \\		\hline
		$\beta$=0.5&    10992      (21.33\%)&  25361      (49.21\%) &   125       (0.24\%)&  12182      (23.64\%)&   2872       (5.57\%)

\\ 
		\hline
		$\alpha\lambda$=0.1 &  8608      (18.48\%)&  33515      (71.94\%)&   1960       (4.21\%)&   2507       (5.38\%)&      0       (0.00\%)

\\
		\hline
		K14		&   52498      (46.28\%)&  23527      (20.74\%)&   3573       (3.15\%)&  14808      (13.05\%) & 19038      (16.78\%)

 \\
		\hline
		FS1&      3362       (2.72\%)&  99291      (80.38\%)&   7264       (5.88\%)&  13607      (11.02\%)&      0       (0.00\%)

 \\  \hline
		FS2&   14581      (99.84\%)&      0       (0.00\%)&     24       (0.16\%)&      0       (0.00\%)&      0       (0.00\%)

\\
		\hline
		
	\end{tabular}
	
\end{table*}

\subsection{{Formation} channels}\label{sec:ECs}
%\tn{Notice that from now on,  the word "BBH" sometimes includes the progenitor binary stars of BBH.}
Table \ref{tab.chan} shows the formation channels of merging BBHs within the Hubble time.
``NoCE'' means that progenitors of BBHs evolve only via mass transfers, but not via any common envelope phase. 
``1CE$_P$'', ``1CE$_S$'', and ``1CE$_D$'' mean that progenitors of BBHs experienced one common envelope phase where
the subscripts ``P'', ``S'', and ``D'' mean that the common envelope phase is caused by the primary giant, the secondary giant, or double giant stars, respectively.
``2CE'' means that progenitors of BBHs experienced two common envelope phases.

For our fiducial model, the main channel is the 1CE$_P$ channel, in which progenitors of BBHs evolves via a primary common envelope phase.
The second channel is the 1CE$_D$ channel.
In this channel, the binary after a double common envelope phase becomes double naked helium star.
The sum of two naked stellar radii is so small that Pop III binaries of this channel can become a closer binary than the other channels.
$\sim80\%$ of merging Pop III BBH progenitors evolve via 1 or 2 common envelope phases in our fiducial model.
However, $\sim15\%$ of the merging Pop III BBH progenitors evolve via the NoCE channel meaning that they evolve only via a mass transfer.
These NoCE merging BBH progenitors do not depend on the common envelope parameter  $\alpha\lambda$.

The fractions of formation channels in Table \ref{tab.chan} depend on the initial binary parameters and the model of binary interactions. 
Low mass progenitors ($M<50~\msun$) of Pop III BBHs tend to evolve only via mass transfer so that the fraction of NoCE for the M100 model is more than that of the fiducial model.
The fraction of NoCE of the $\beta$=0.5 model is also more than that of the fiducial model, because the mass loss during  mass transfer makes the separation  wide,
and the binary will be slightly harder to experience a common envelope phase than the fiducial model (Eq. \eqref{eq.stMT}).

In the fiducial model in Table \ref{tab.chan}, the fraction of NoCE is much less than that of the K14 model because the mass transfer rate of the fiducial models (Eq. \eqref{eq:MTrate1}) is much higher than that of K14 (Eq. \eqref{eq:m1dotf}), which is our previous model.
If the mass transfer rate is high and the accretor is a black hole, the separations in later evolution phases tend to be large due to the mass loss from the binary system so that {the binary stars are} hard to merge.

{In the cases of the $\alpha\lambda$=0.1, FS1, and FS2 models, no 2CE channel exists. For the former case, the small common envelope parameter of $\alpha\lambda=0.1$ makes the binaries easier to merge during the common envelope phase.}
In the case of the FS1 model, almost all BBH progenitors evolve via a common envelope phase.
{This is due to the IMF of FS1 model.}
In this case, the typical mass of Pop III stars is so massive $\sim100~\msun$ that they tend to evolve via a red giant {phase to a common envelope phase.}
In the case of the FS2 model, almost all merging BBH progenitors evolve via the NoCE channel. This is due to the initial separation. 
The typical separation of the FS2 model is $90~\rsun$
which is too close to evolve for BBH progenitors via a common envelope phase.
Almost all binaries which have common envelope phases tend to merge without forming BBH systems.

\subsection{BBH chirp mass distribution}\label{cmass}
\begin{figure}
\begin{center}
\includegraphics[width=\hsize]{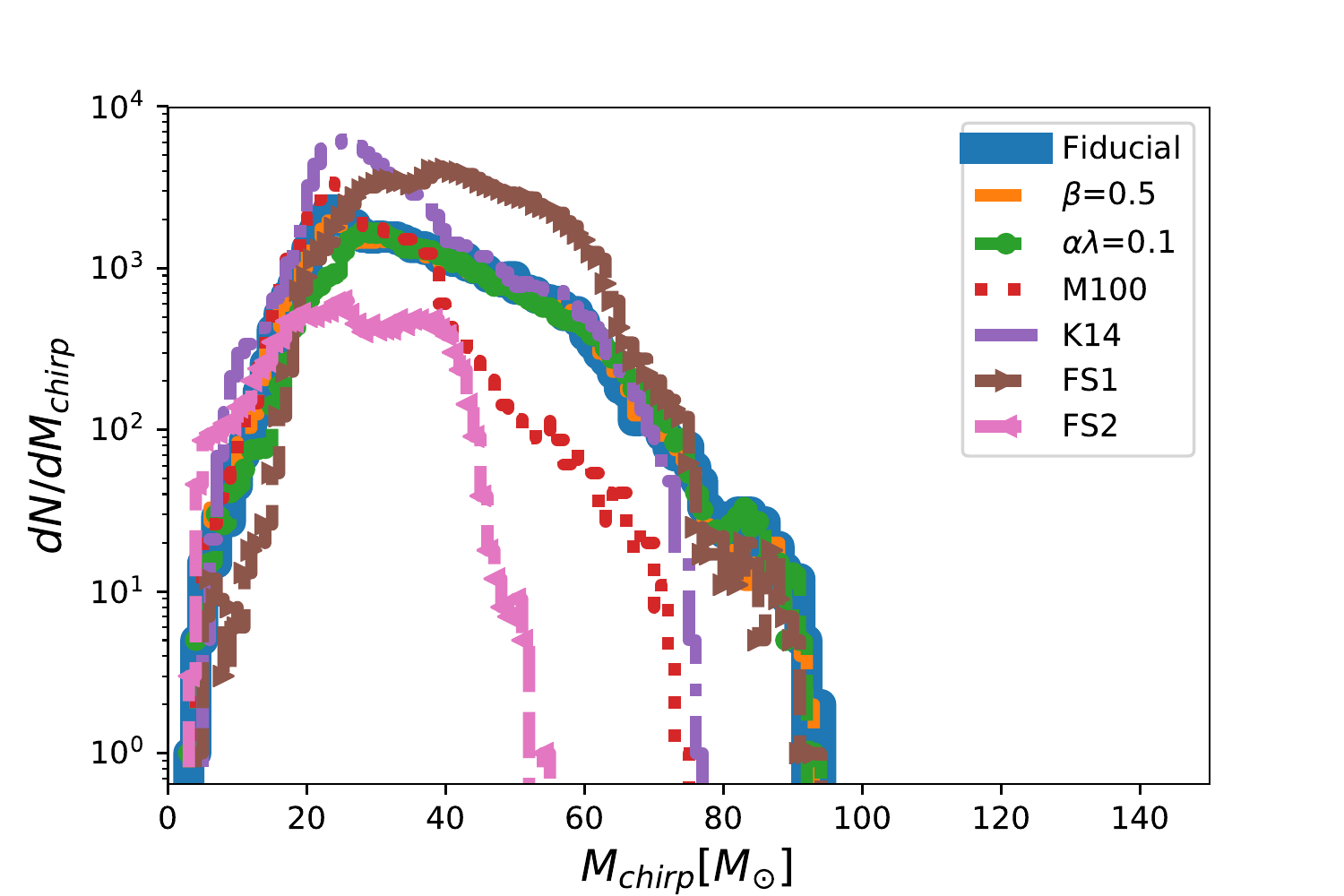}
\end{center}
\caption{Chirp mass distribution of BBHs.
Our fiducial model is denoted by the thick blue line.
The $\beta=0.5$ and $\alpha\lambda=0.1$ models
are shown by the thin orange (dashed) and green (with filled circles) lines, respectively. Notice that these lines almost overlap with that of the fiducial model.
The red dotted line is for the M100 model.
The K14 model is denoted by the purple dash-dotted line.
The FS1 and FS2 models are shown by 
the brown and the pink dashed lines
with the right and left-pointing triangles, respectively.
The detailed description of each model is summarized in Sec.~\ref{sec:IandC} with Tables~\ref{tab.init} and~\ref{tab.bin}.}  
\label{mass}
\end{figure}

Figure \ref{mass}  shows the chirp mass
($M_{\rm chirp}=(M_1 M_2)^{3/5}/M_{\rm total}^{1/5}$) distribution  of merging Pop III BBHs for various models. We first notice that
the peak chirp mass of merging Pop III BBHs is more or less $\sim30~\msun$ for all models, {which} does not depend on the initial conditions and the binary evolution parameters so much.
The reason for this tendency comes from the characteristic of  the evolution pass of Pop III stars.
Pop III stars with mass $M<50~\msun$ do not pass through  the common envelope phase so much.
They typically evolve via stable RLOF phases and their mass loss is smaller than that in the evolution passes via a common envelope phase.
They tend to lose $1/10$--$1/3$ of their mass so that their chirp mass tend to be $\sim 30~\msun$. 
Pop III stars with $M > 50~\msun$ are likely to have the common envelope phase, and they lose $1/2$--$2/3$ of their mass so that they also tend to be $\sim 30~\msun$ BHs.
These two BBH formation channels lead the peak value of chirp mass as $\sim30~\msun$.
{In our previous study \citep{Kinugawa2016} and this paper, models with various initial parameter distribution functions and binary prescription models are calculated. However, the results show that the peak value of chirp mass is similar $\sim30~\msun$ for each model.
Thus, even if the initial conditions or binary evolution parameters change or the one channel is inactive, the remaining channels seem to keep the peak value of chirp mass at $\sim30~\msun$.}
Therefore, the typical chirp mass of Pop III BBHs is almost independent of the initial conditions and the binary parameter uncertainties.

However, the maximum possible mass of Pop III BBHs depends on the initial parameters and binary interaction models.
For  M100, K14 and FS2 models, the maximum chirp masses of BBH are smaller than the other models, {although they can make massive BBHs whose total masses are more than $100~\msun$}. 
In the cases of M100 and K14 models, the maximum initial mass is smaller than that of the other models, so that the maximum mass of Pop III BBHs reflects this difference. 
On the other hand, in the case of FS2 model, the value of the maximum mass is determined by the evolution channel since 
almost all of merging Pop III BBH progenitors in the FS2 model evolve via the NoCE channel.
These progenitors tend to be a blue giant whose mass is $\lesssim50~\msun$, so that the maximum mass of BBHs is much less than that of the other models.

%\begin{figure}
%\begin{center}
%\includegraphics[width=\hsize]{BBCtotalmass.pdf}
%\end{center}
%\caption{\tk{Total mass distributions of BBHs.
%The legends are the same as Fig.~\ref{mass}.}}  
%\label{total mass}
%\end{figure}

\subsection{Merger time distribution}\label{sec:mergertime}
\begin{figure}
\begin{center}
\includegraphics[width=\hsize]{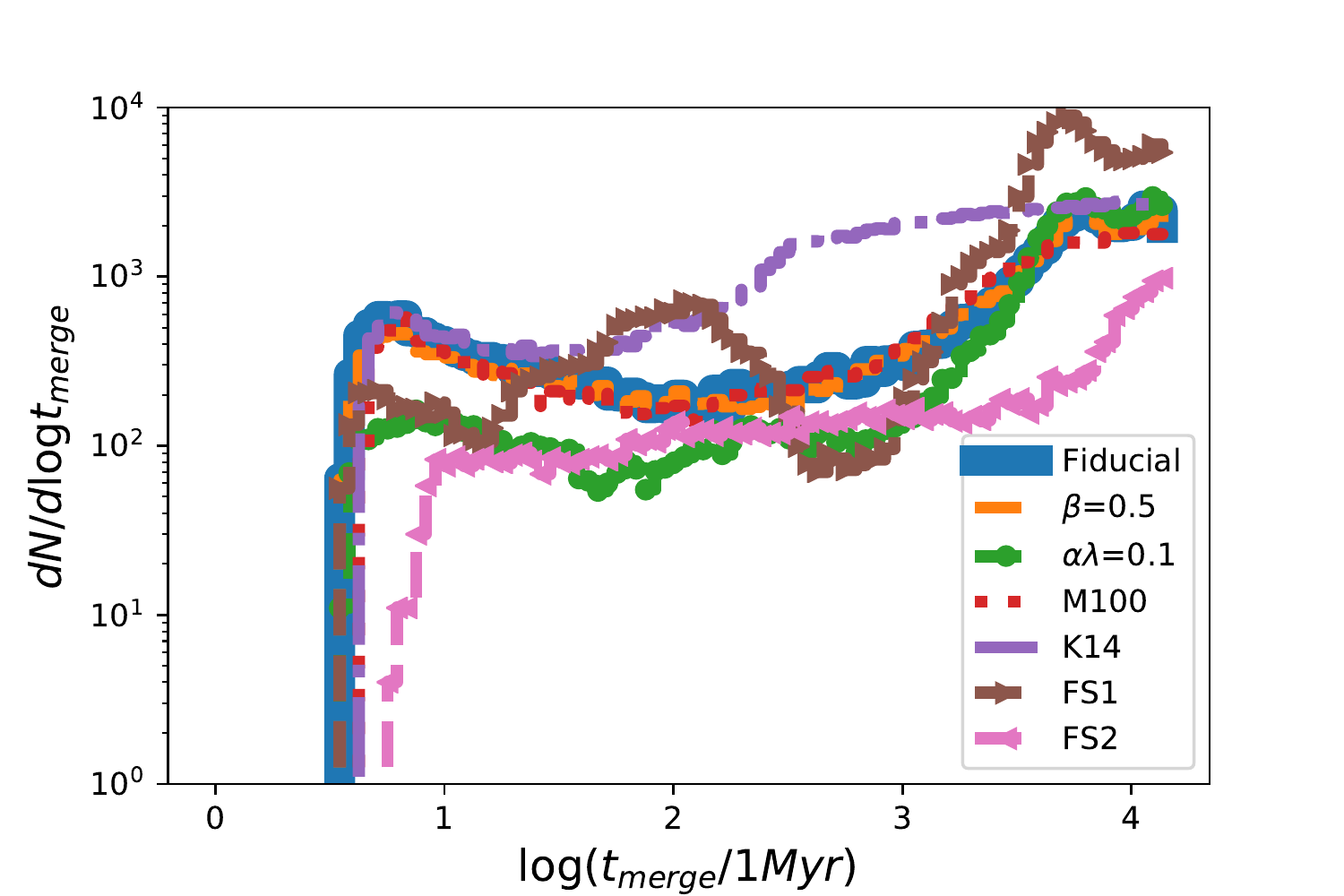}
\end{center}
\caption{Merger time distribution of merging BBHs.
The legends are the same as Fig.~\ref{mass}.}  
\label{mergertime}
\end{figure}

\begin{figure*}
\begin{tabular}{cc}
      %---- 最初の図 ---------------------------
      \begin{minipage}[t]{0.4\hsize}
\includegraphics[width=\hsize]{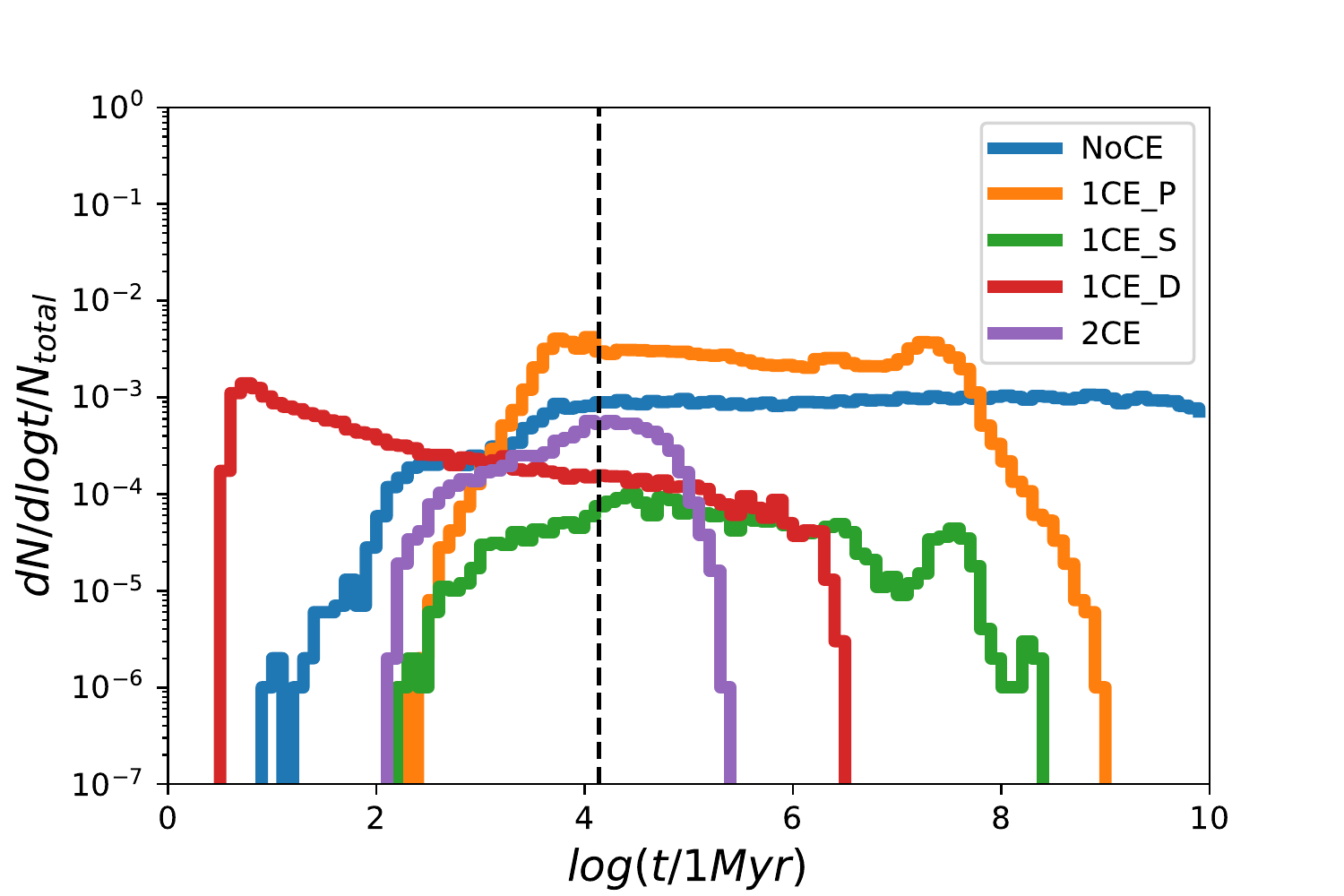}
  \subcaption{fiducial model}
\label{mergertime_fidutial}
\end{minipage} 
      \begin{minipage}[t]{0.4\hsize}
        \centering
\includegraphics[width=\hsize]{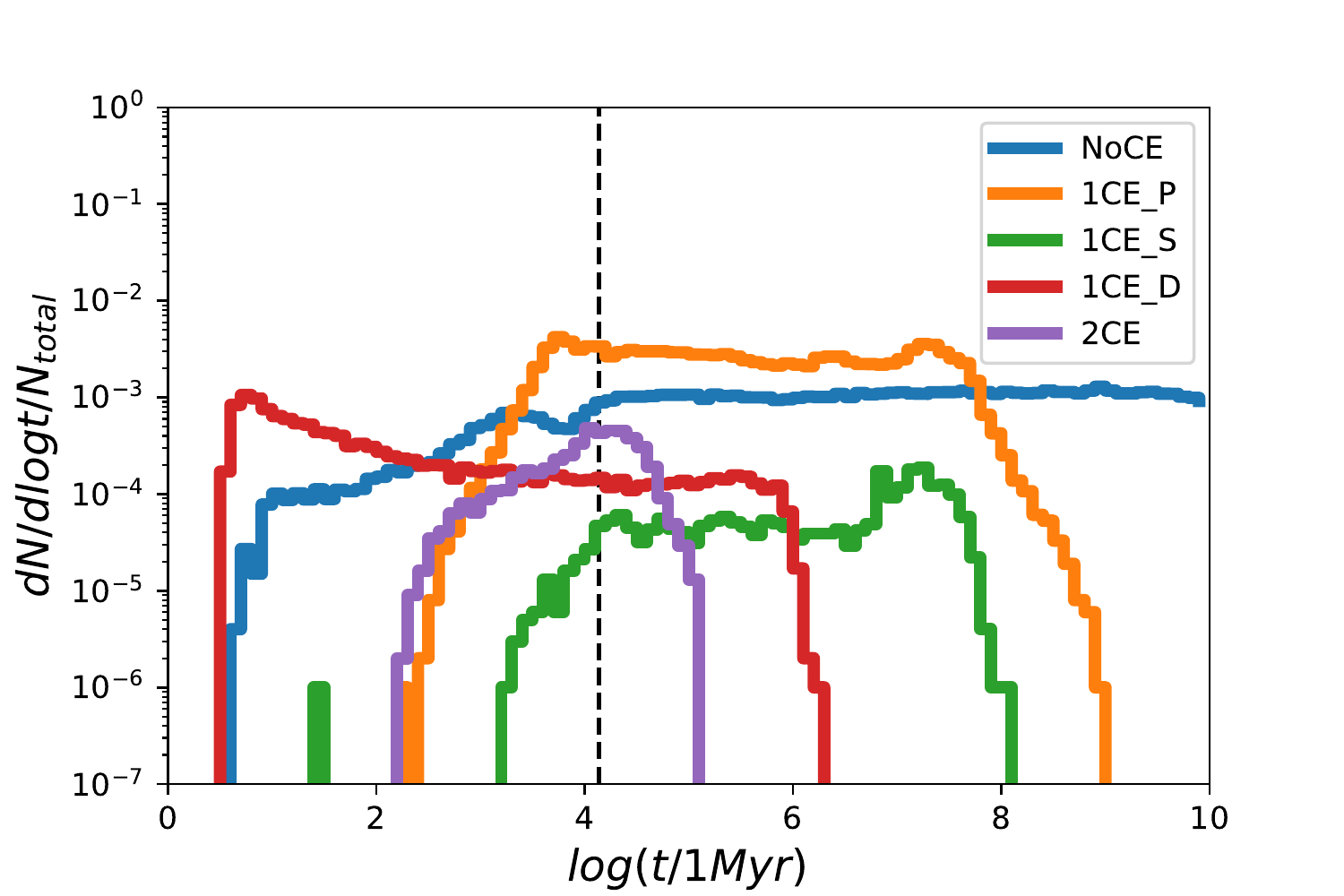}
\subcaption{$\beta$=0.5 model}  
\label{mergertime_MT05}
\end{minipage} \\
      \begin{minipage}[t]{0.4\hsize}
        \centering
\includegraphics[width=\hsize]{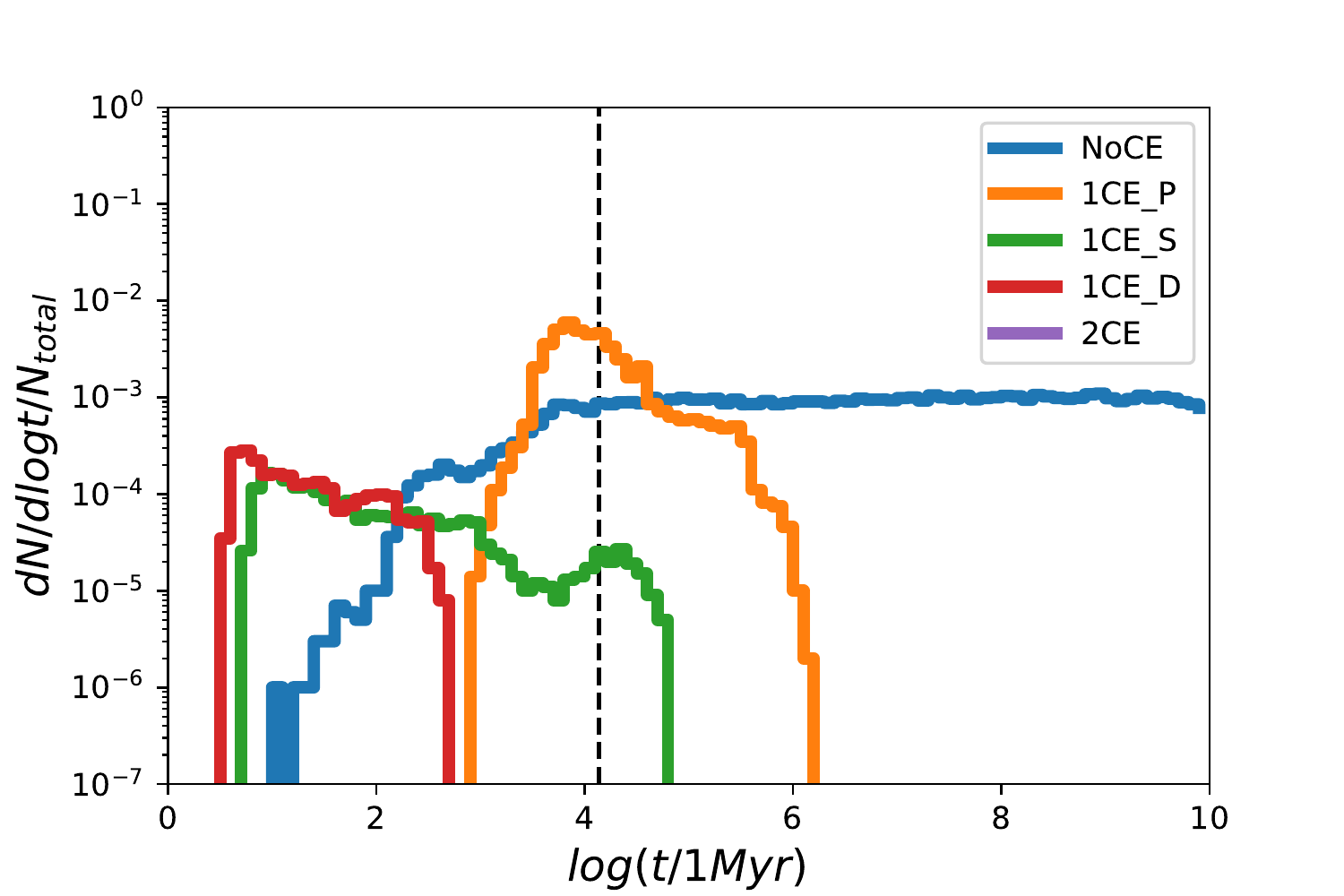}
\subcaption{$\alpha\lambda$=0.1 model}  
\label{mergertime_al01}
\end{minipage} 
      \begin{minipage}[t]{0.4\hsize}
        \centering
\includegraphics[width=\hsize]{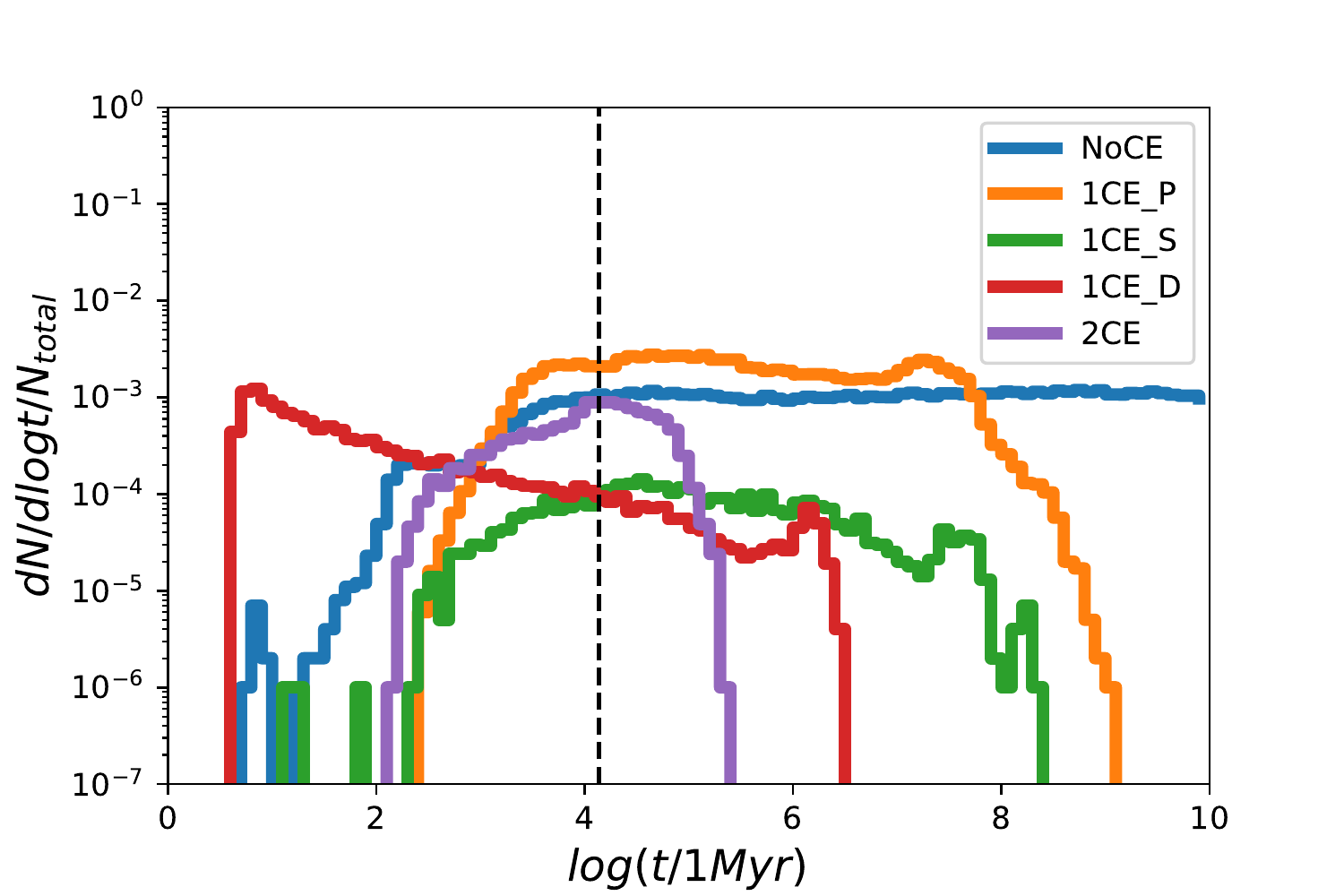}
\subcaption{M100 model}  
\label{mergertime_M100}
\end{minipage} \\
      \begin{minipage}[t]{0.4\hsize}
        \centering
\includegraphics[width=\hsize]{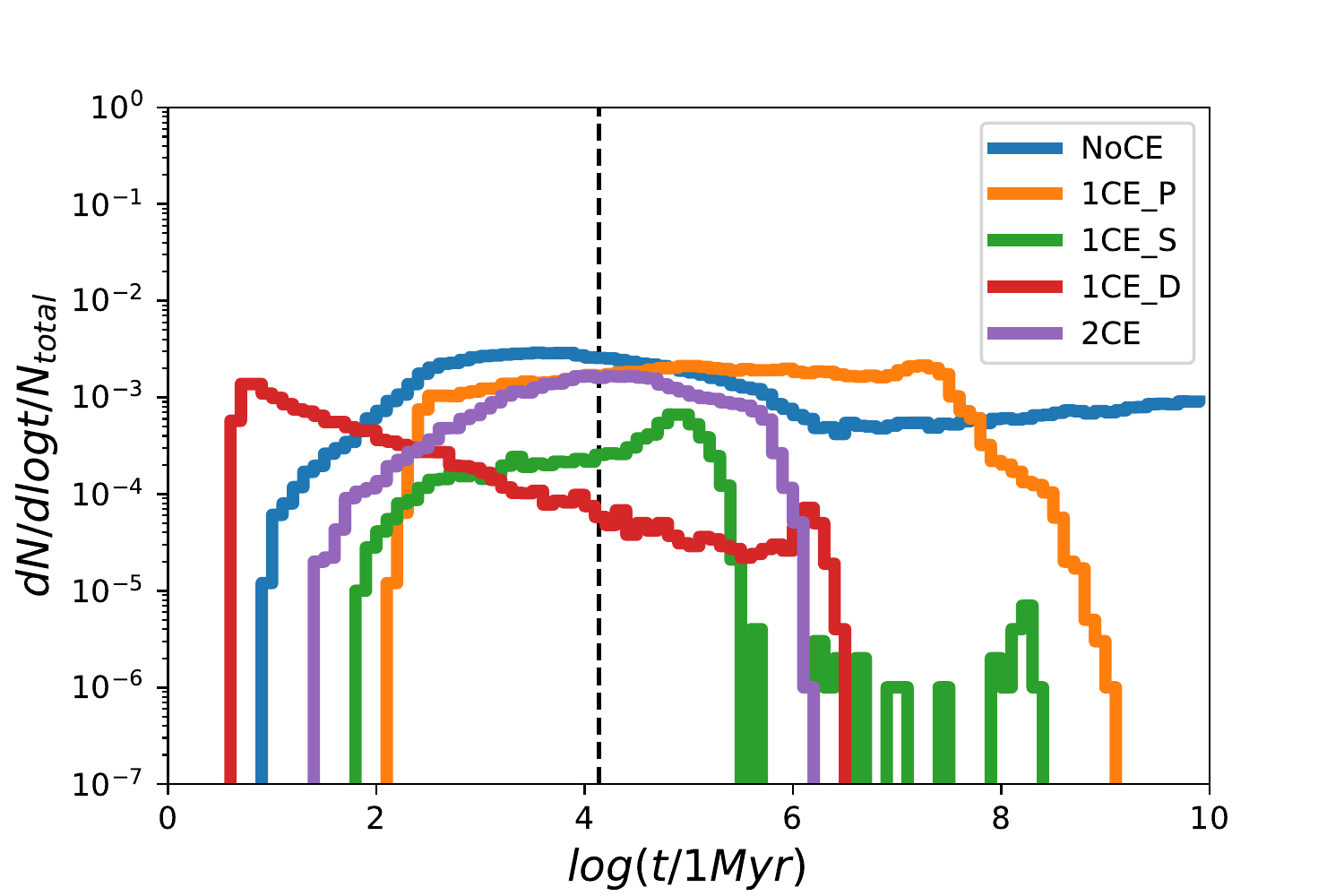}
\subcaption{K14 model}  
\label{mergertime_K14}
\end{minipage} 
      \begin{minipage}[t]{0.4\hsize}
        \centering
\includegraphics[width=\hsize]{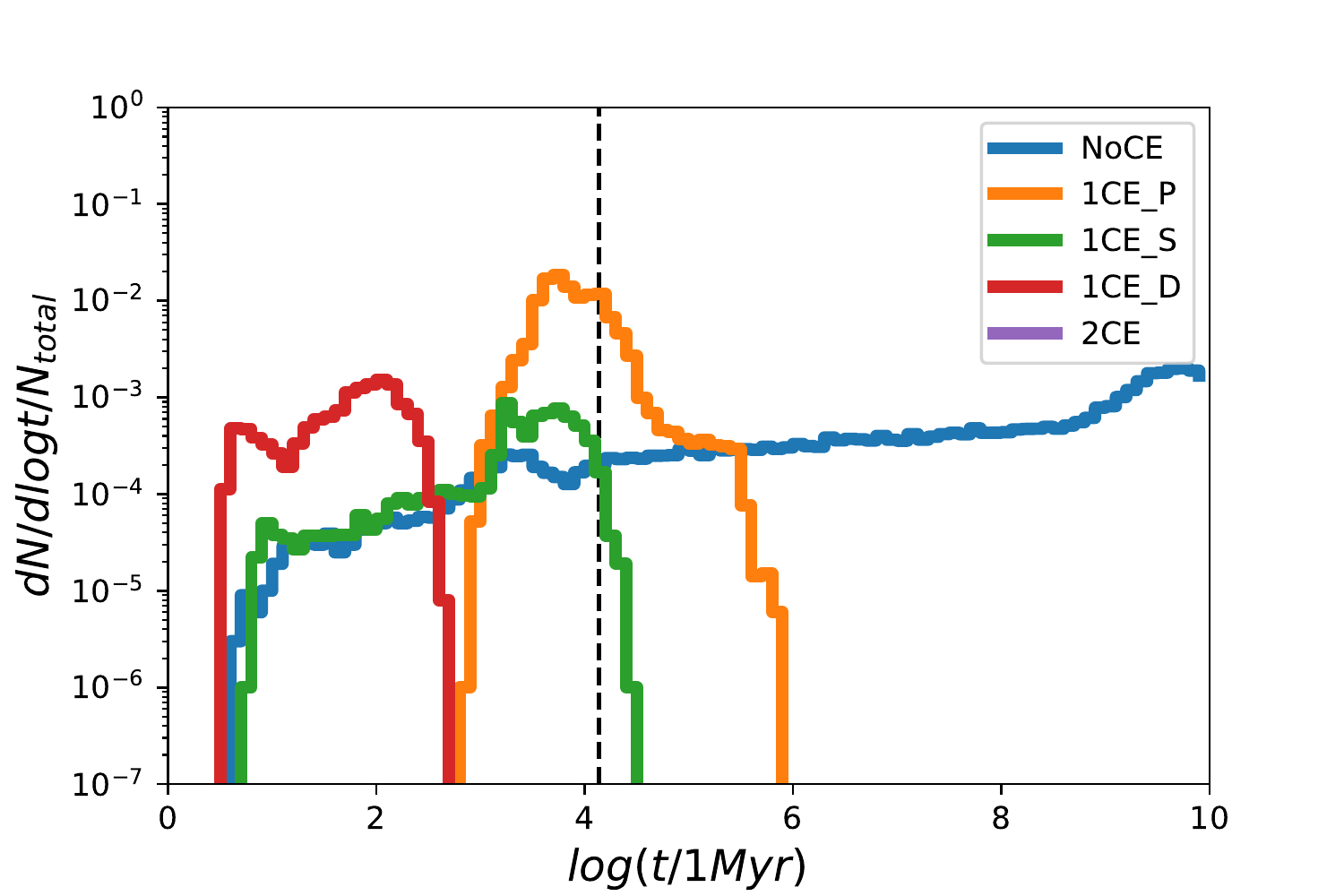}
\subcaption{FS1 model}  
\label{mergertime_FS1}
\end{minipage} \\
      \begin{minipage}[t]{0.4\hsize}
        \centering
\includegraphics[width=\hsize]{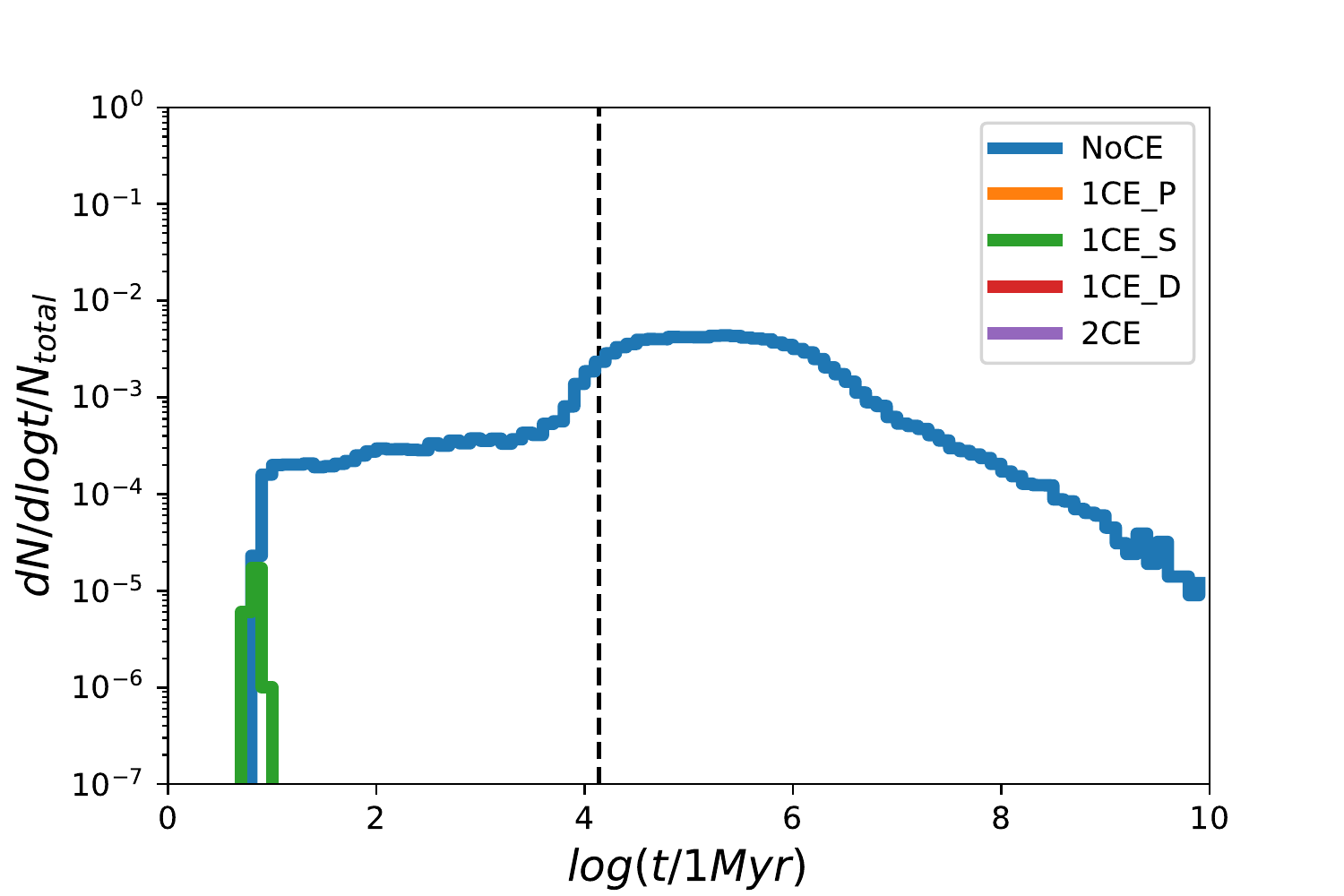}
\subcaption{FS2 model.} 
\label{mergertime_FS2}
\end{minipage}
      %---- 図はここまで ----------------------
    \end{tabular}
    \caption{Merger time distribution of BBHs for each model. 
Note that the horizontal axis shows the merger time as ($\log(t_{\rm merge}/1\,{\rm Myr})$).
$t=0$ means the time of the BBH formation, but not the time of the Big Bang.
The black dashed vertical line means the Hubble time.
The blue, orange, green, red and purple lines
show the NoCE, 1CE$_P$, 1CE$_S$, 1CE$_D$ and 2CE channels,
respectively. The detailed definitions of the channels
are described in Sec.~\ref{sec:ECs} and Table~\ref{tab.chan}.}
\label{mergertime_channel}
\end{figure*}

Figure \ref{mergertime} shows the merger time distribution of merging Pop III BBHs for each model, and Figure \ref{mergertime_channel} shows the Pop III BBH merger time distribution of each channel for each model.
$t=0$ means the time of the BBH formation.
{The horizontal axis is the merger time ($\log(t_{\rm merge}/1~{\rm Myr})$), 
while the vertical axis is the number of merging BBHs per unit logarithmic time.
The black dashed vertical line shows the merger time equal to the Hubble time.
In order to show the channel dependence of the merger time, we describe the merger time from 1 Myr to $10^{10}$ Myr. Please do not regard ``10'' in the horizontal axis as the Hubble time but ``10'' means $10^{16}$ yr.
The purpose to argue the merger rate distribution at far future is to show the physical mechanism more clearly. }

From Fig.~\ref{mergertime}, we see that distribution functions of merging time for almost all models are roughly logflat, which reflect the initial distribution function of separation. However, in more details for almost all models, the number of merging BBHs is increasing for $t_{\rm merge}>10^3~{\rm Myr}$ because the 1CE$_P$ channel becomes effective from $t_{\rm merge} \sim 10^3$~Myr, which can be confirmed from the following argument. 

Let us notice  Fig. \ref{mergertime_fidutial}. In the case of the 1CE$_P$ channel (the orange line), Pop III binaries evolve via a common envelope phase with the primary giant and the secondary main sequence star. Since the typical radius of Pop III main sequence stars is $\sim10~\rsun$, the separation just after the common envelope phase has to be about twice more than the main sequence radius to avoid a stellar merger before the binary becomes a BBH.
The merger time by the gravitational radiation is described by Eq. \eqref{eq:semimajor} as $10^3\,$Myr$~(a/20\,\rsun)^4(M_1/30\,\msun)^{-1}(M_2/30\,\msun)^{-1}$ $(M_{\rm total}/60\,\msun)^{-1}$. 
Thus, typical merger time of 1CE$_P$ is more than $10^3$~Myr. This is the reason for the starting time in the increase of the merger rate after $t \sim 10^3$~Myr.

The cases of $\beta=0.5$ model (Fig. \ref{mergertime_MT05}) and M100 model (Fig. \ref{mergertime_M100}) are almost the same as the fiducial model, although the number of BBHs from the NoCE channel is slightly larger than that of the fiducial model.
In the $\beta=0.5$ model, the mass transfer is much more stable than that of fiducial model due to the mass loss of transferred material (Eq. \eqref{eq.stMT}).
In the M100 model, since the maximum mass of initial mass distribution is smaller than that of the fiducial model, the mass transfer is much more stable than that of the fiducial model, too.

In the $\alpha\lambda$=0.1 (Fig. \ref{mergertime_al01}) and FS1 (Fig. \ref{mergertime_FS1}) models, they have a large peak before $10^4$~Myr.
Small common envelope parameter makes the separation just after the common envelope  small so that   the number of stellar mergers during common envelope increases.
Thus, separations of BBH progenitors evolved via a common envelope phase tends to be much more shrunken than the other model so that the fraction of binaries with long merger time decreases.
On the other hand, the number of binaries whose merger time is smaller than $\sim10^3$ Myr, is small because of the same reason as the fiducial model.
Therefore, the Pop III BBHs mergers from the 1CE$_P$ channel are localized at $t_{\rm merge}\sim10^3-10^4$~Myr (Figs. \ref{mergertime_al01} and \ref{mergertime_FS1}).

In the case of the K14 model of Fig. \ref{mergertime_K14}, the number of BBHs in the NoCE channel whose merger time is less than $10^6$~Myr, is more than those of the other models.
The reason is that the mass transfer rate of the K14 model is much less than that of the other models.
In the case of the mass transfer of BH binaries, if the mass transfer rate is small, the binaries are not easier to become wide due to the small mass loss from the binary systems.
{Thus, BBHs made by the NoCE channel are easier to merge in a small merger time than those of the other models (Fig. \ref{mergertime_K14}).}

In the case of FS2 model of Fig. \ref{mergertime_FS2}, there is a peak at $t\sim10^4$--$10^6$ Myr.
It reflects the Gaussian peak of initial separation distribution.

\subsection{Merger rate}

\begin{figure}
\begin{center}
\includegraphics[width=\hsize]{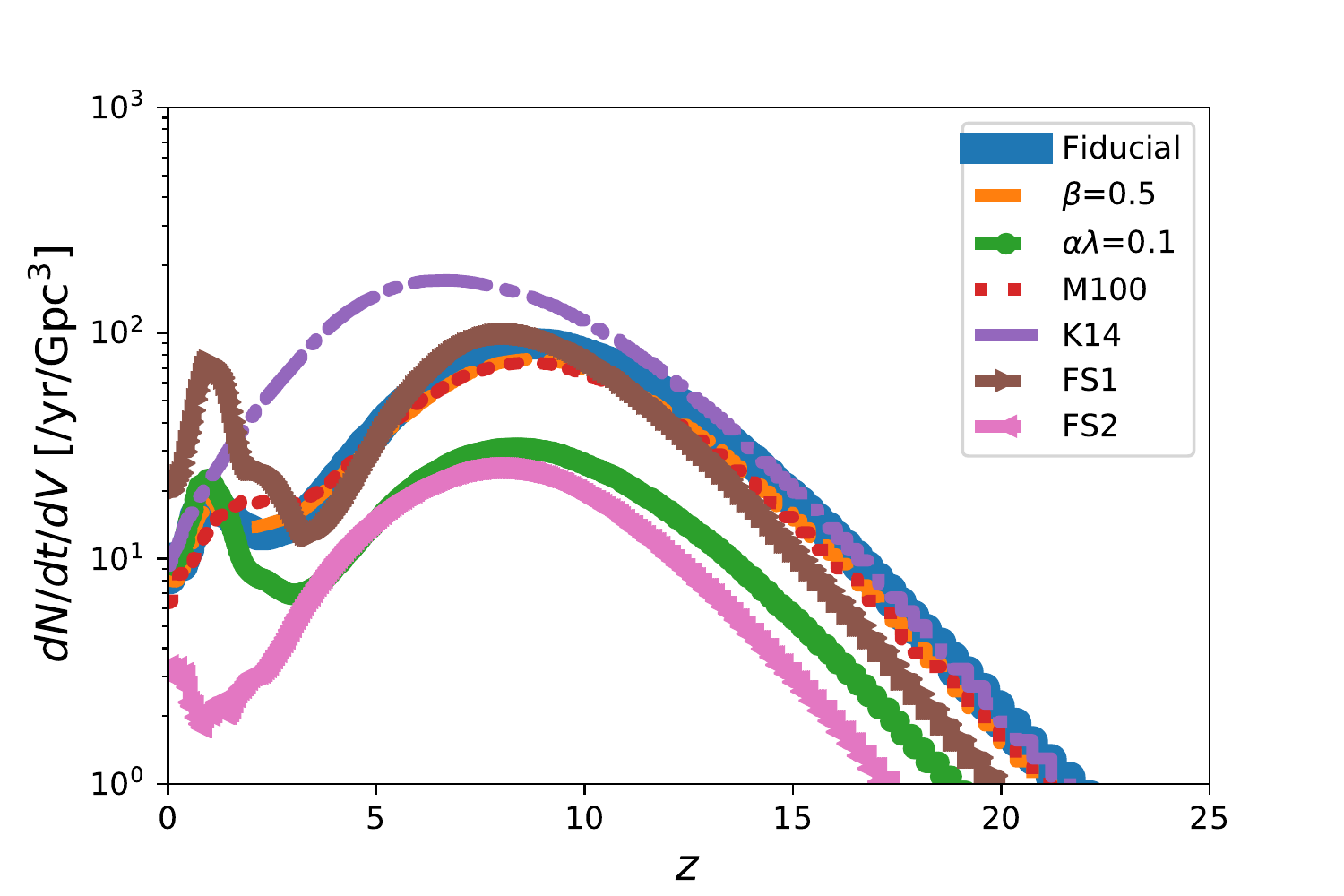}
\end{center}
\caption{Merger rate densities of Pop III BBHs. 
The horizontal axis shows the redshift.
The legends are the same as Fig.~\ref{mass}.}  
\label{mergerratez}
\end{figure}

Figure \ref{mergerratez} and Table \ref{tab.rate} show
that the merger rate densities for each model.
The merger rate at $z=0$ is 3.34--21.2 $\rm yr^{-1}~Gpc^{-3}$.
This value is consistent with the lower bound of aLIGO/aVIRGO result 9.7--101~$\rm yr^{-1}~Gpc^{-3}$ \citep{Abbott_2019}.

The merger rate densities have a major peak at $z\sim10$ for all models.
This peak reflects the peak of the Pop III SFR.
Except for the K14 and FS2 models, the merger rate densities have a second peak at $z\sim1$.
This peak is made by the contribution of the 1CE$_P$ channel (see Figs. \ref{mergertime_fidutial}, \ref{mergertime_MT05},
\ref{mergertime_al01}, \ref{mergertime_M100} and \ref{mergertime_FS1}).
Especially, for the $\alpha\lambda$=0.1 and FS1 models, the contribution of the 1CE$_P$ channel at $z\sim1$ is so high that the peak of merger rate density at $z\sim1$ is higher than the other models.
In the case of K14 model, the NoCE channel is more effective than the other models, so that the effect of the 1CE$_P$ channel is hidden at the low redshift region (see Fig. \ref{mergertime_K14}). 
In the case of FS2 model shown in Fig. \ref{mergertime_FS2},
they have no peak at $z\sim1$ because there is no BBH from the 1CE$_P$ channel.
But, the merger rate density decreases from $z\sim0$ to $z\sim1$ by the merging BBH of the NoCE channel due to the initial separation distribution.

\begin{table}
	\caption{Merger rate densities of merging Pop~III BBHs [$\rm yr^{-1}~Gpc^{-3}$]. We have picked the numerical values at some redshifts from Fig.~\ref{mergerratez}.}
	\label{tab.rate}
		\centering
		\begin{tabular}{c| c c c c c c}
		\hline
		model & z=0 & z=0.1&z=1 & z=5 & z=10 & z=20\\
		\hline
		Fiducial&8.13 &  10.1 & 19.0   & 36.7 &   80.7 & 1.86

 \\ \hline

		$\beta$=0.5&  8.09    & 7.94 &  18.0 &33.0  &  68.6 & 1.52

\\ 
		\hline
		$\alpha\lambda$=0.1 & 10.3 & 9.68 & 22.2  & 14.5 & 26.2&  0.545  

\\
		\hline
		
    	M100&  6.36  &  6.96&  13.8  & 34.0 &   65.6 & 1.45

 \\		\hline
		K14		& 9.30 &10.5&22.6 &  147 & 112 & 1.89

 \\
		\hline
		FS1&  21.2  & 19.8 & 75.2 & 32.1 & 78.4 &   0.901

 \\  \hline
		FS2& 3.34 & 3.16     &  2.02 & 15.2  &  19.0  &0.233 
\\
		\hline
		
	\end{tabular}
	
\end{table}

\subsection{Spin distribution}\label{sec.spin}
{Table \ref{tab.spin} shows  the averages of the effective spin at each redshift for each model.
This table shows that Pop III low spin BBHs are easy to merge at low redshift, and Pop III {high spin} BBHs tend to merge at high redshift.
Almost all BBHs are born at $z\sim10$ and Pop III BBHs whose merger time is short can merge near $z\sim 10$.
Progenitors of BBHs with short merger time are so close that they are {easily spun up by the tidal interaction.  Table \ref{tab.spin} shows that except for K14, the mean effective spin $\left\langle \chi_{\rm eff} \right\rangle$ at $z=0$ ranges
$0.02$--$0.3$ while at $z=10$ it does $0.16$--$0.64$. While in K14 model the average value of the effective spin ($\left\langle \chi_{\rm eff} \right\rangle$) is almost constant and varying only from 0.57 (z=0) to 0.69 (z=10)}. 
}

\begin{figure*}
\begin{tabular}{cc}
      %---- 最初の図 ---------------------------
      \begin{minipage}[t]{0.4\hsize}
\includegraphics[width=\hsize]{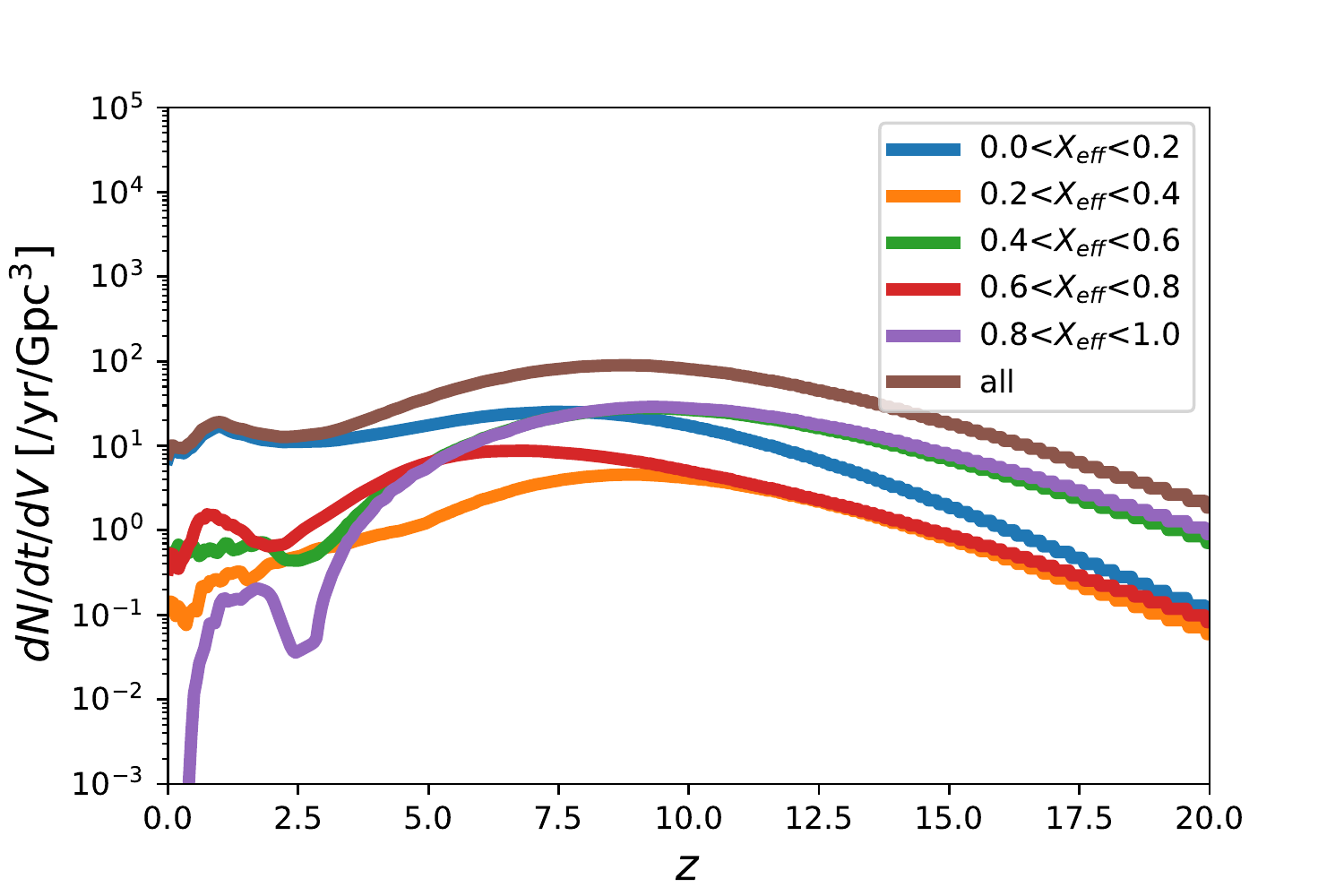}
  \subcaption{fiducial model}
\label{spinz_fidutial}
\end{minipage} 
      \begin{minipage}[t]{0.4\hsize}
        \centering
\includegraphics[width=\hsize]{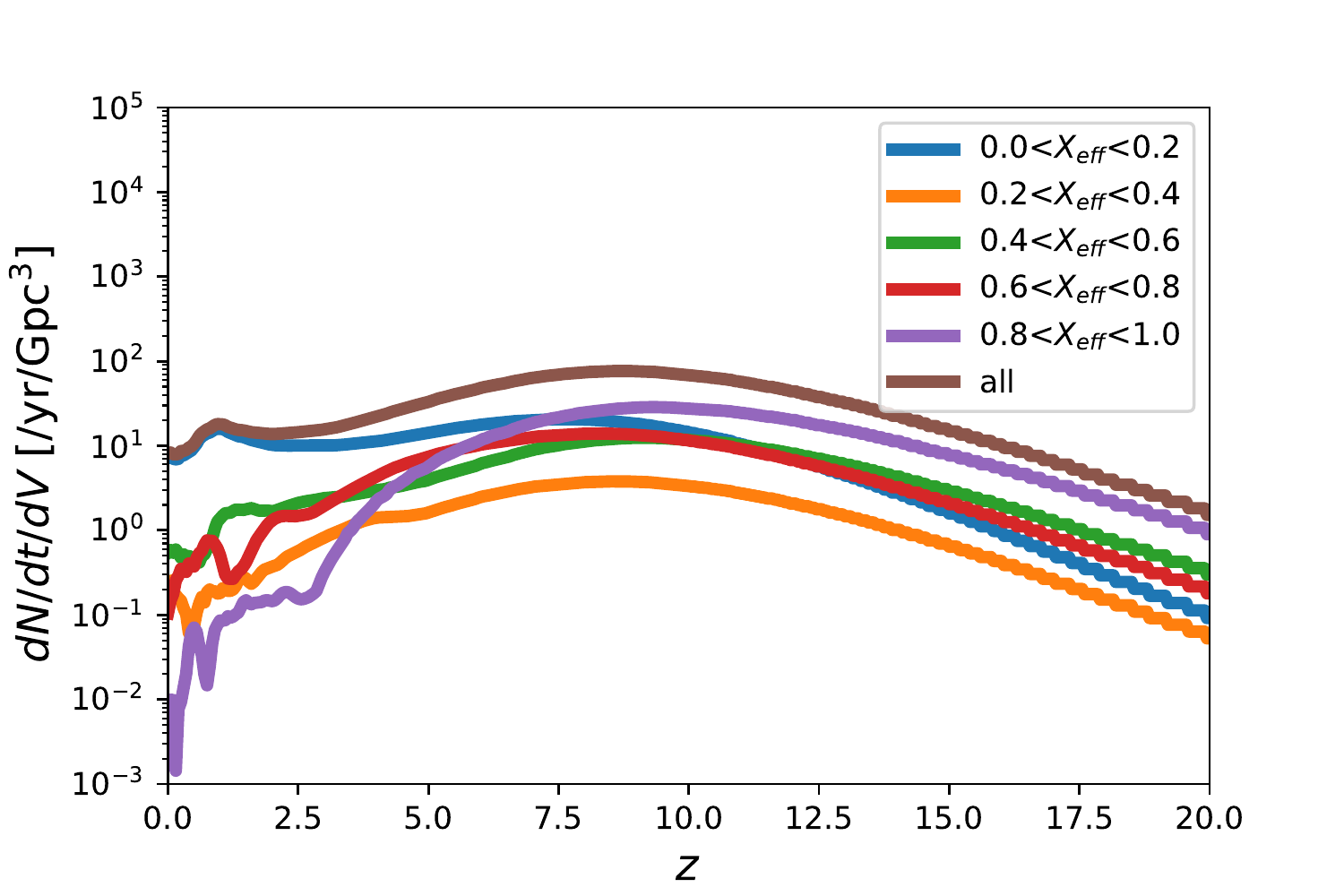}
\subcaption{$\beta$=0.5 model}  
\label{spinz_MT05}
\end{minipage} \\
      \begin{minipage}[t]{0.4\hsize}
        \centering
\includegraphics[width=\hsize]{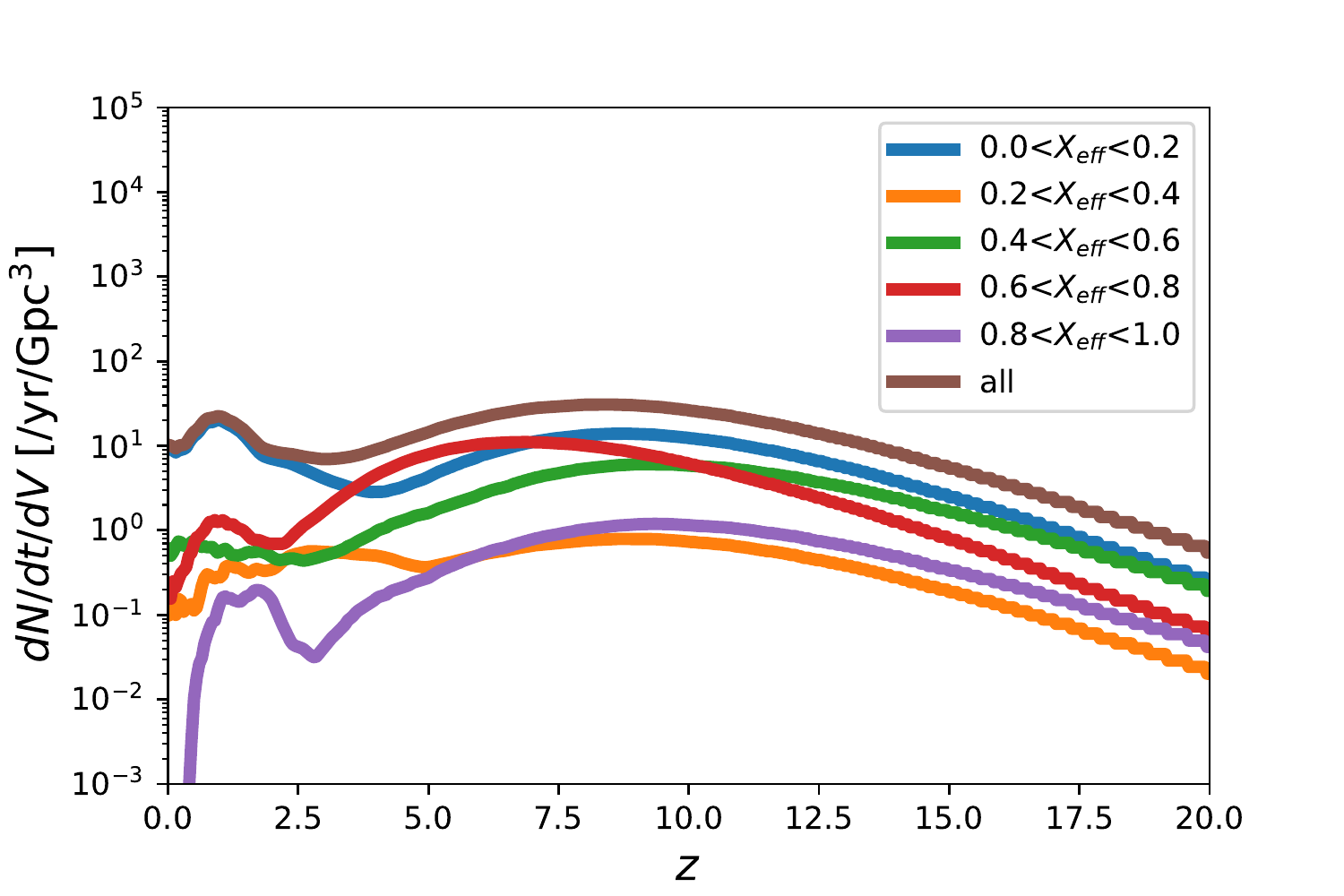}
\subcaption{$\alpha\lambda$=0.1 model}  
\label{spinz_al01}
\end{minipage} 
      \begin{minipage}[t]{0.4\hsize}
        \centering
\includegraphics[width=\hsize]{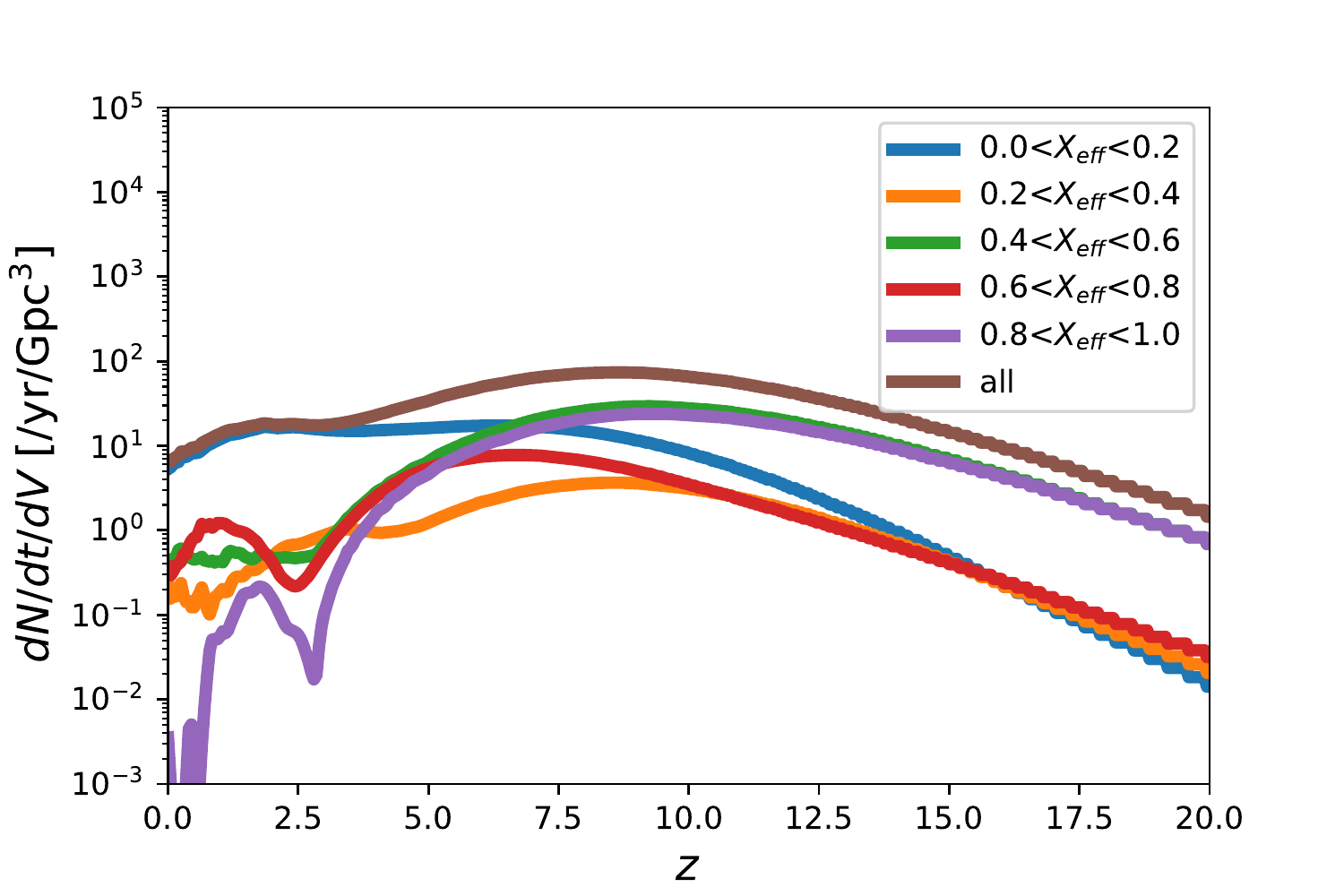}
\subcaption{M100 model}  
\label{spinz_M100}
\end{minipage} \\
      \begin{minipage}[t]{0.4\hsize}
        \centering
\includegraphics[width=\hsize]{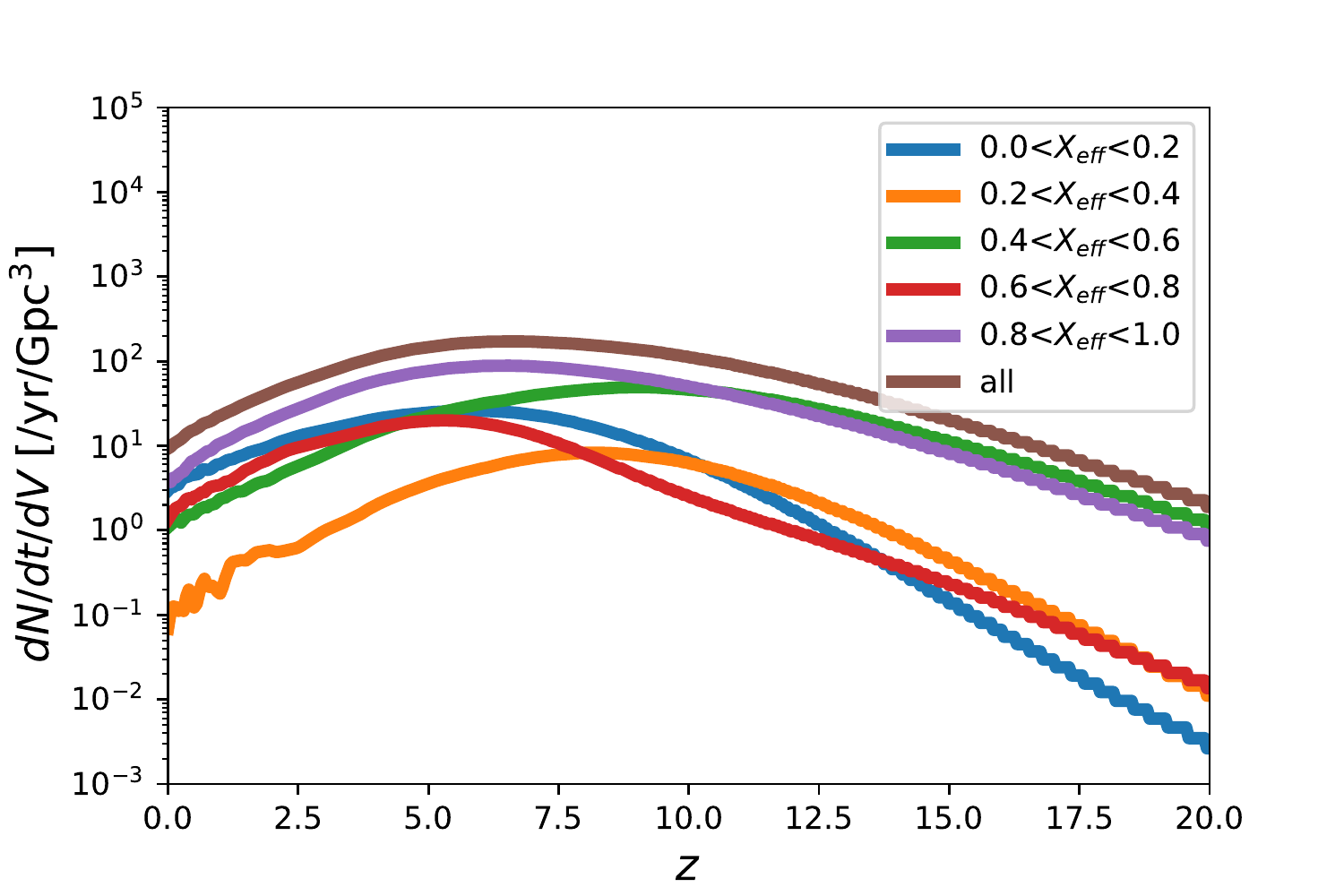}
\subcaption{K14 model}  
\label{spinz_K14}
\end{minipage} 
      \begin{minipage}[t]{0.4\hsize}
        \centering
\includegraphics[width=\hsize]{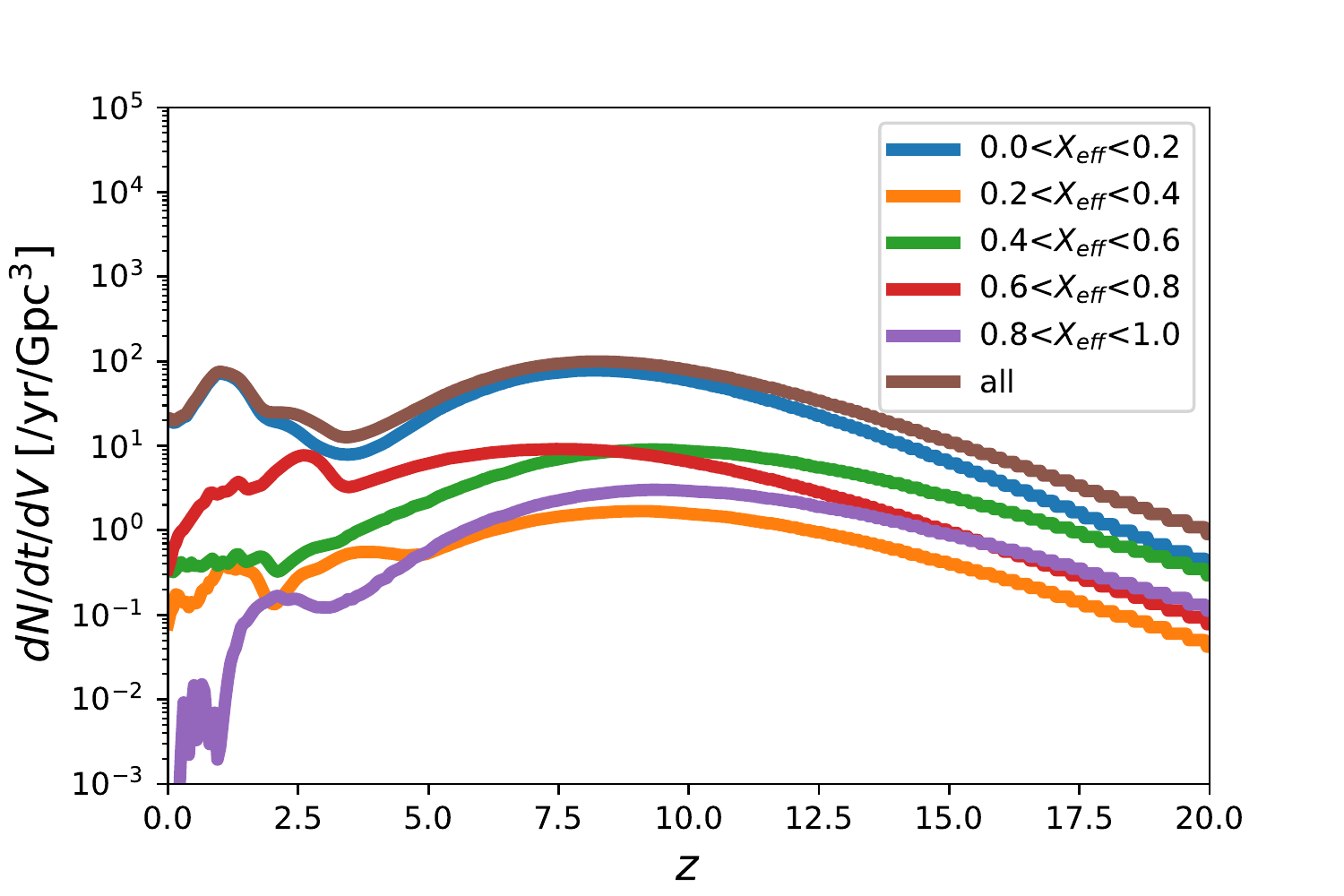}
\subcaption{FS1 model}  
\label{spinz_FS1}
\end{minipage} \\
      \begin{minipage}[t]{0.4\hsize}
        \centering
\includegraphics[width=\hsize]{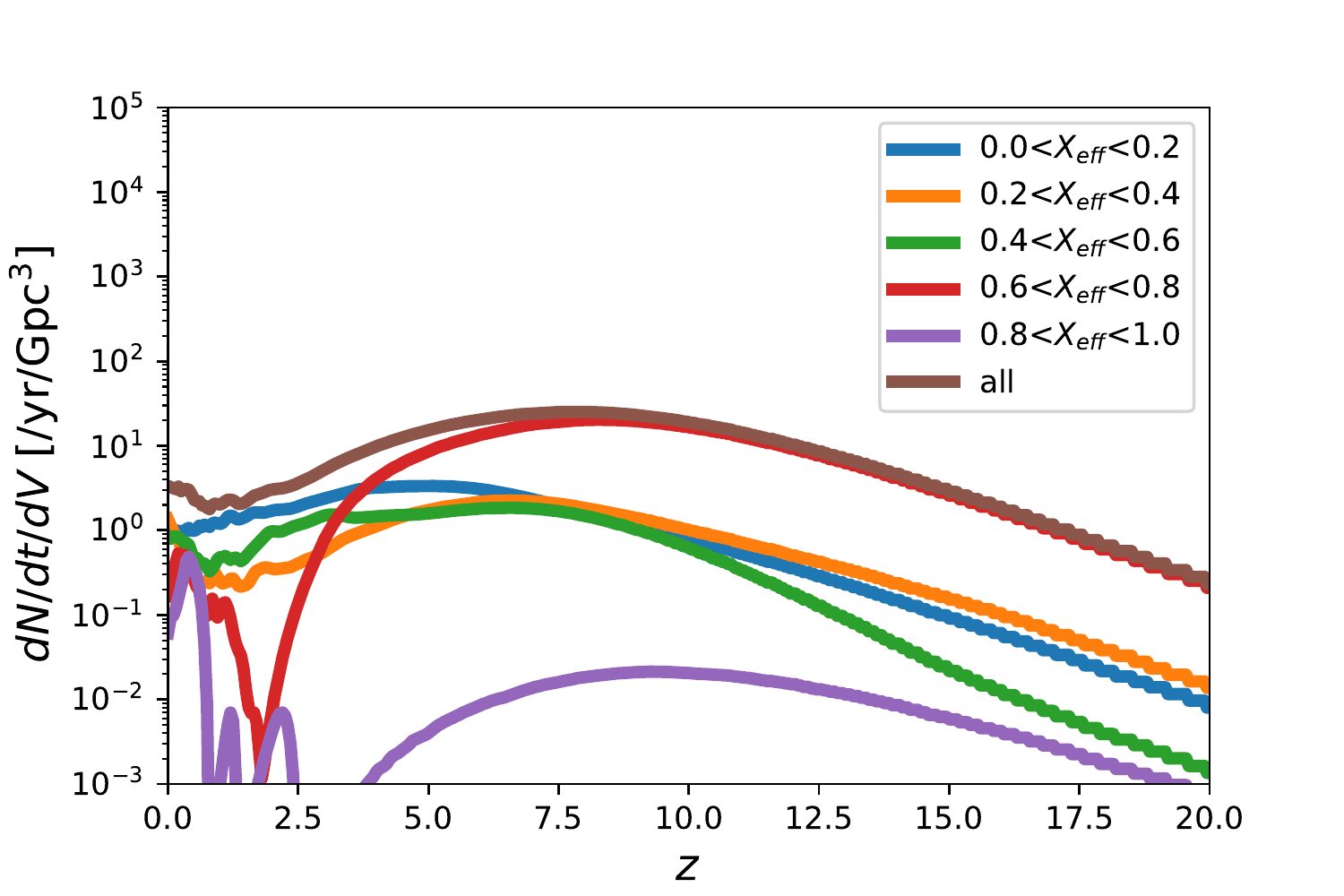}
\subcaption{FS2 model.} 
\label{spinz_FS2}
\end{minipage}
\end{tabular}
\caption{Merger rate for different spin interval as a function of the redshift for each model. Each colored line denotes
a different range of the effective spin $\chi_{\rm eff}$.
The total merger rate is shown as the brown line.} 
\label{spinz}
\end{figure*}

Figure \ref{spinz_fidutial} shows the merger rate for different spin interval as a function of the redshift for our fiducial model.
In practice, we use the effective spin parameter $\chi_{\rm eff}=(M_1 \chi_1 + M_2 \chi_2)/M_{\rm total}$
where $\chi_1$ and $\chi_2$ are the nondimensional spin parameters of each BH.
Each line describes the merger rate density for each $\chi_{\rm eff}$ region.
The merger rate density of the lowest spin region $\chi_{\rm eff}$ (blue line) is dominant at the low redshift.
On the other hand, the merger rate density of the highest spin region $\chi_{\rm eff}$ (purple line) is almost zero in low redshift values ($z \lesssim 1 $), but it is dominant or sub-dominant in high redshift values ($z\gtrsim10$).
This feature comes from the tidal effect, that is, whether the tidal interaction is effective or not.
Usually stars in a binary system tend to lose the spin angular momentum by binary interactions in the mass loss of the Roche lobe overflow or the common envelope phase.
However, if the tidal interaction is effective, they can get the spin angular momentum from the orbital one.
The smaller separation the binary has, the more effective the tidal interaction is.
Thus, BBHs whose merger times are very short tend to have large spins.
Because the Pop III SFR has peak at $z\sim10$, such short merge time BBHs tend to merge near $z\sim10$
Therefore, highest spin BBHs tend to merge at $z\sim 10$.

Figure \ref{spinz} shows the merger rate for different spin interval as a function of the redshift for the other models.
%Except for K14 model (that is, models in Figs. \ref{spinz_MT05}, \ref{spinz_al01}, \ref{spinz_M100}, \ref{spinz_FS1}, and \ref{spinz_FS2}), 

In the case of $\beta=0.5$ (Fig. \ref{spinz_MT05}) and M100 (Fig. \ref{spinz_M100}) models,
BBHs which merge at low redshift have low spins, but at high redshift, BBHs have similar high spins as that of the fiducial model (Fig. \ref{spinz_fidutial}).

In the case of $\alpha\lambda=0.1$ (Fig. \ref{spinz_al01}) and FS1 (Fig. \ref{spinz_FS1}) models, the merger rate densities of the highest $\chi_{\rm eff}$ (the purple line) are not large compared to other models at high redshift.
BBH progenitors via the 1CE$_D$ channel have the short merger time (see the red line in Fig. \ref{mergertime_fidutial}).
In this case, if two naked helium star binaries which are the remnant of the double common envelope is close, the tidal force (Eq.\ref{eq:dodt}) is so strong that they are easy to become high spin BBHs. 
They are the main source of high spin BBHs which merge at the high redshift in the fiducial model.
On the other hand, in the $\alpha\lambda=0.1$ model, the faction of Pop III BBHs of the 1CE$_D$ channel whose merger time $\lesssim10^2$ Myr is smaller than that of the fiducial model (compare Fig. \ref{mergertime_al01} with Fig. \ref{mergertime_fidutial}).
The number of high spin BBHs made by 1CE$_D$ is small in $\alpha\lambda=0.1$ model.
This is the reason for that the merger rate densities of the highest $\chi_{\rm eff}$ (the purple line) are not large.
In the case of the FS1 model, the reason is different from the $\alpha\lambda=0.1$ model. 
In the FS1 model, Pop III BBHs of the 1CE$_D$ channel are dominant on merger time $<10^2$~Myr (see Fig. \ref{mergertime_FS1}), but Pop III BBHs of the 1CE$_D$ channel in the FS1 model tend not to have high spins unlike the fiducial model.
Progenitors of the 1CE$_D$ channel in the FS1 model are  more massive than those of the fiducial model so that they tend to collapse to black holes just after the common envelope phase.
Thus, they cannot spin up via the tidal lock after the common envelope phase.

In the case of K14 model (Fig. \ref{spinz_K14}), the highly spinning BBHs are dominant for all redshift value. 
Since in K14 model, we have used the old prescription of the tidal interaction (Eq. \eqref{eq:E2,1}),
the tidal interaction efficiency is strong enough to become tidal lock easier than the other models \citep{Kinugawa2016b,Kinugawa2016c}.
Thus, the high spin BBHs are majority in the whole redshift value in K14 model.
In the case of FS2 model, at $z\gtrsim3$, the Pop III BBHs whose effective spin is $0.6<\chi_{\rm eff}<0.8$ are dominant.

\begin{table}
	\caption{Averages of the effective spin, $\left\langle \chi_{\rm eff} \right\rangle$ of merging Pop III BBHs at some redshifts.}
	\label{tab.spin}
		\centering
		\begin{tabular}{c| c c c c c c}
		\hline
		model & z=0 & z=0.1&z=1 & z=5 & z=10 & z=20\\
		\hline
		Fiducial& 0.068&  0.068 & 0.075  & 0.36 &   0.57 & 0.72

 \\ \hline

		$\beta$=0.5&  0.052   & 0.058 & 0.066  & 0.40  &  0.62 &0.78

\\ 
		\hline
		$\alpha\lambda$=0.1 & 0.044 & 0.051 & 0.063  & 0.47 & 0.33&  0.34  

\\
		\hline
		
    	M100&  0.087  &  0.074&   0.086 & 0.35 &   0.61 & 0.74

 \\		\hline
		K14		& 0.57 & 0.57    &  0.62  &  0.70 & 0.69 & 0.70

 \\
		\hline
		FS1&  0.020  & 0.032 & 0.030 & 0.19 & 0.16 &  0.36 

 \\  \hline
		FS2&0.30  & 0.33     &  0.19 & 0.48  &  0.64  & 0.65
\\
		\hline
		
	\end{tabular}
	
\end{table}

\subsection{Detection rate}
\begin{figure}
\begin{center}
\includegraphics[width=\hsize]{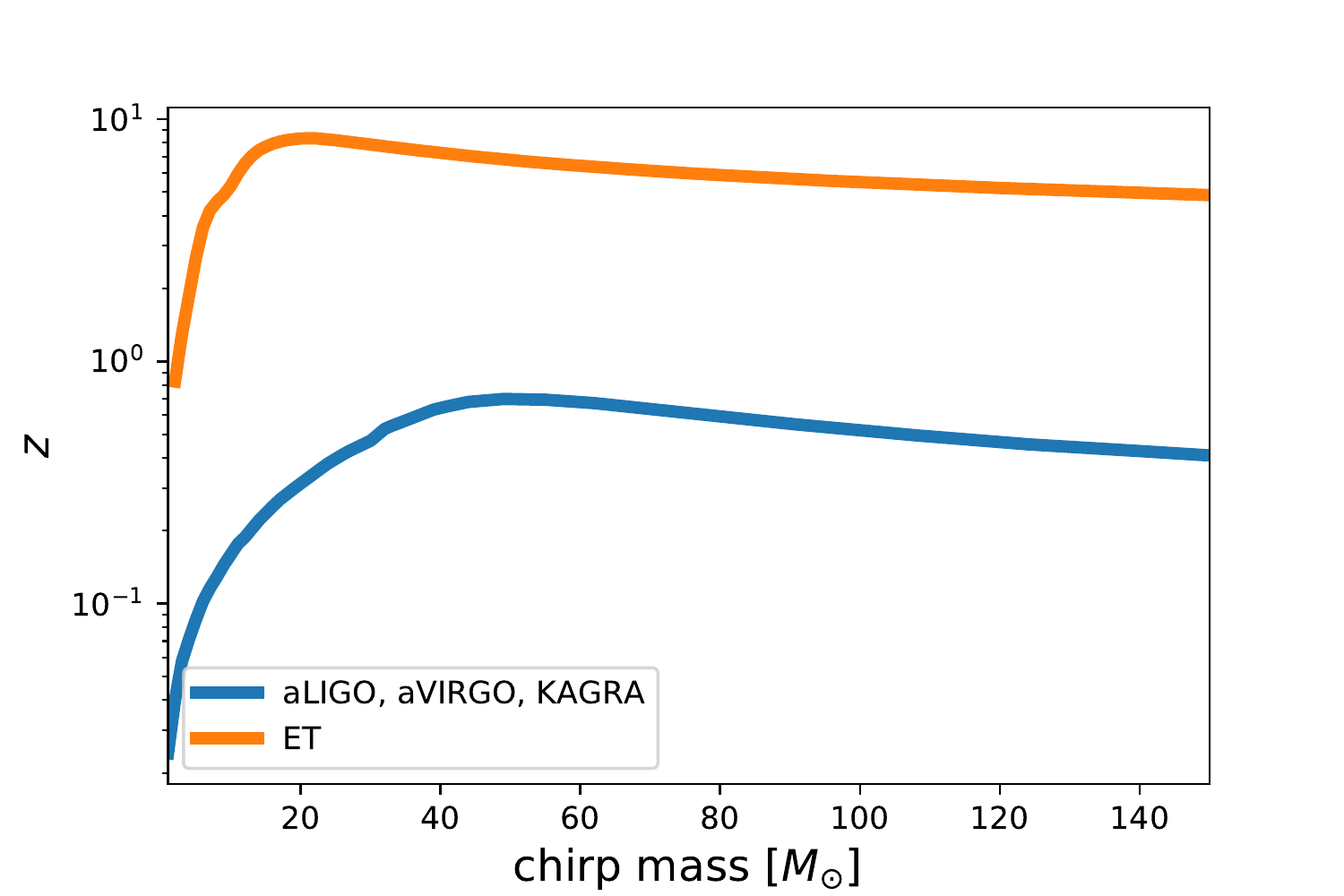}
\end{center}
\caption{Detection range of the second-generation GW detectors
such as aLIGO, aVIRGO, and KAGRA (blue), and the third-generation detector, ET (orange).
The horizontal axis shows the chirp mass, 
{but not the redshifted chirp mass}.}  
\label{detection range}
\end{figure}

\begin{figure*}
    \begin{tabular}{cc}
      \begin{minipage}[t]{0.5\hsize}
        \centering
        \includegraphics[width=\hsize]{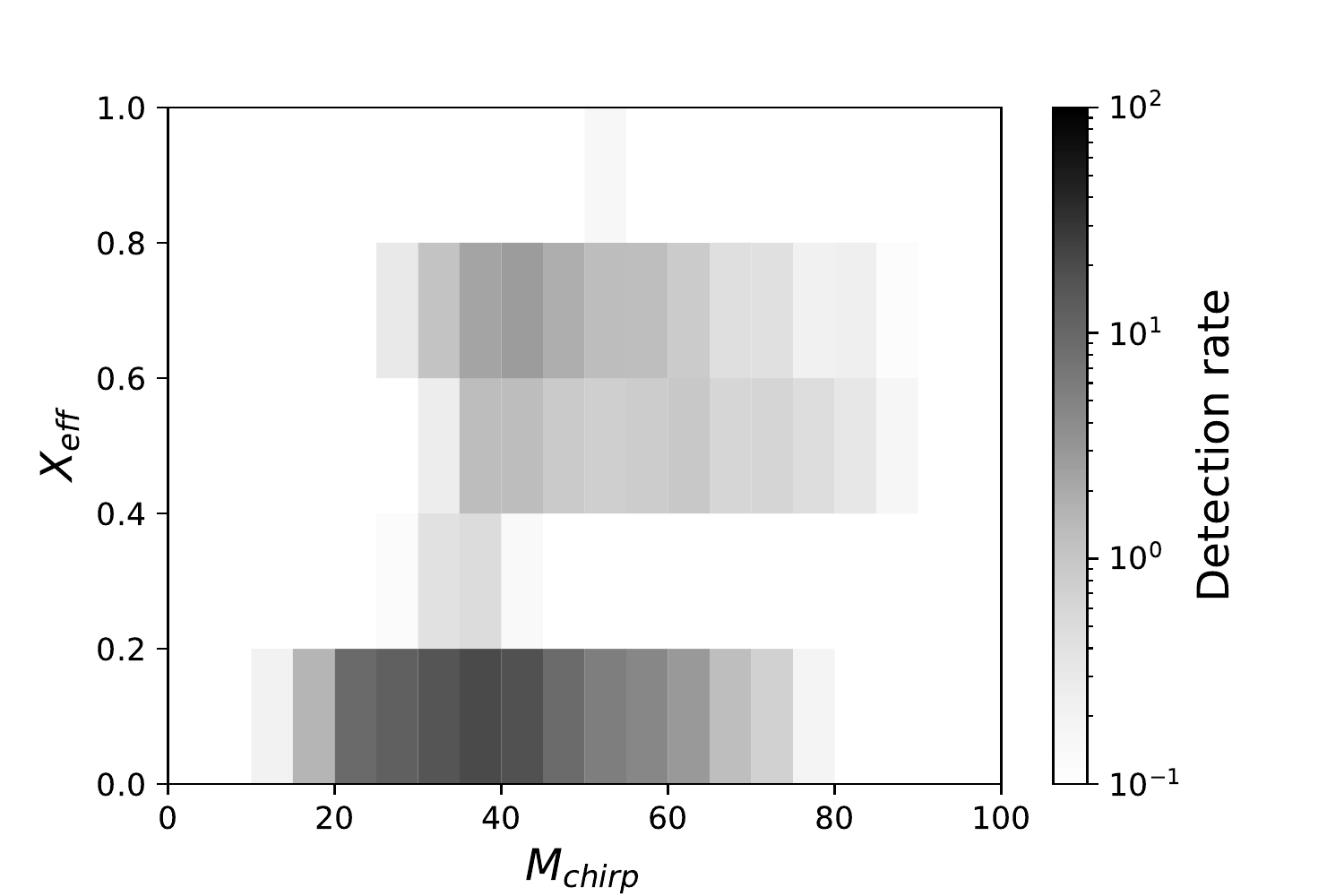}
        \subcaption{fiducial model}
        \label{fidutial_KAGRA}
      \end{minipage} 
      \begin{minipage}[t]{0.5\hsize}
        \centering
        \includegraphics[width=\hsize]{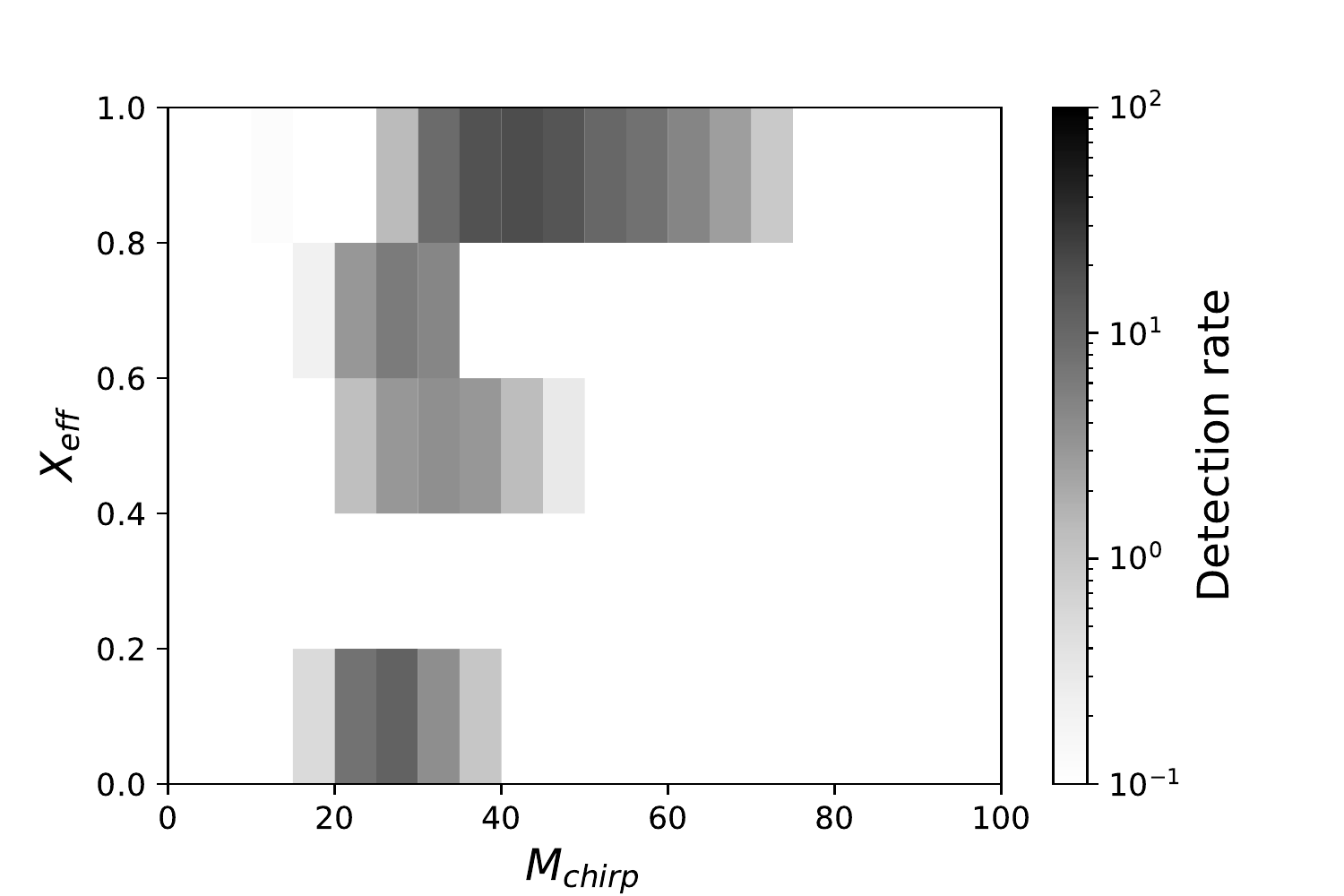}
        \subcaption{K14 model}
        \label{K14_KAGRA}
      \end{minipage}
    \end{tabular}
    \caption{Pop III BBH detection rate of aLIGO, aVIRGO, and KAGRA [/yr]. The horizontal axis shows the chirp mass $M_{\rm chirp}$, and the vertical axis is the effective spin parameter $\chi_{\rm eff}$. The darker region has a higher detection rate.}
    \label{KAGRAdetect1}
  \end{figure*}

\begin{figure*}
    \begin{tabular}{cc}
      \begin{minipage}[t]{0.5\hsize}
        \centering
        \includegraphics[width=\hsize]{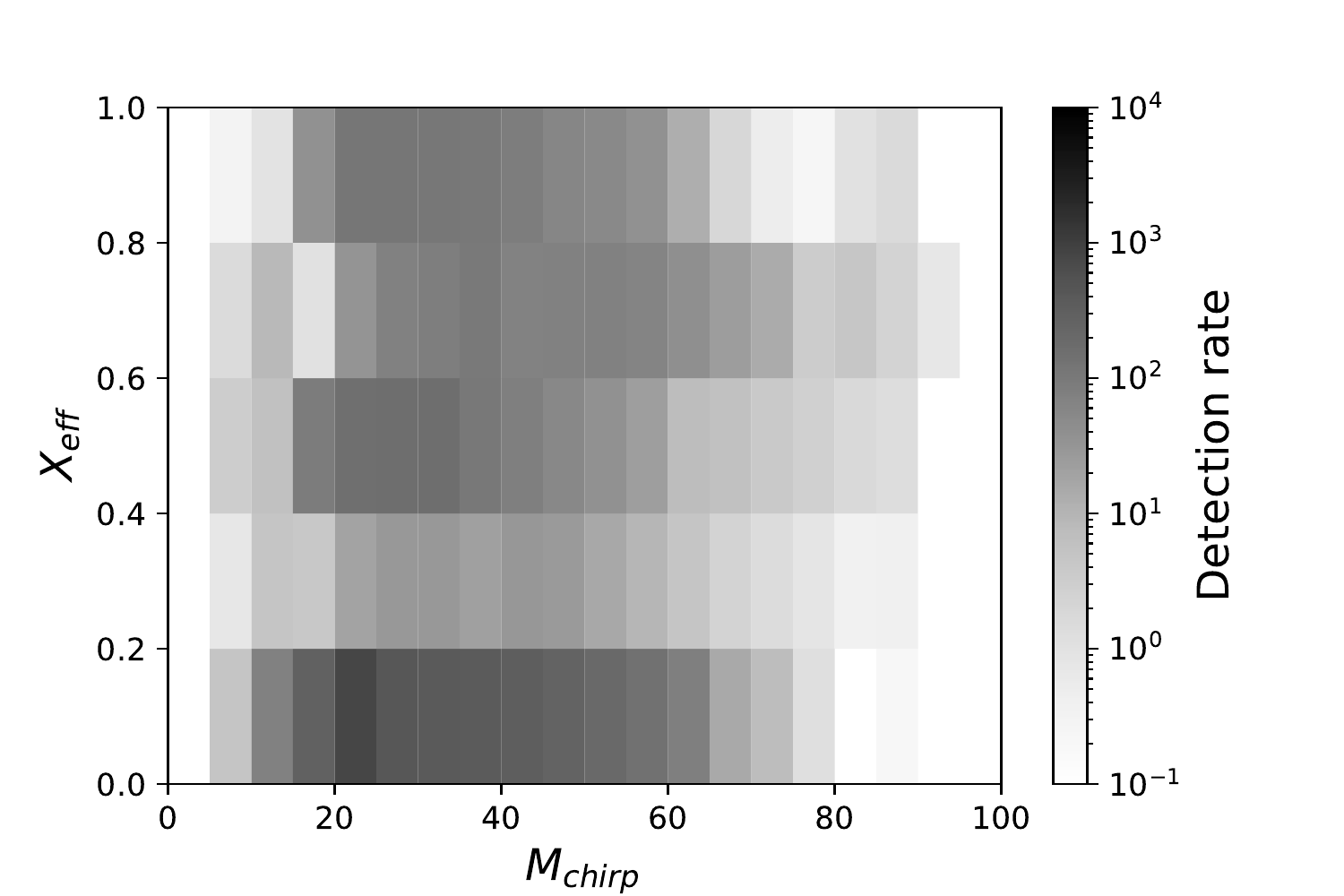}
        \subcaption{fiducial model}
        \label{fidutial_ET}
      \end{minipage} 
      \begin{minipage}[t]{0.5\hsize}
        \centering
        \includegraphics[width=\hsize]{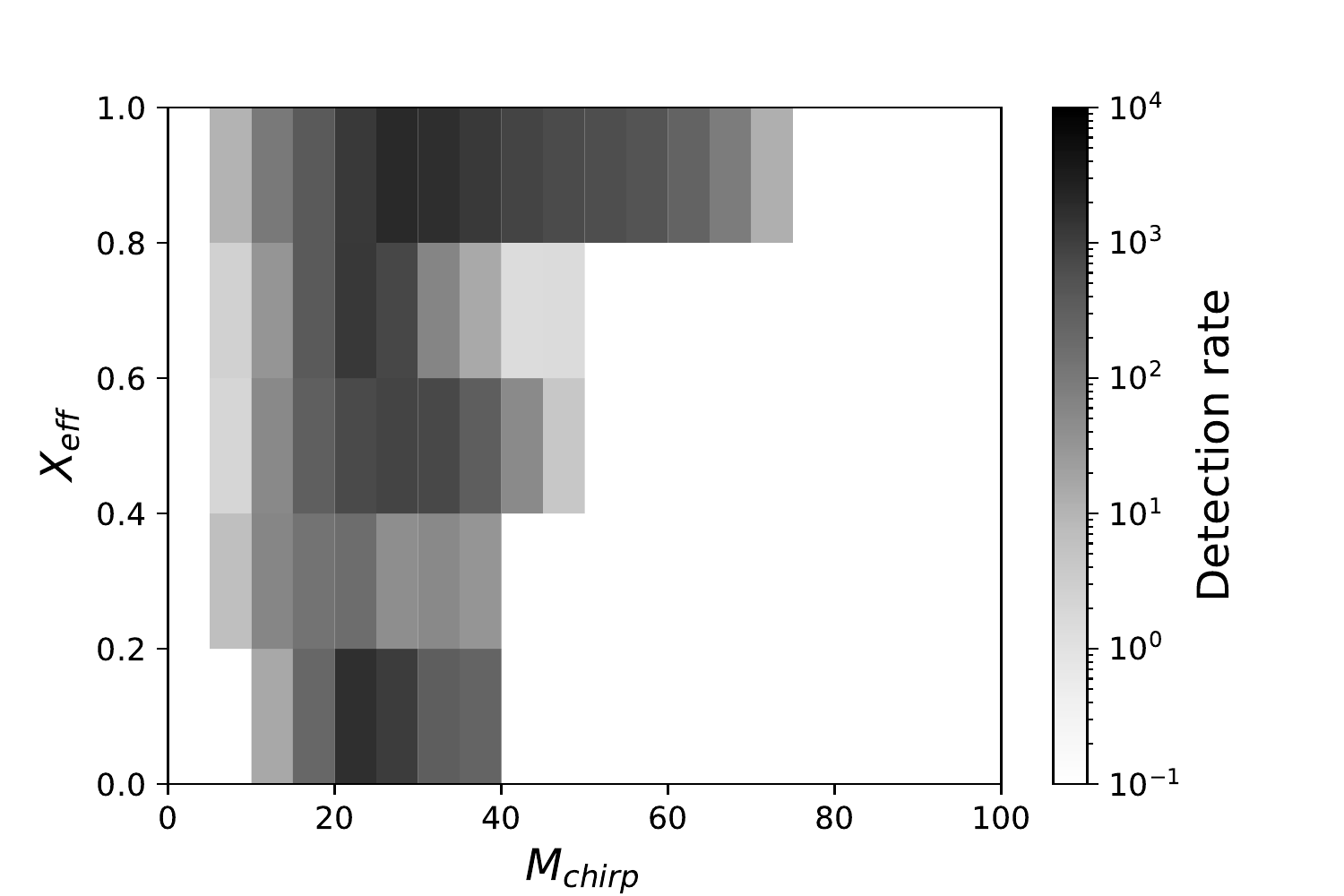}
        \subcaption{K14 model}
        \label{K14_ET}
      \end{minipage}
    \end{tabular}
    \caption{Pop III BBH detection rate of ET [/yr].
        The style is the same as Fig.~\ref{KAGRAdetect1}.
        Note that the scale of the detection rate is different from Fig.~\ref{KAGRAdetect1}.}
        \label{ETdetect1}
  \end{figure*}

\begin{figure*}
    \begin{tabular}{cc}
      \begin{minipage}[t]{0.5\hsize}
        \centering
        \includegraphics[keepaspectratio, scale=0.5]{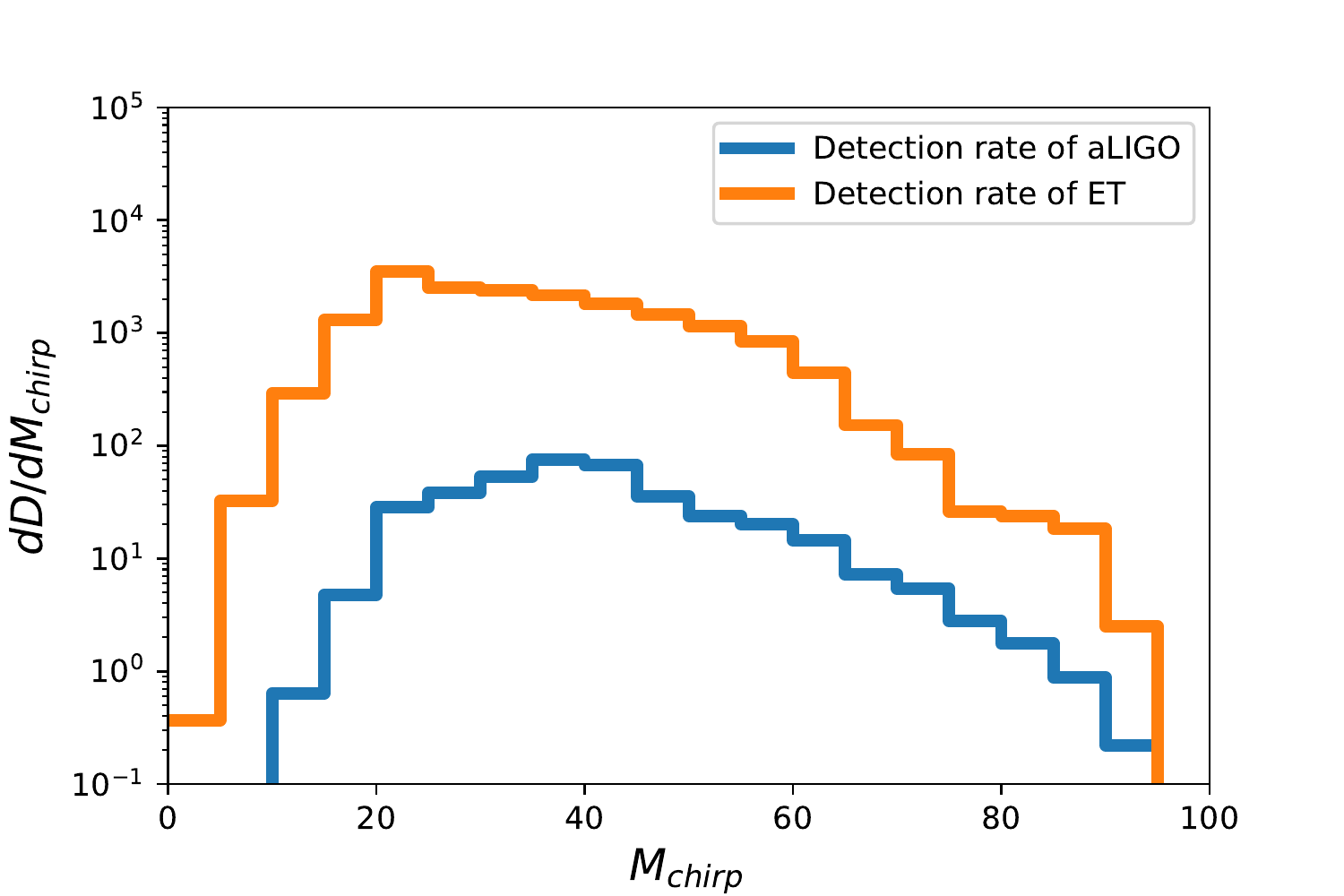}
        \subcaption{fiducial model}
        \label{fidutial_Mass}
      \end{minipage} &
        \begin{minipage}[t]{0.5\hsize}
        \centering
        \includegraphics[keepaspectratio, scale=0.5]{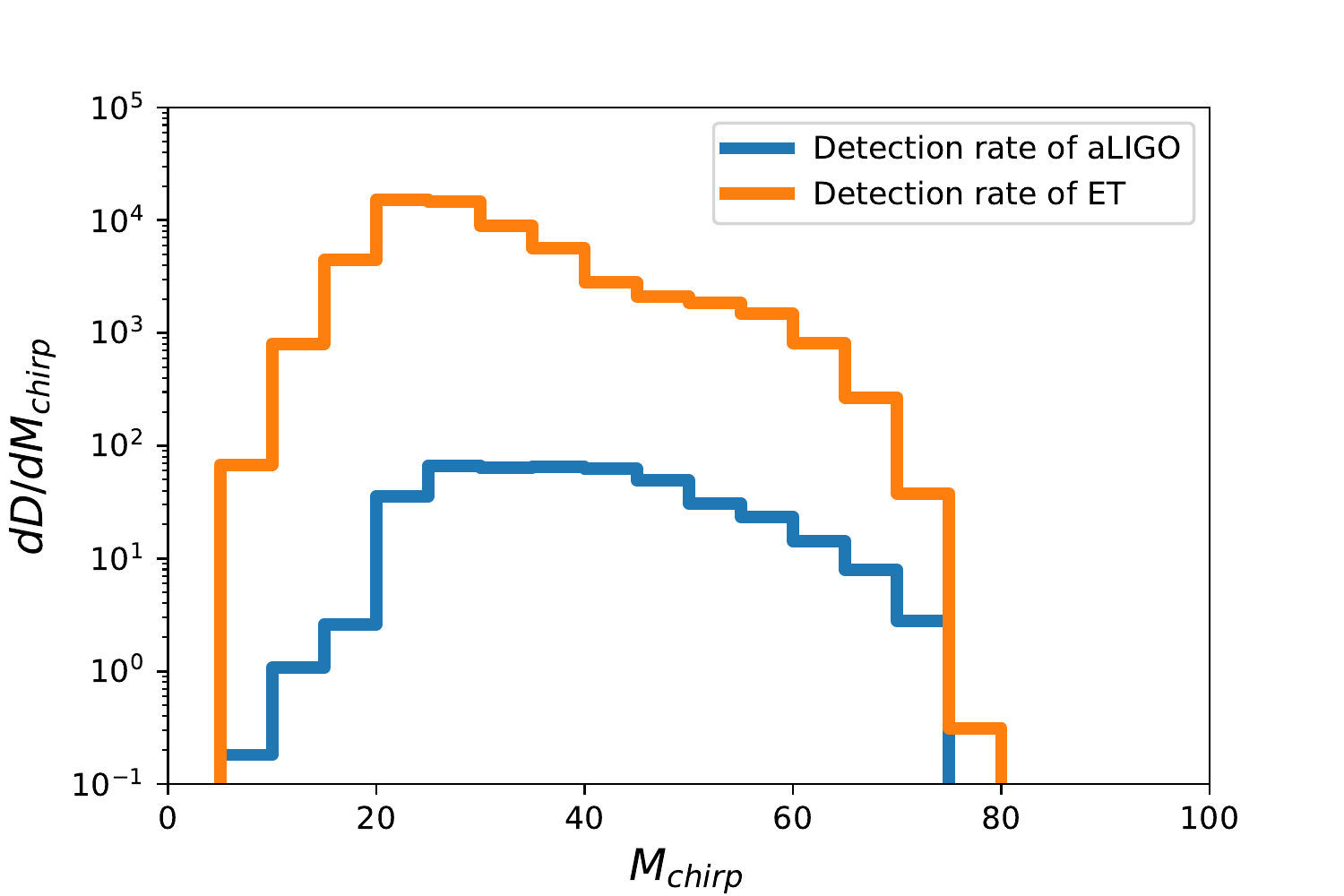}
        \subcaption{K14 model}
        \label{K14_CMass}
      \end{minipage}
      
    \end{tabular}
     \caption{Chirp mass ($M_{\rm chirp}$) distribution of detectable Pop III BBHs in the fiducial model for aLIGO (blue) and ET (orange).
        The horizontal axis is the chirp mass.}
        \label{CMass1}
  \end{figure*}

\begin{figure*}
    \begin{tabular}{cc}
\begin{minipage}[t]{0.5\hsize}
        \centering
        \includegraphics[keepaspectratio, scale=0.5]{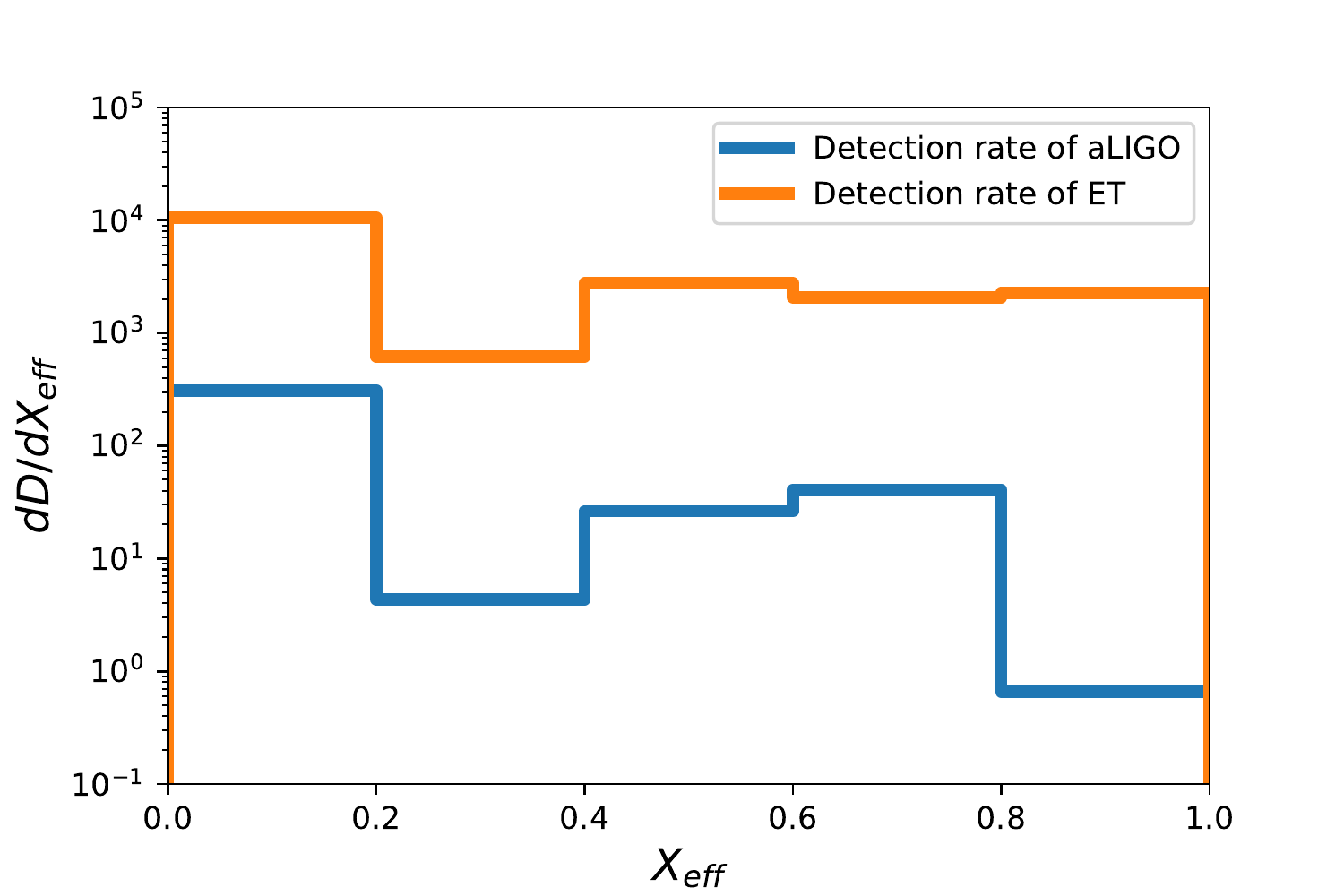}
        \subcaption{fiducial model}
        \label{fidutial_Spin}
      \end{minipage} &
      
      \begin{minipage}[t]{0.5\hsize}
        \centering
        \includegraphics[keepaspectratio, scale=0.5]{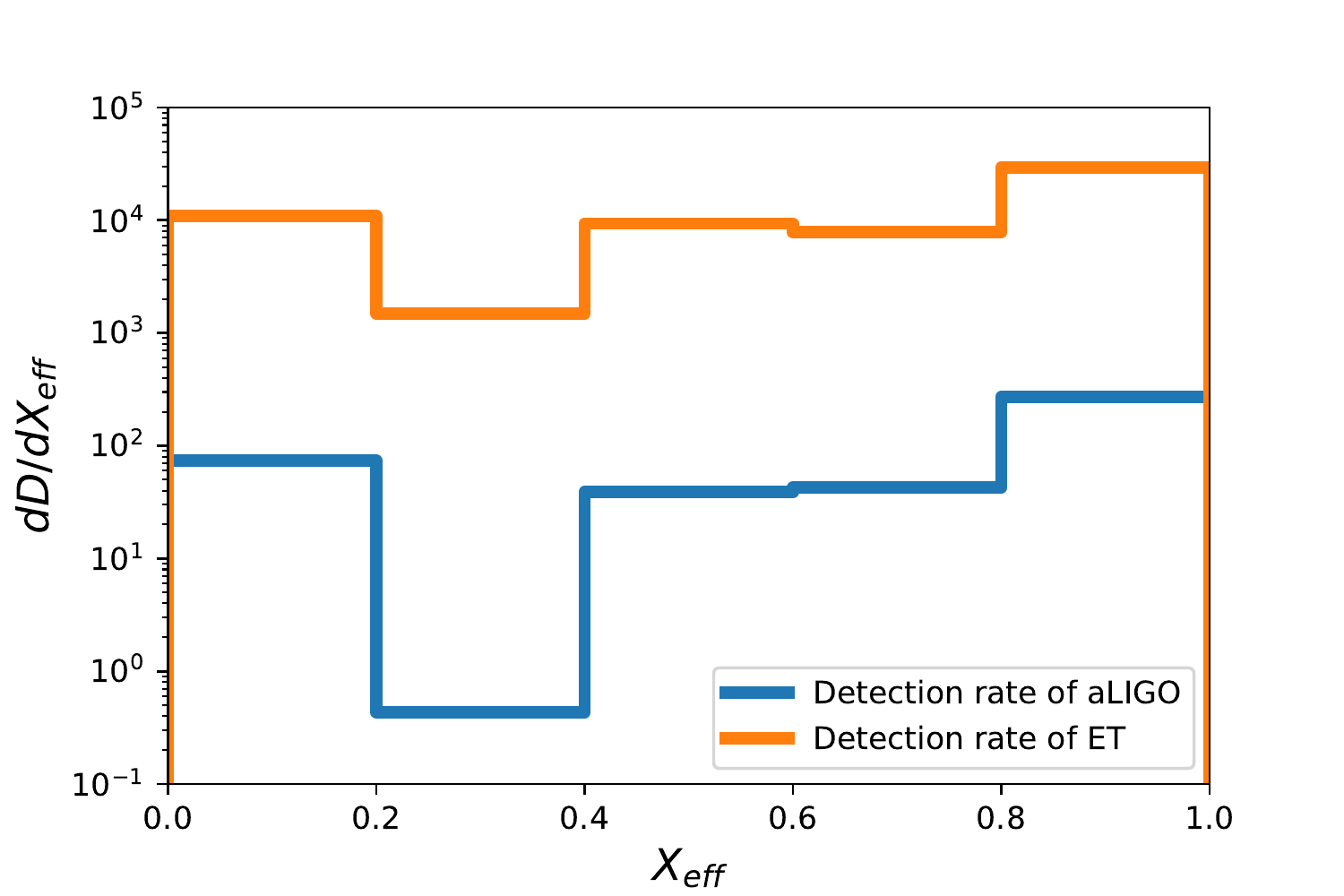}
        \subcaption{K14 model}
        \label{K14_Spin}
      \end{minipage}
      
    \end{tabular}
    \caption{Effective spin ($\chi_{\rm eff}$) distribution of detectable Pop III BBHs for aLIGO (blue) and ET (orange).
        The horizontal axis is the effective spin parameter.}
        \label{Spin1}
  \end{figure*}

Figure \ref{detection range} shows the detection rate [/yr] of second-generation GW detectors such as aLIGO, aVIRGO, and KAGRA \citep{KAGRA} as a function of the chirp mass of BBH, and that of ET \citep{ET} which is a third-generation ground-based GW observatory.
The detection rate of Pop III BBHs in the fiducial model are shown in Fig. \ref{fidutial_KAGRA} for the second-generation GW detectors
and Fig. \ref{fidutial_ET} for ET as a function of the spin parameter and the chirp mass.
Figures \ref{fidutial_Mass} and \ref{fidutial_Spin}
show the chirp mass and the effective spin distributions of detectable Pop III BBHs in the fiducial model for the second-generation detectors and ET.
Detectable Pop III BBHs of the second-generation detectors have two peak regions at $M_{\rm chirp}\sim$ 30--40 $\msun$ and ~$\chi_{\rm eff}\sim0$ as well as  $M_{\rm chirp}\sim$ 30--40 $\msun$ and~$\chi_{\rm eff}\sim0.5$.
The second-generation detectors can detect only the low redshift ($z < 1$) BBH mergers.
Most Pop III BBHs merging at the low redshift tend to have low spin (see Sec. \ref{sec.spin} and Fig. \ref{spinz_fidutial}).
On the other hand, some of the Pop III BBHs evolved via the 1CE$_P$ channel which merge at low redshift can have large secondary spin, and their spins can be $\chi_{\rm eff}\sim0.5$.
%\tn{The meaning of next three sentences are not clear. They  seem wrong. BBH mass distribution does not depend on the detectors. } 
%\tk{The difference between Pop III BBHs mass distributions  for the second-generation detectors and those for ET is the peak of the chirp mass distribution. The peak of detectable Pop III BBHs chirp mass for the second-generation detectors is $M_{\rm chirp}\sim$ 30--40 $\msun$, while that for ET is $M_{\rm chirp}\sim20~\msun$.This difference comes from difference of the peak sensitivity location of GW detectors as a function of the chirp mass (Fig. \ref{detection range}). For the spin distribution of detectable Pop III BBHs (Fig. \ref{fidutial_Spin}), ET can detect higher spin BBHs than the second-generation detectors.}
{Fig. \ref{detection range} shows the detectable range in $z$ of GWs from BBHs as a function of the chirp mass by the second-generation detectors (blue) and ET (orange). By ET, the maximum observable redshift of merger of BBH is $z\sim 10$ for $\mchirp \sim 20~\msun$.
Pop III BBHs which merge at high redshift tend to have high spins because they are easily spun up by the tidal interaction as we know in Table \ref{tab.spin}.
Thus, ET can detect BBH mergers at such high redshift so that the detectable Pop III BBHs are expected to have high spins (Fig. \ref{fidutial_Spin}).}


In order to compare present results with those of the previous paper, we show the detection rate of K14 models in Fig. \ref{K14_KAGRA}, while 
as for the other models, the figures of detection rates are showed in Appendix.
Figures \ref{K14_KAGRA}, and \ref{K14_ET} show the detection rates of Pop III BBHs in K14 model for the second-generation GW detectors and ET, respectively.
Figure \ref{K14_CMass} shows that the chirp mass distributions of detectable Pop III BBHs in K14 model for the second-generation detectors and ET.
Figure \ref{K14_Spin} shows the effective spin distribution of detectable Pop III BBHs in K14 model for the second-generation detectors and ET.
Differences between our fiducial model and K14 model are the mass transfer rate and the tidal coefficient factor $E$ (Eq. \eqref{eq:E2,1}).
Especially, the difference of tidal coefficient factor $E$ makes the large difference of the spin distribution of detectable Pop III BBHs.
In K14 model, we use the same $E$ (Eq. \eqref{eq:E2,1}) in our previous works, and the $E$ (Eq. \eqref{eq:E2,1}) is too easier to make tidal lock binaries than the new $E$ (Eq. \eqref{eq:E2new}) in the fiducial model.
Thus, detectable Pop III BBHs in K14 model tend to have large spins.
Therefore, the spin of detectable BBHs strongly depends on the tidal coefficient factor $E$.

%In Apendix, we summarize
%the Pop III BBH detection rate
%and the effective spin distributions
%for the other models.

\section{Summary and Discussion}
In this paper, we have calculated Pop III binary evolutions for seven models.
Our results show that Pop III BBHs tend to be $M_{\rm chirp}\sim 30~\msun$ BBHs and they can merge at present day due to a long merger time.
The merger rate densities of Pop III BBHs at $z=0$ are 3.34--21.2 $\rm /yr/Gpc^3$ which is consistent with aLIGO/aVIRGO result 9.7--101 $\rm /yr/Gpc^3$.
Pop III binaries might contribute the massive BBHs of aLIGO/aVIRGO detections.
If the BBH chirp mass distribution has bimodal peaks at $\sim10~\msun$ and at $\sim30~\msun$ {which might be suggested by Fig.1}, the massive peak might be made from Pop III sources.
For the spin of BBH mergers, Pop III BBH mergers at $z\sim0$ tend to have low spin parameters $\chi_i$.
This feature is consist with the aLIGO/aVIRGO analysis.
On the other hand, Pop III BBH mergers at the high redshift parameter
tend to have high spin parameters $\chi_i$.
Future plans of GW detectors such as ET, CE \citep{Reitze:2019iox} and DECIGO \citep{Seto:2001qf,Nakamura_2016} can detect massive BBH mergers with high spin parameters $\chi_i$ at high redshift values, $z > 10$.
These detectors can check the redshift dependence of BBH merger rate and spin parameter distribution.
Therefore, massive BBH detections by GWs will be important for the stellar evolution study at the early universe. 
{Note that there are still uncertainty of spin evolution and  it is still not possible to compute fully consistent rotating models.
Thus, the spin dependence on the separation can be qualitatively correct, but spin values might be quantitatively incorrect.
We must calibrate the spin evolution of massive stars by using the future observation results.}

{The main differences between the fiducial model in this paper and our previous works \citep{Kinugawa2014, Kinugawa2016,Kinugawa2016b,Kinugawa2016c} are the mass transfer rate (Sec.~\ref{stable MT}) and the treatment of tidal interaction (Sec.~\ref{tidal}).
We have used much higher mass transfer rate than that of our previous works, but the features of merging Pop III BBHs such as peak of chirp mass, and the merger rate do not change so much.
On the other hand, the treatment of tidal interaction makes the large difference.
The tidal coefficient factor $E$ of the dynamical tide  is less effective than that of our previous works, so the average spin of the fiducial model is much smaller than that of our previous works.
}

{In comparison with \cite{Hartwig_2016}, they suggested that the merger rate of Pop III BBHs in their model is lower than that of our fiducial model by using almost the same SFR.
Because they used a low metal binary evolution model ($Z=10^{-1}\zsun$) \citep{demink2015} which is much harder to make massive BBHs than those from Pop III binaries.
We show that the Pop III BBHs are consistent with the aLIGO/aVIRGO result by using the new SFR.
In comparison with \cite{Belczynski_2017},
they calculated low metal binary evolution using a modified low metal stellar evolution ($Z=5\times10^{-3}\zsun$), and suggested that most Pop III BBHs merged at the early universe and the merger rate at present day is much smaller than the aLIGO/aVIRGO result.
However, the Pop III stellar evolution is completely different from low metal stellar evolution, especially the envelope structure at the giant phase.
Thus, the difference in treatment of stellar models make the differences between our results and \cite{Belczynski_2017}.}

%\tk{Recently, \cite{Liu2019} have considered another possibility of Pop III BBH mergers whose origin by dynamical capture.
%They calculated the merger rate of Pop III BBHs made by dynamical capture in cosmological hydrodynamic simulations.
%Although their merger rate ($0.04 \rm yr^{-1}~Gpc^{-3}$) is much smaller than that for our field binary case, the Pop III BBHs made by dynamical capture might have different features from Pop III BBHs from field binaries such as large eccentricity, different mass spectrum, and so on.
%These differences might be observed by ET or another future GW observations. 
%}

In this paper, we focus only on the BBHs of first star remnants.
However, neutron star - black hole binaries (NSBHs) of first star remnants can also be detected by aLIGO and ET.
In our previous study \citep{Kinugawa2017}, we calculated Pop III NSBHs detection rate and the chirp mass distribution.
Pop III NSBHs tend to be more massive than Pop I, II NSBHs.
The typical chirp mass of Pop III NSBHs is $\sim6~\msun$ which consist of $1.4~\msun$ NS and $\sim50~\msun$ BH.
The chirp mass distribution of NSBHs might become the evidence of Pop III orgin like Pop III BBHs.
{If the spin evolution of BHs in Pop III NSBHs is the same as that of Pop III BBHs}, the spin distribution of detectable NSBHs might be different in each redshift band like the case of BBHs.
Furthermore, it depends on the BH spin whether the massive NSBH has a electromagnetic counter part or not.
Massive NSBHs generally do not have a electromagnetic counter part because the massive BH can absorb the NS without the tidal disruption.
However,  tidal disruption might occur around high spin BHs even if the BH is massive \citep{Lovelace2013}. 
Thus massive NSBHs which merge at high redshift might tend to have electromagnetic counter parts.

Recently, aLIGO/aVIRGO published a new BBH event, GW190412
whose chirp mass is around $13.2~\msun$~\citep{2020arXiv200408342T}.
This event is focused because of the mass ratio $q\sim0.3$ meaning that the BBHs is the pair of a massive stellar-mass BH ($\sim30~\msun$) and the stellar-mass BH ($\sim10~\msun$),
and they have also nonzero effective spin parameter ($\chi_{\rm eff}\sim0.2$).
\cite{2020arXiv200408342T} have considered the case in which the effective spin parameter comes from that of the primary massive BH.
However, \cite{2020arXiv200409288M} have given alternative interpretation of the effective spin parameter of GW190412 
by using astrophysically motivated prior 
with a negligible spin parameter of the more massive BH 
and an aligned high spin parameter of the less massive BH.
In order to discuss this point, we calculate the formation of Pop III BBH like GW190412
(we assume $M_{\rm chirp}$ = 10--16 $\msun$, $q$ = 0.15--0.45, and $\chi_{\rm eff}$ = 0--0.5)
to find that Pop III binaries can form such mass ratio BBHs.
However, in order to satisfy such effective spin parameter, the primary BH must have $\chi_1\sim0$.
In our result, a BH spin parameter tends to become $\chi_i \sim0$ or $\chi_i \sim0.998$, because the BH spin is determined if the tidal interaction is effective or not.
When the tidal interaction is effective at the primary stellar evolution, the primary BH spin parameter tends to become $\chi_1\sim0.998$ so that the effective spin parameter of BBH like GW190412 must be a higher value.
Thus, in order to form BBHs like GW190412, the secondary has to contribute the effective spin parameter.
This feature is the same as the Pop II case \citep{2020arXiv200409288M}.
The merger rate of Pop III BBH like GW190412 at $z$=0 is $\sim0.01\rm~yr^{1}~Gpc^{-3}$.
This value is less than that of Pop II population synthesis result ($\sim0.1\rm~yr^{-1}~Gpc^{-3}$) \citep{Olejak2020}.

{Merging Pop III BBHs can be efficient sources of the gravitational wave background (GWB).
\cite{Inayoshi_2016} show that the GWB from Pop III BBH mergers can be detected by the O5 of aLIGO/aVIRGO, using our previous result \citep{Kinugawa2014} and the Pop III SFR with constraint of Plank result \citep{Planck_2015}.
However, aLIGO might detect just a mere part of it,
and ET, CE, and DECIGO can detect the GWB from Pop III BBH mergers more clearly.
Furthermore, the shape of GWB depends on chirp mass and redshift distributions of BBH mergers so that we can get the information of BBH mergers from GWB. }

\section*{Acknowledgements}
We thank Kenta Hotokezaka, Kohei Inayoshi, Kyohei Kawaguchi, Ataru Tanikawa, and Takashi Yoshida for useful discussion.
T. K. acknowledges support from University of Tokyo Young Excellent researcher program.
T. N. acknowledges support from
JSPS KAKENHI Grant No. JP15H02087.
H. N. acknowledges support from
JSPS KAKENHI Grant Nos. JP16K05347 and JP17H06358.

\bibliographystyle{mnras}

\bibliography{ref}

\section*{Appendix}  
  
In this appendix, we present
the Pop III BBH detection rate,
the chirp mass and the effective spin distributions
of detectable Pop III BBHs for 5 models,
except for the fiducial and K14 models in the 7 models.

\begin{figure*}
    \begin{tabular}{cc}
      \begin{minipage}[t]{0.5\hsize}
        \centering
        \includegraphics[keepaspectratio, scale=0.5]{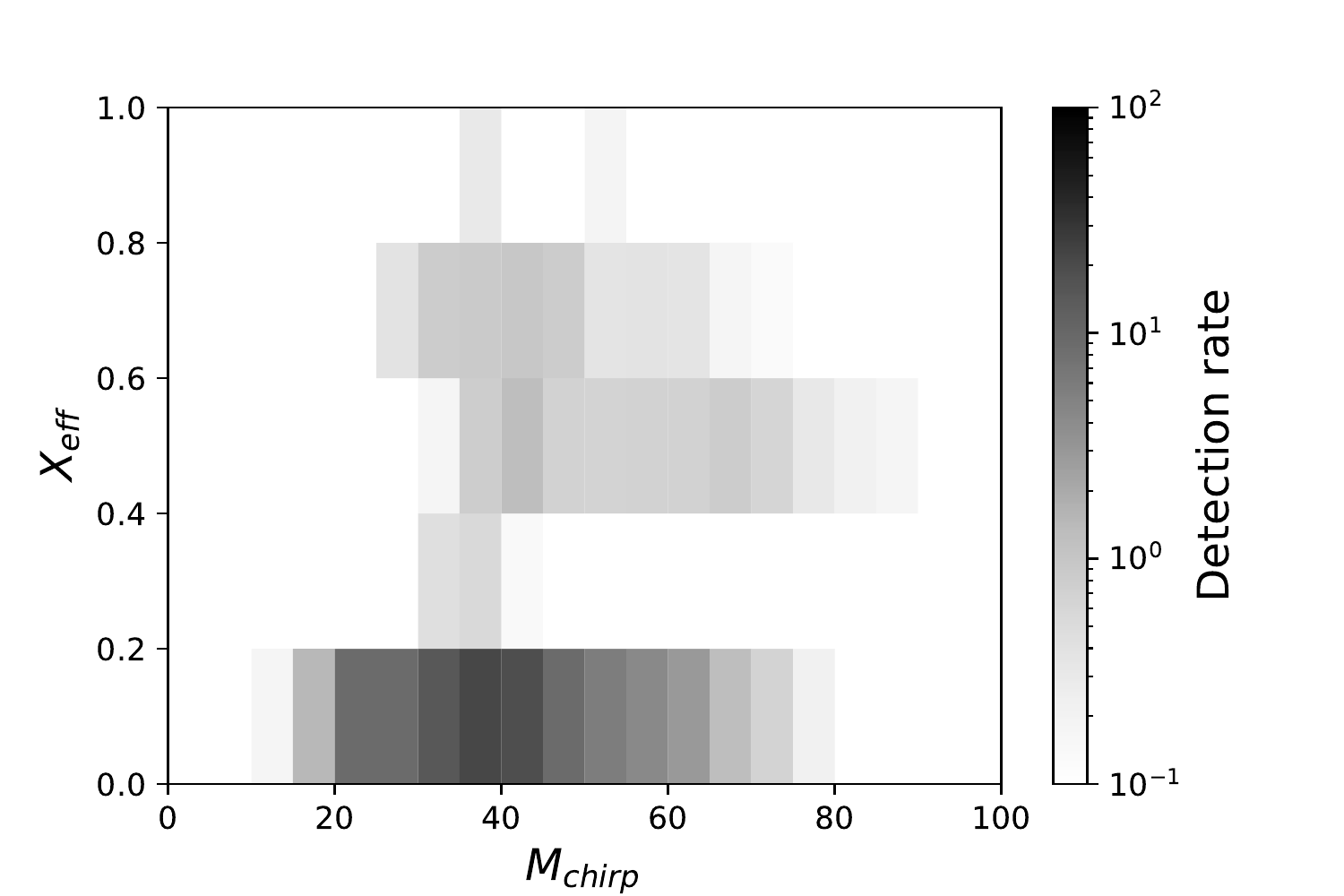}
        \subcaption{$\beta$=0.5 mode}
        \label{MT05_KAGRA}
      \end{minipage} 
      \begin{minipage}[t]{0.5\hsize}
        \centering
        \includegraphics[keepaspectratio, scale=0.5]{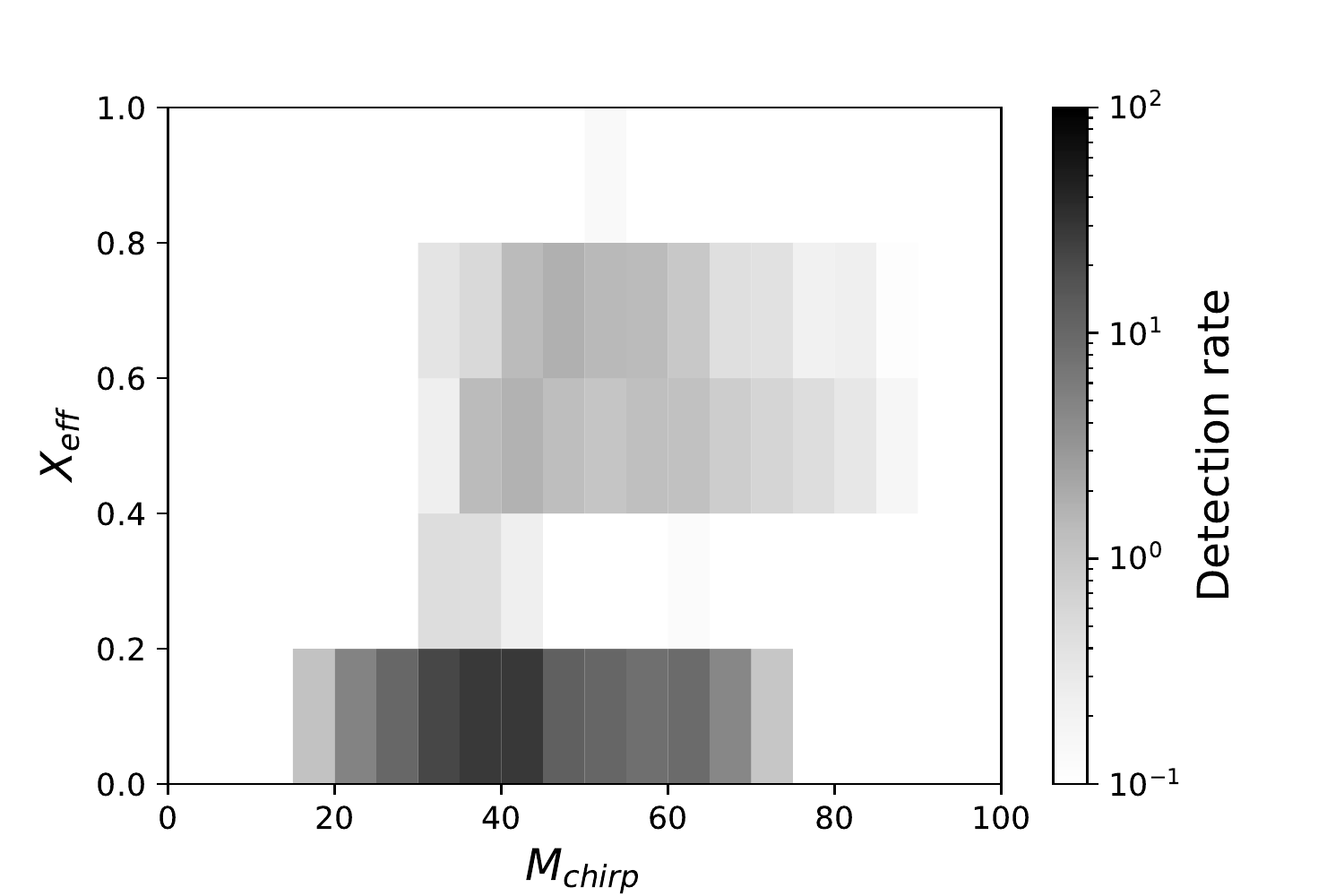}
        \subcaption{$\alpha\lambda$=0.1 model}
        \label{al01_KAGRA}
      \end{minipage} \\
      \begin{minipage}[t]{0.5\hsize}
        \centering
        \includegraphics[keepaspectratio, scale=0.5]{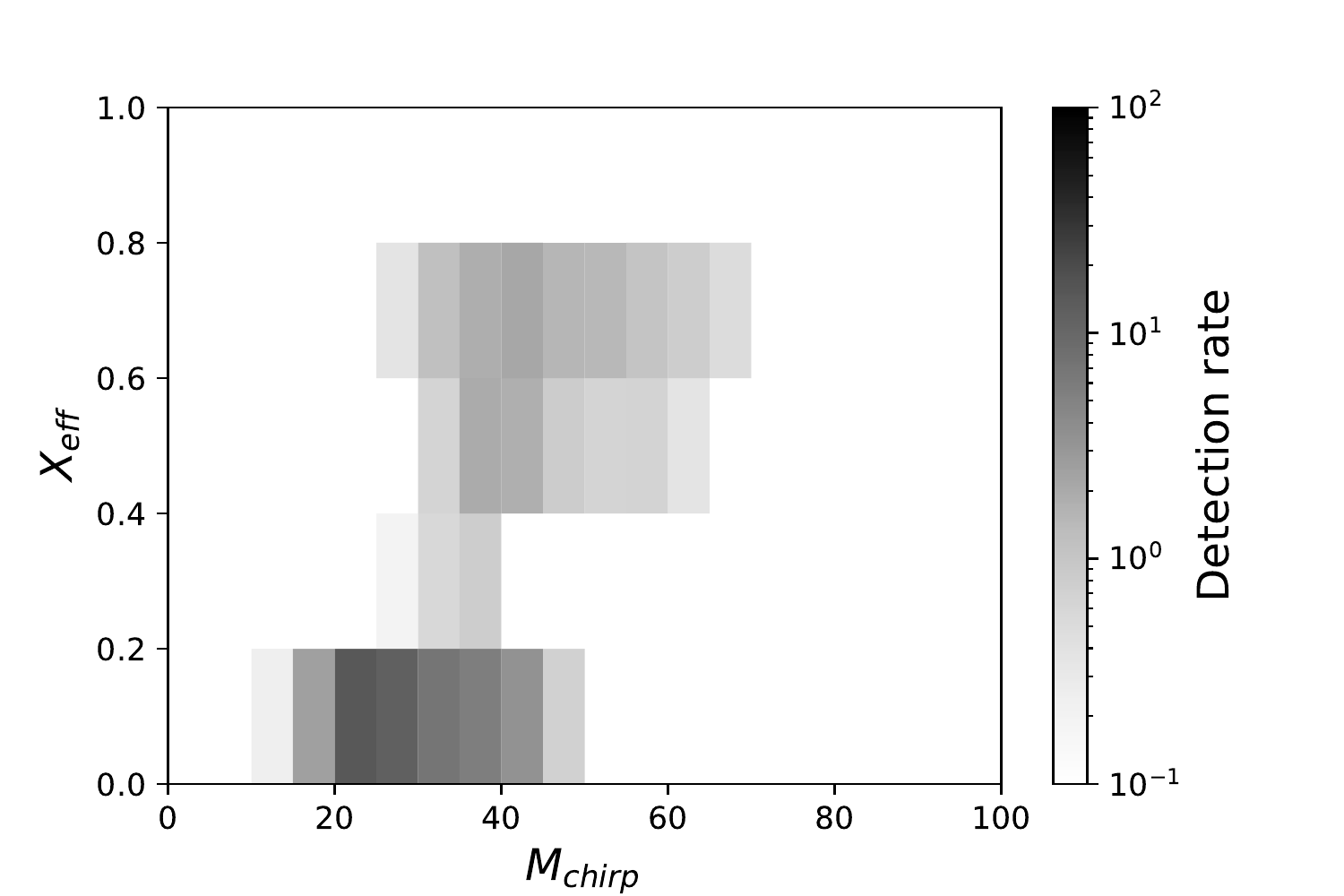}
        \subcaption{M100 model}
        \label{M100_KAGRA}
      \end{minipage} 
      \begin{minipage}[t]{0.5\hsize}
        \centering
        \includegraphics[keepaspectratio, scale=0.5]{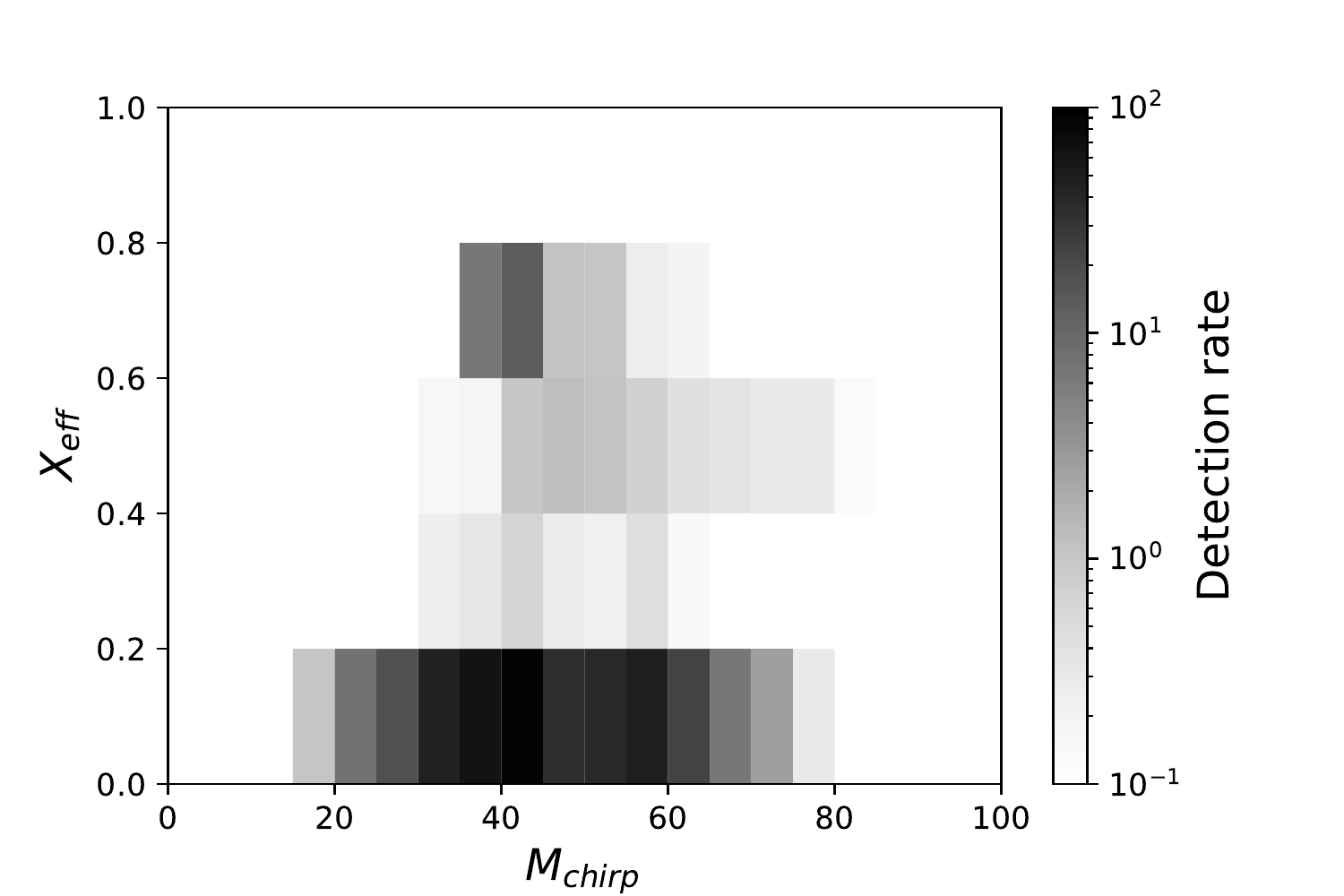}
        \subcaption{FS1 model}
        \label{FS1_KAGRA}
      \end{minipage} \\
            \begin{minipage}[t]{0.5\hsize}
        \centering
        \includegraphics[keepaspectratio, scale=0.5]{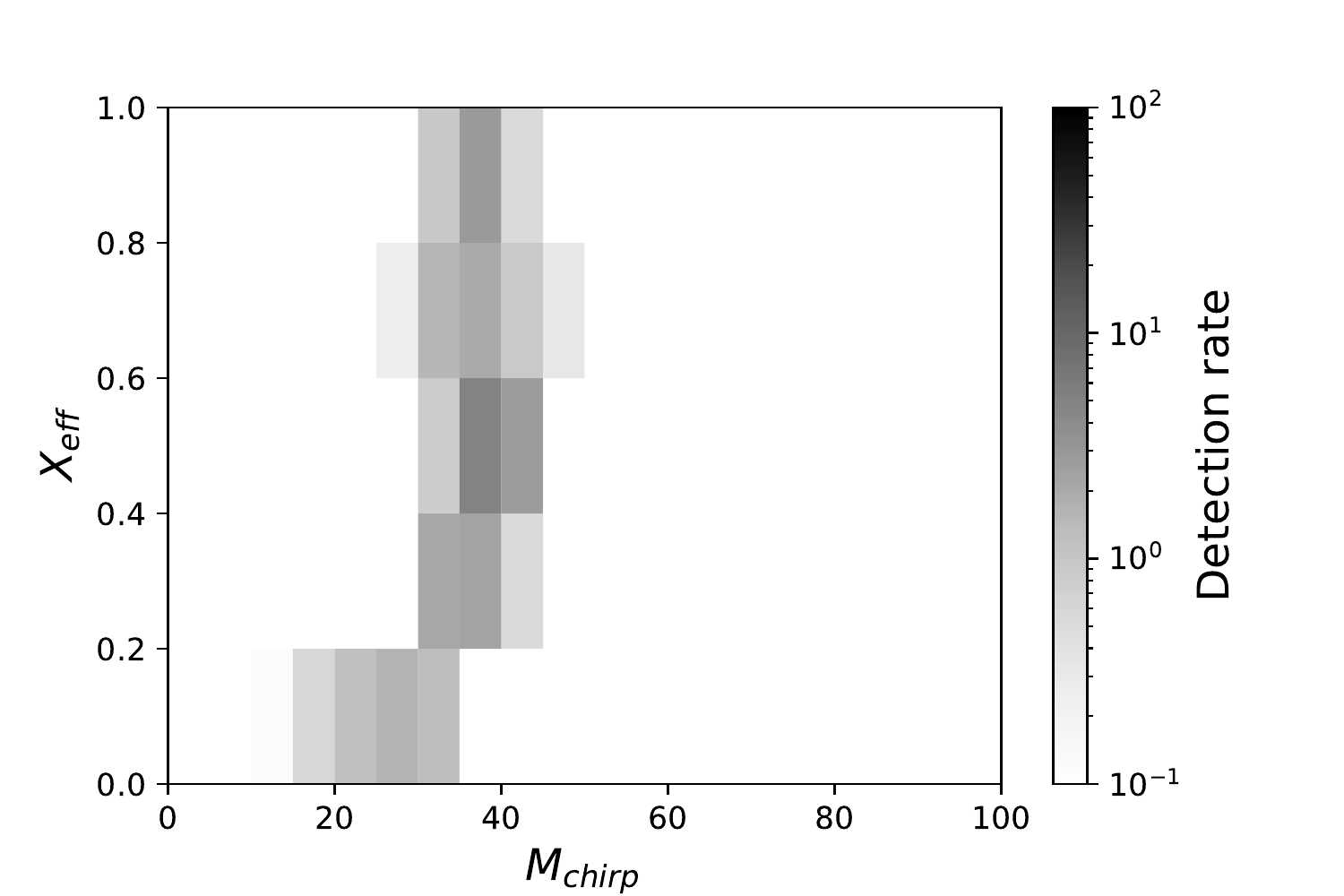}
        \subcaption{FS2 model}
        \label{FS2_KAGRA}
      \end{minipage}

    \end{tabular}
    \caption{Pop III BBH detection rate of aLIGO, aVIRGO, and KAGRA.
        The style is the same as Fig.~\ref{KAGRAdetect1}.}
        \label{KAGRAdetect2}
  \end{figure*}

Figure \ref{KAGRAdetect2} shows the detection rate of Pop III BBHs for the second-generation GW detectors.

\begin{figure*}
    \begin{tabular}{cc}
      \begin{minipage}[t]{0.5\hsize}
        \centering
        \includegraphics[keepaspectratio, scale=0.5]{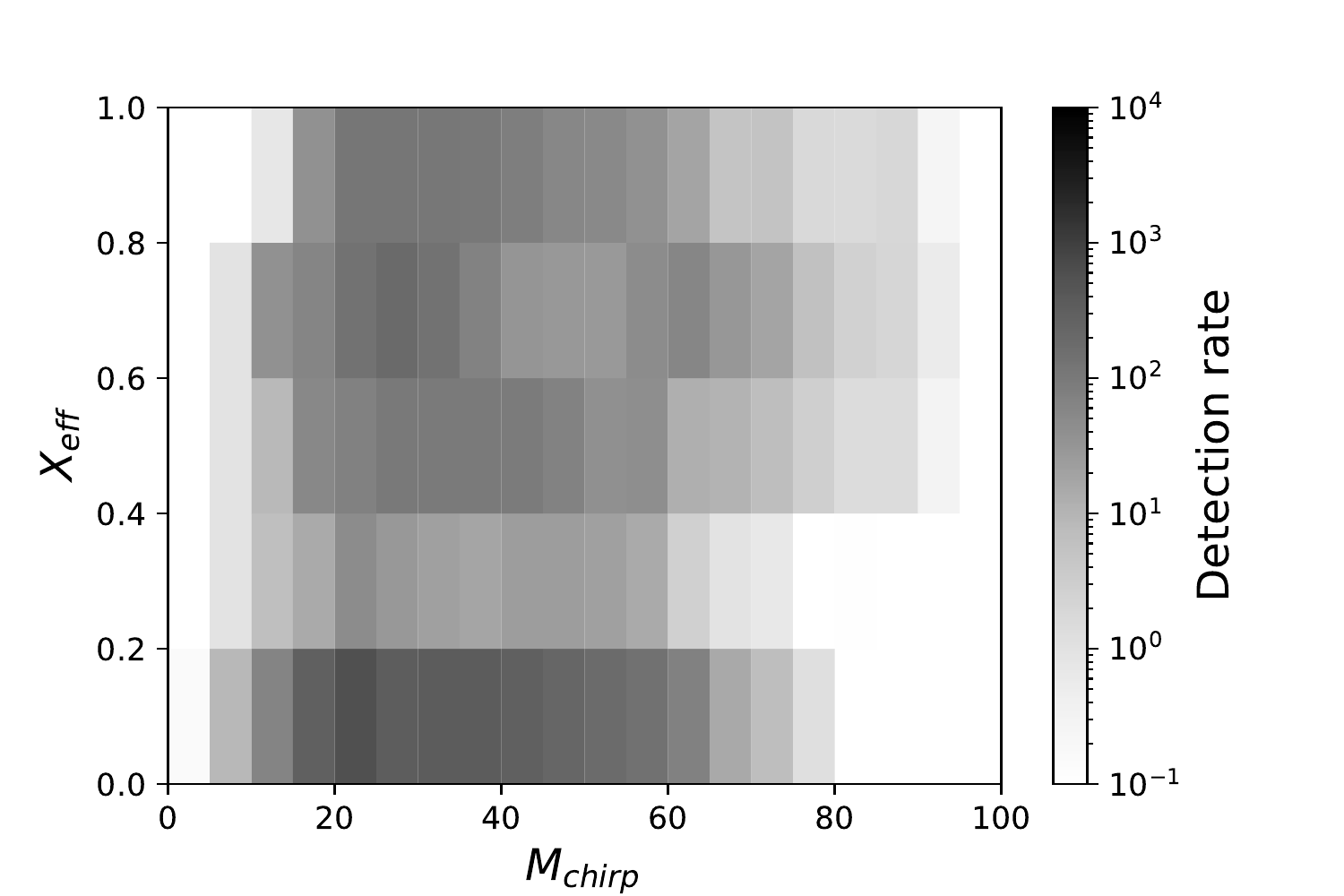}
        \subcaption{$\beta$=0.5 model}
        \label{MT05_ET}
      \end{minipage}
            \begin{minipage}[t]{0.5\hsize}
        \centering
        \includegraphics[keepaspectratio, scale=0.5]{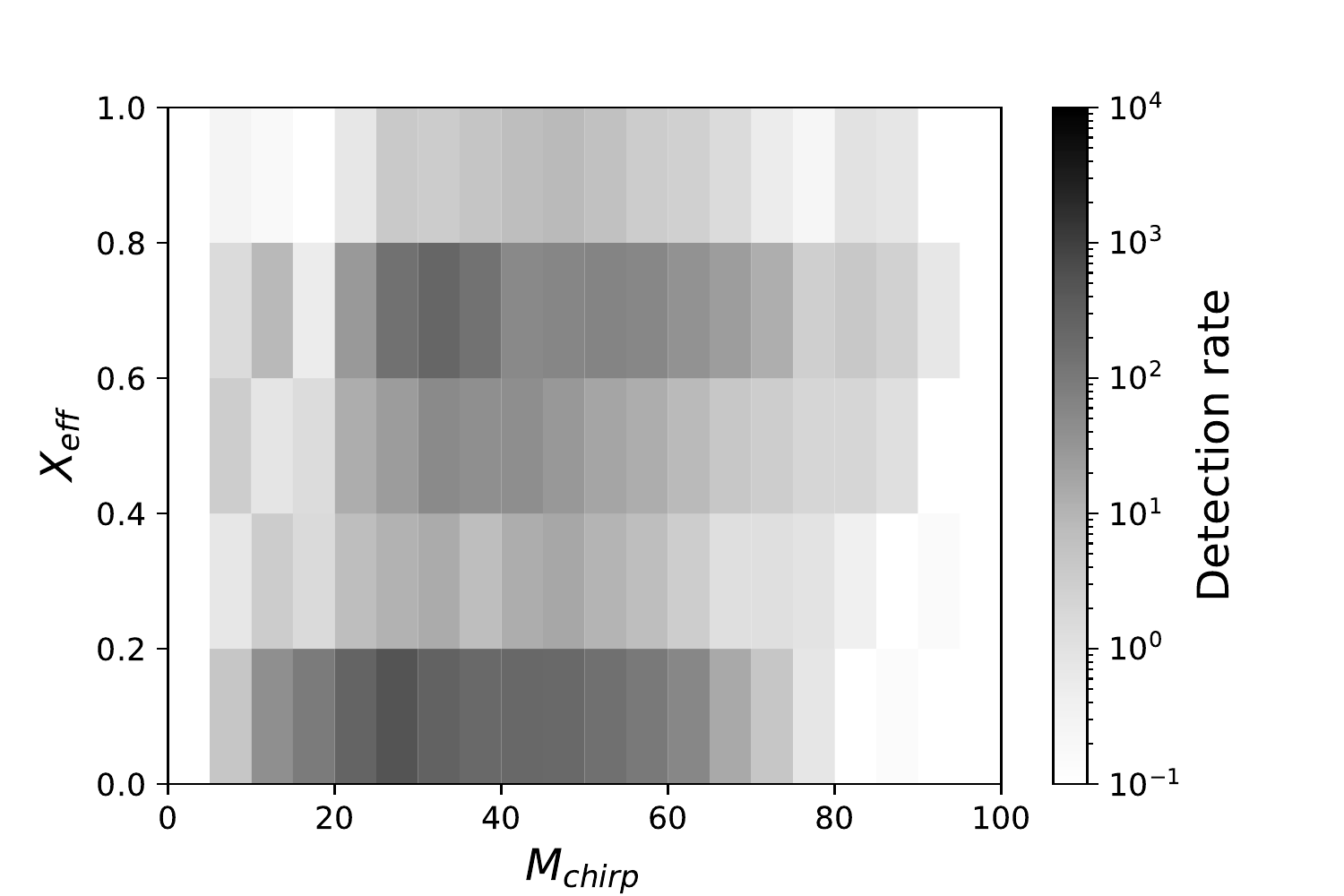}
        \subcaption{$\alpha\lambda$=0.1 model}
        \label{al01_ET}
      \end{minipage}\\
        \begin{minipage}[t]{0.5\hsize}
        \centering
        \includegraphics[keepaspectratio, scale=0.5]{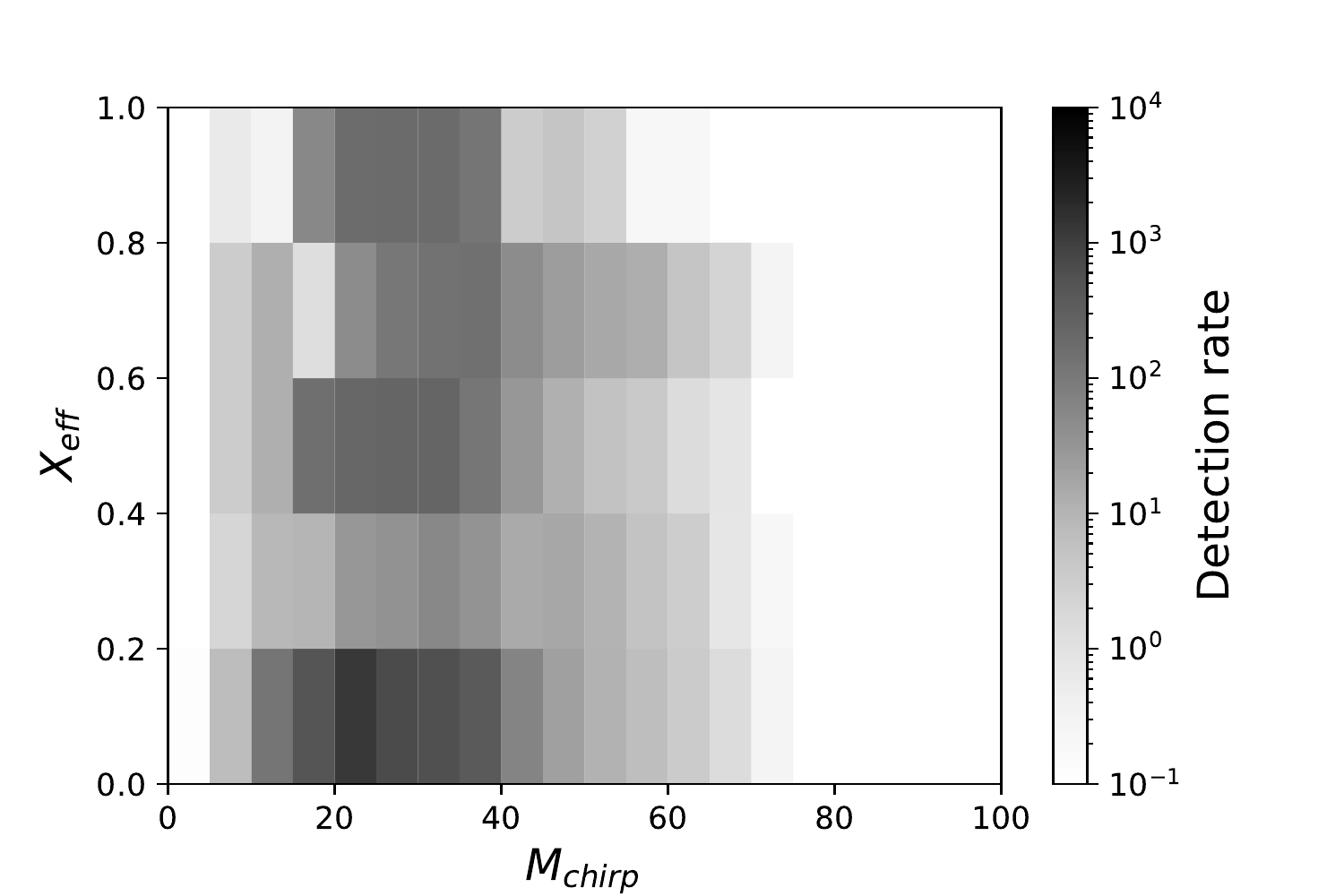}
        \subcaption{M100 model}
        \label{M100_ET}
      \end{minipage}
          \begin{minipage}[t]{0.5\hsize}
        \centering
        \includegraphics[keepaspectratio, scale=0.5]{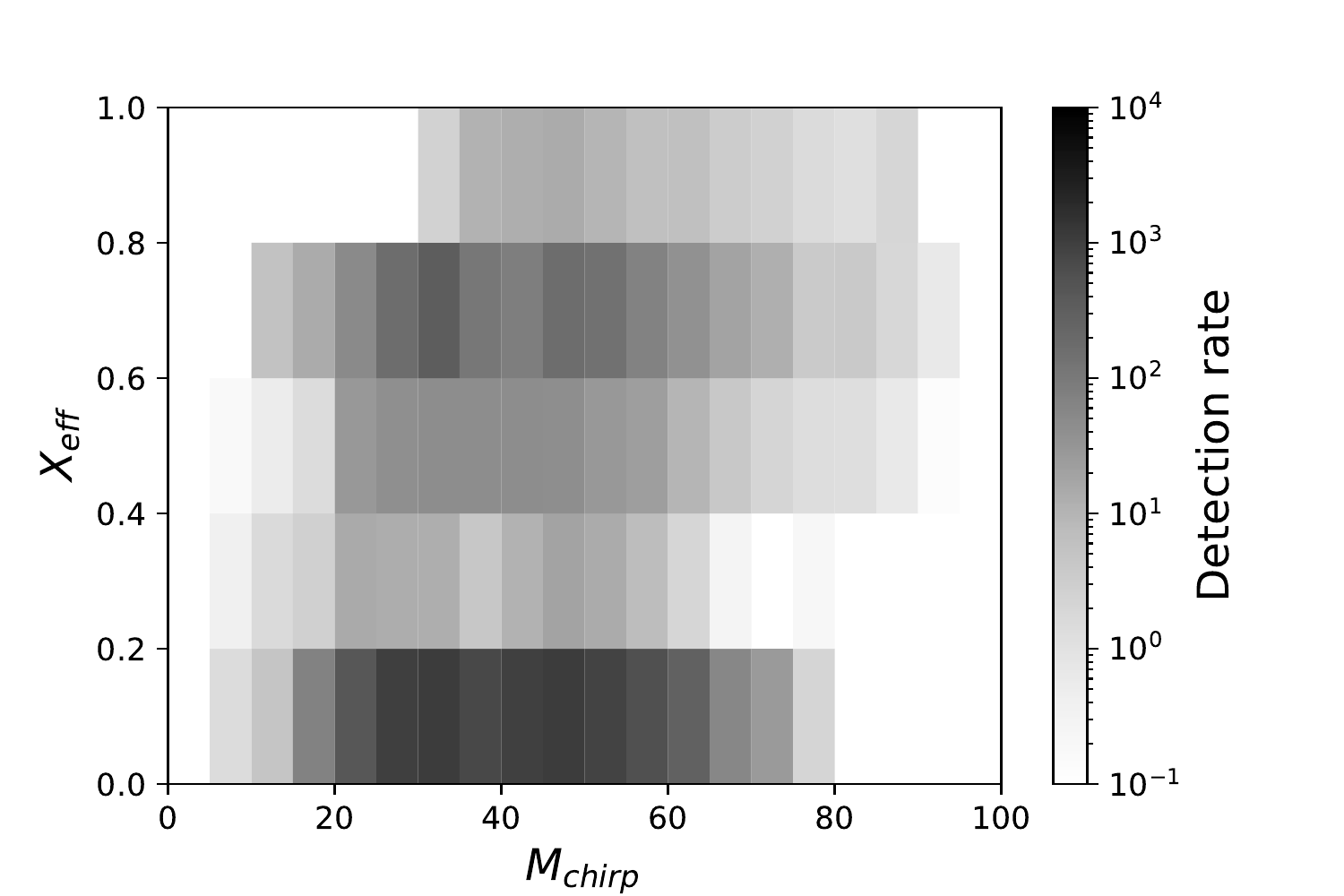}
        \subcaption{FS1 model}
        \label{FS1_ET}
      \end{minipage}\\
            \begin{minipage}[t]{0.5\hsize}
        \centering
        \includegraphics[keepaspectratio, scale=0.5]{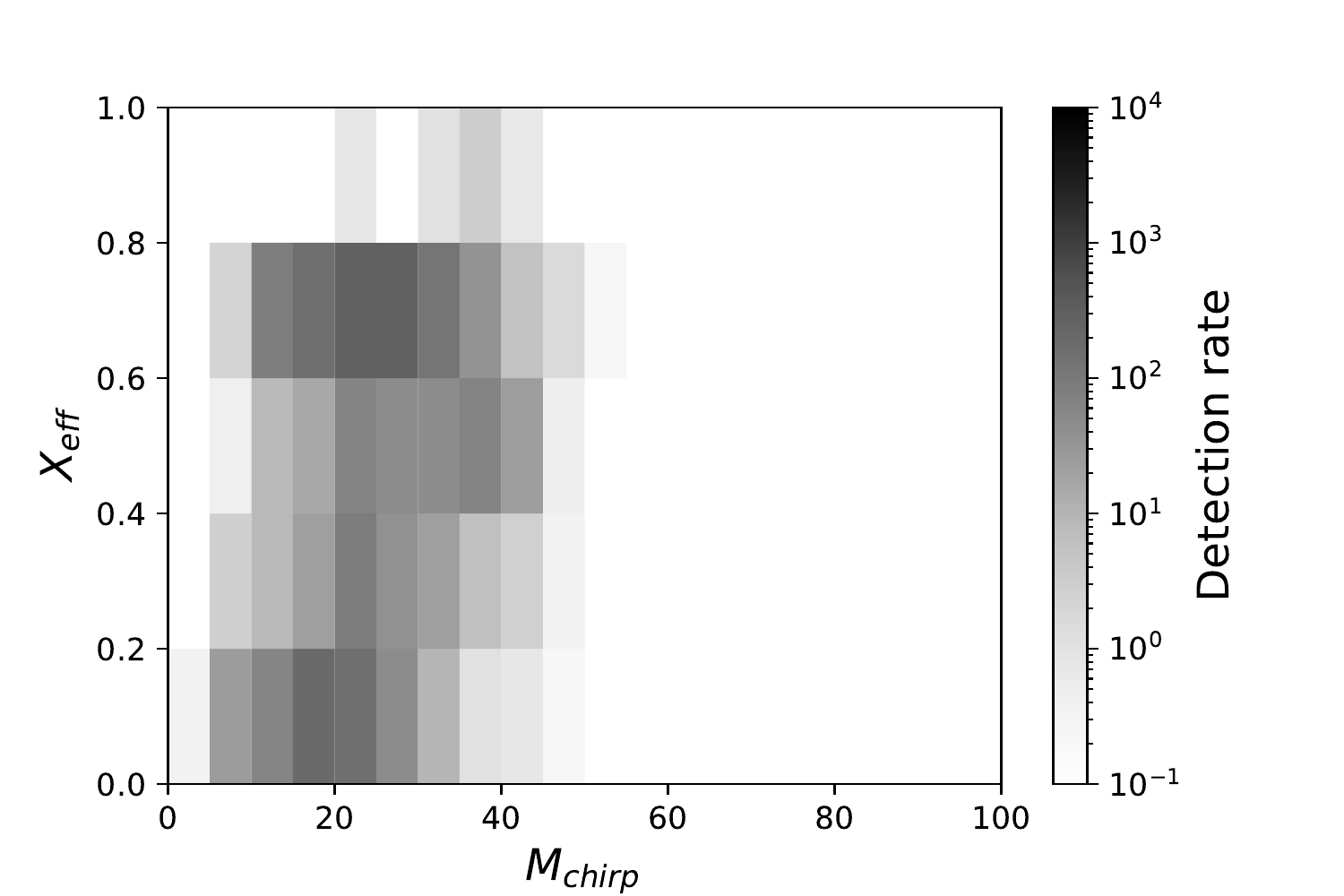}
        \subcaption{FS2 model}
        \label{FS2_ET}
      \end{minipage}
      
    \end{tabular}
    \caption{Pop III BBH detection rate of ET.
        The style is the same as Fig.~\ref{ETdetect1}.
        Note that the scale of the detection rate is different from Fig.~\ref{KAGRAdetect2} again.}
        \label{ETdetect2}
  \end{figure*}

%such as aLIGO, aVIRGO, and KAGRA.
Figure \ref{ETdetect2} shows the detection rate of Pop III BBHs for ET.

\begin{figure*}
    \begin{tabular}{cc}
      \begin{minipage}[t]{0.5\hsize}
        \centering
        \includegraphics[keepaspectratio, scale=0.5]{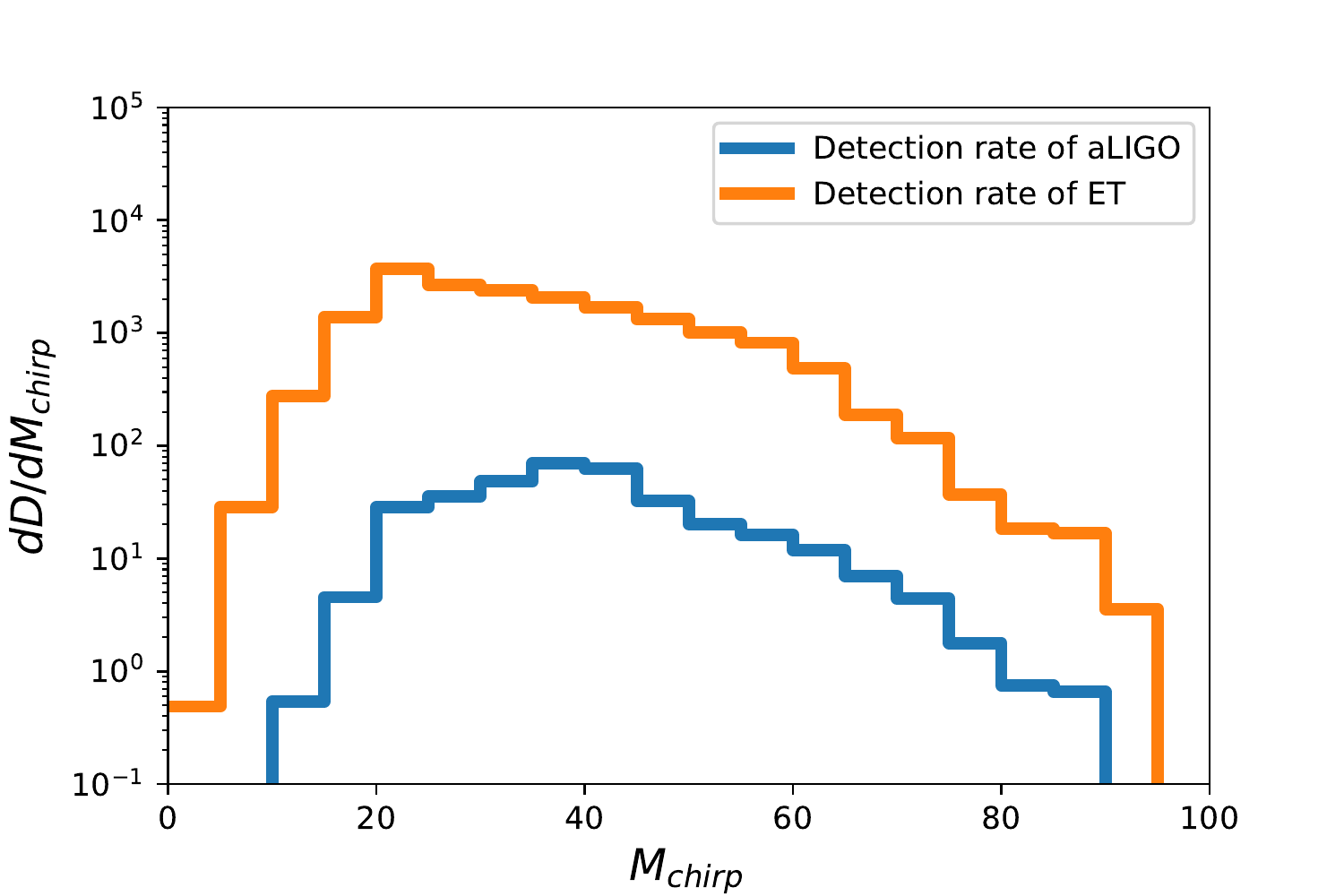}
        \subcaption{$\beta$=0.5 model}
        \label{MT05_CMass}
      \end{minipage} 
        
      \begin{minipage}[t]{0.5\hsize}
        \centering
        \includegraphics[keepaspectratio, scale=0.5]{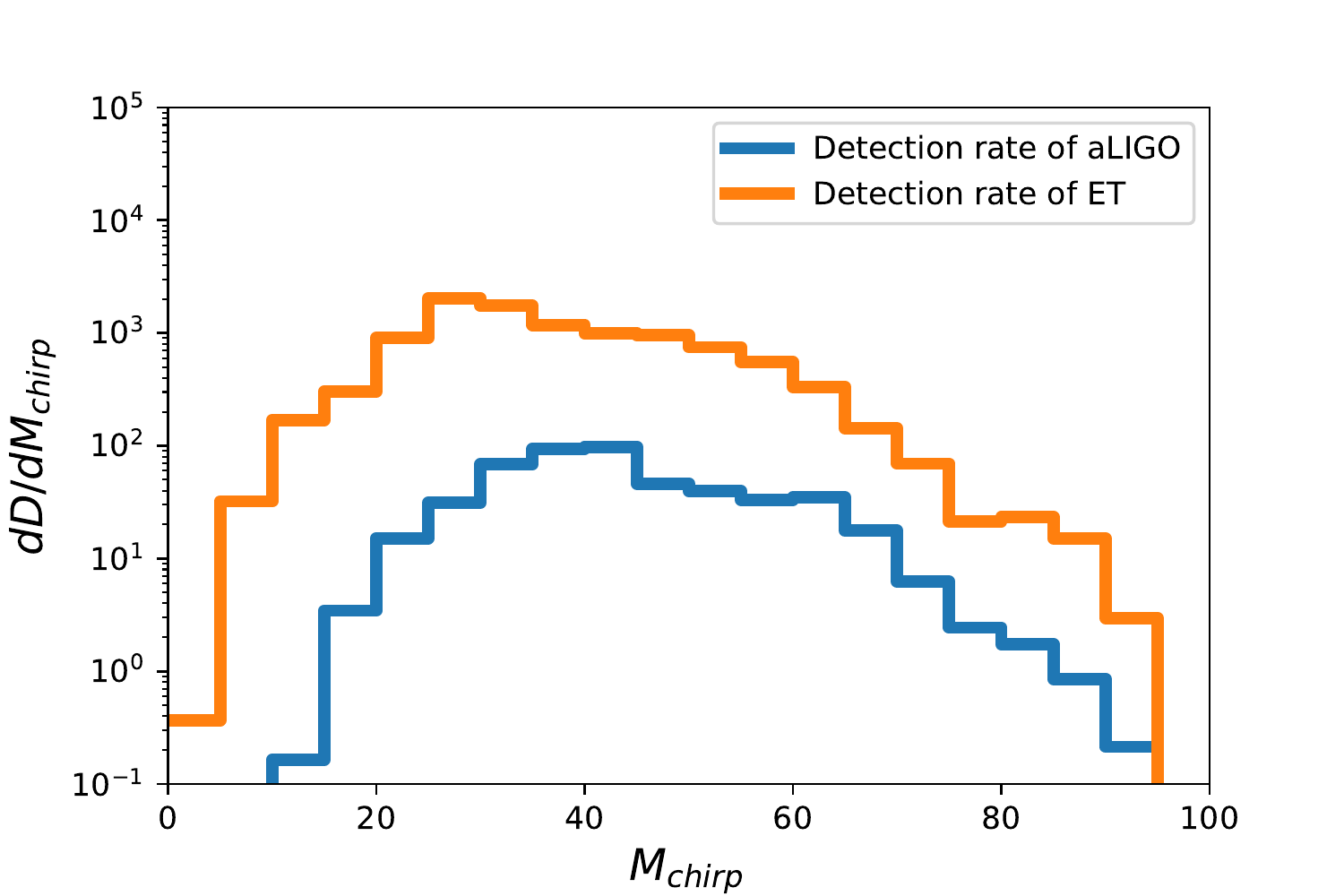}
        \subcaption{$\alpha\lambda$=0.1 model}
        \label{al01_CMass}
      \end{minipage} \\

            \begin{minipage}[t]{0.5\hsize}
        \centering
        \includegraphics[keepaspectratio, scale=0.5]{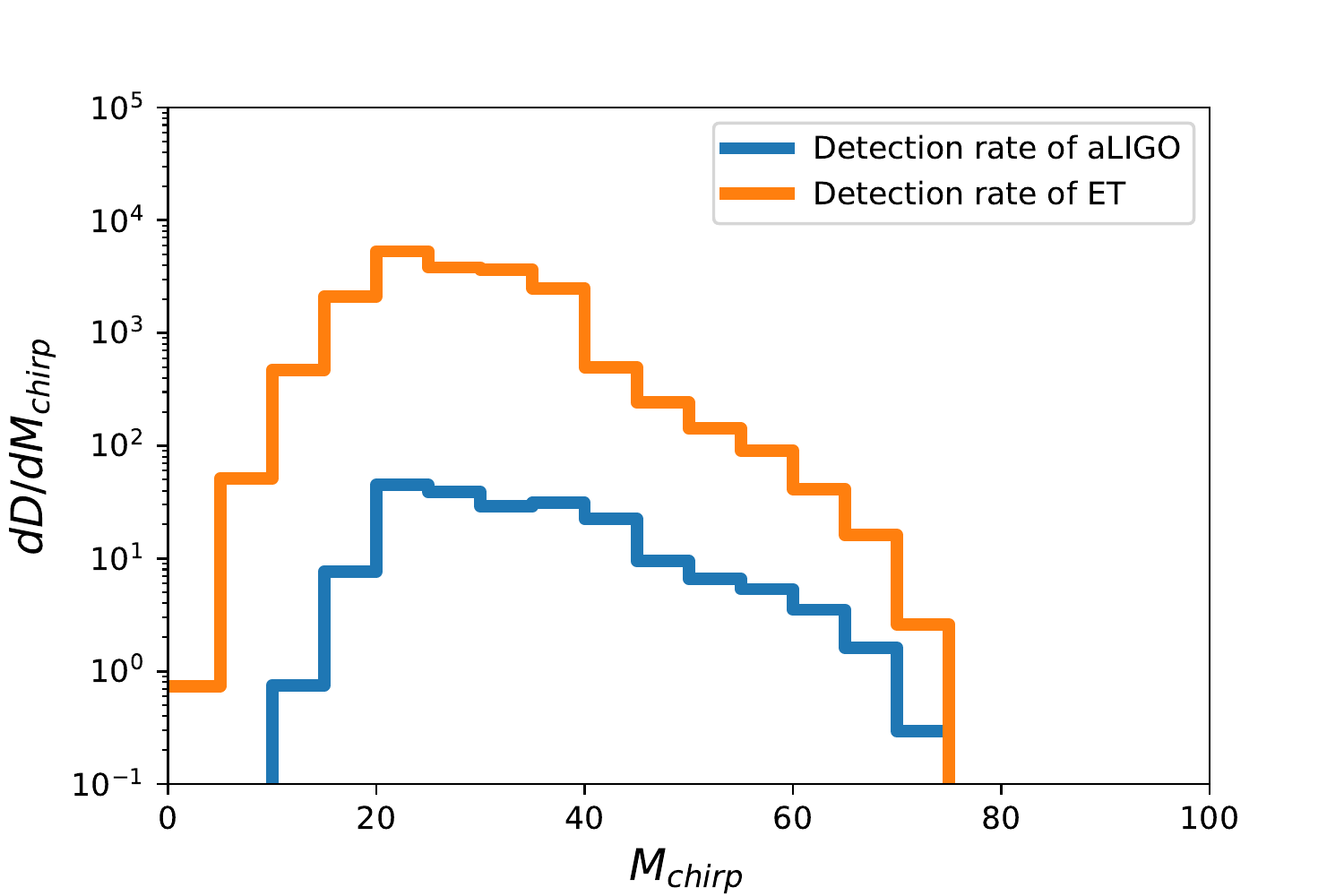}
        \subcaption{M100 model}
        \label{M100_CMass}
      \end{minipage} 
  
            \begin{minipage}[t]{0.5\hsize}
        \centering
        \includegraphics[keepaspectratio, scale=0.5]{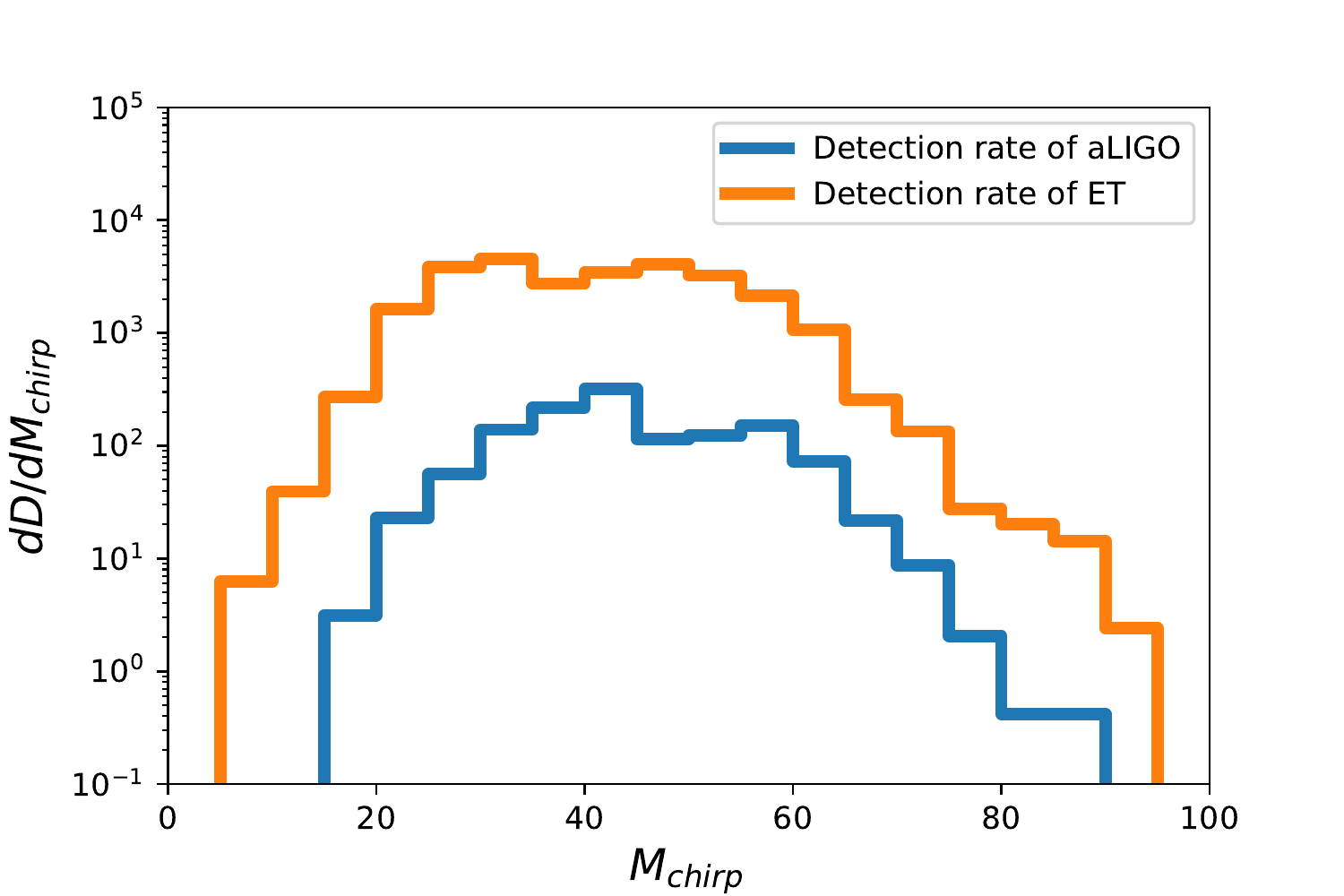}
        \subcaption{FS1 model}
        \label{FS1_CMass}
      \end{minipage} \\
      \begin{minipage}[t]{0.5\hsize}
        \centering
        \includegraphics[keepaspectratio, scale=0.5]{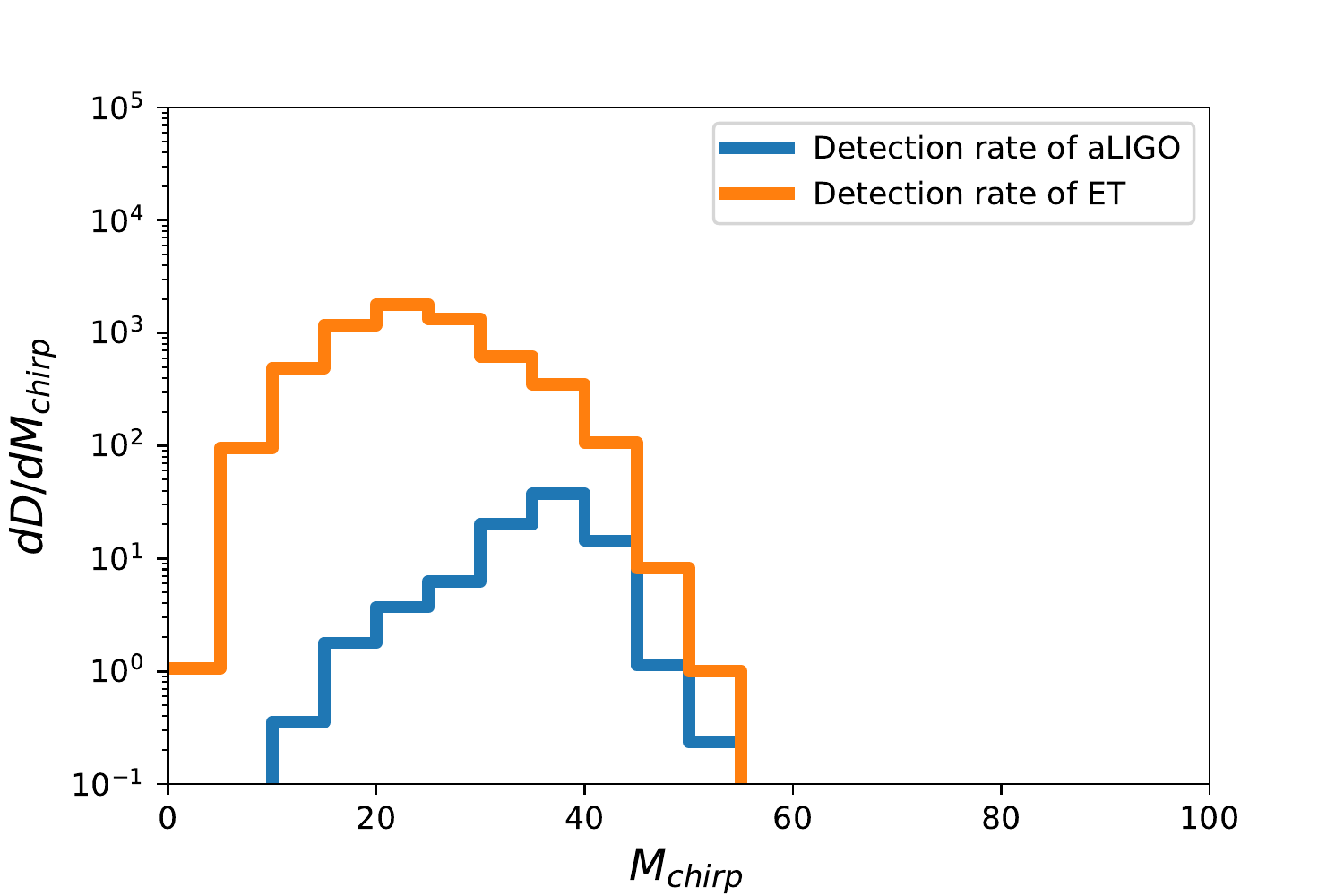}
        \subcaption{FS2 model}
        \label{FS2_CMass}
      \end{minipage} 
      
    \end{tabular}
    \caption{Chirp mass ($M_{\rm chirp}$) distribution of detectable Pop III BBHs  for aLIGO and ET.
        The line styles are the same as Fig.~\ref{CMass1}.}
        \label{CMass2}
  \end{figure*}

Figure \ref{CMass2} shows the chirp mass distributions of detectable Pop III BBHs for the second-generation detectors and ET.

\begin{figure*}
    \begin{tabular}{cc}
        \begin{minipage}[t]{0.5\hsize}
        \centering
        \includegraphics[keepaspectratio, scale=0.5]{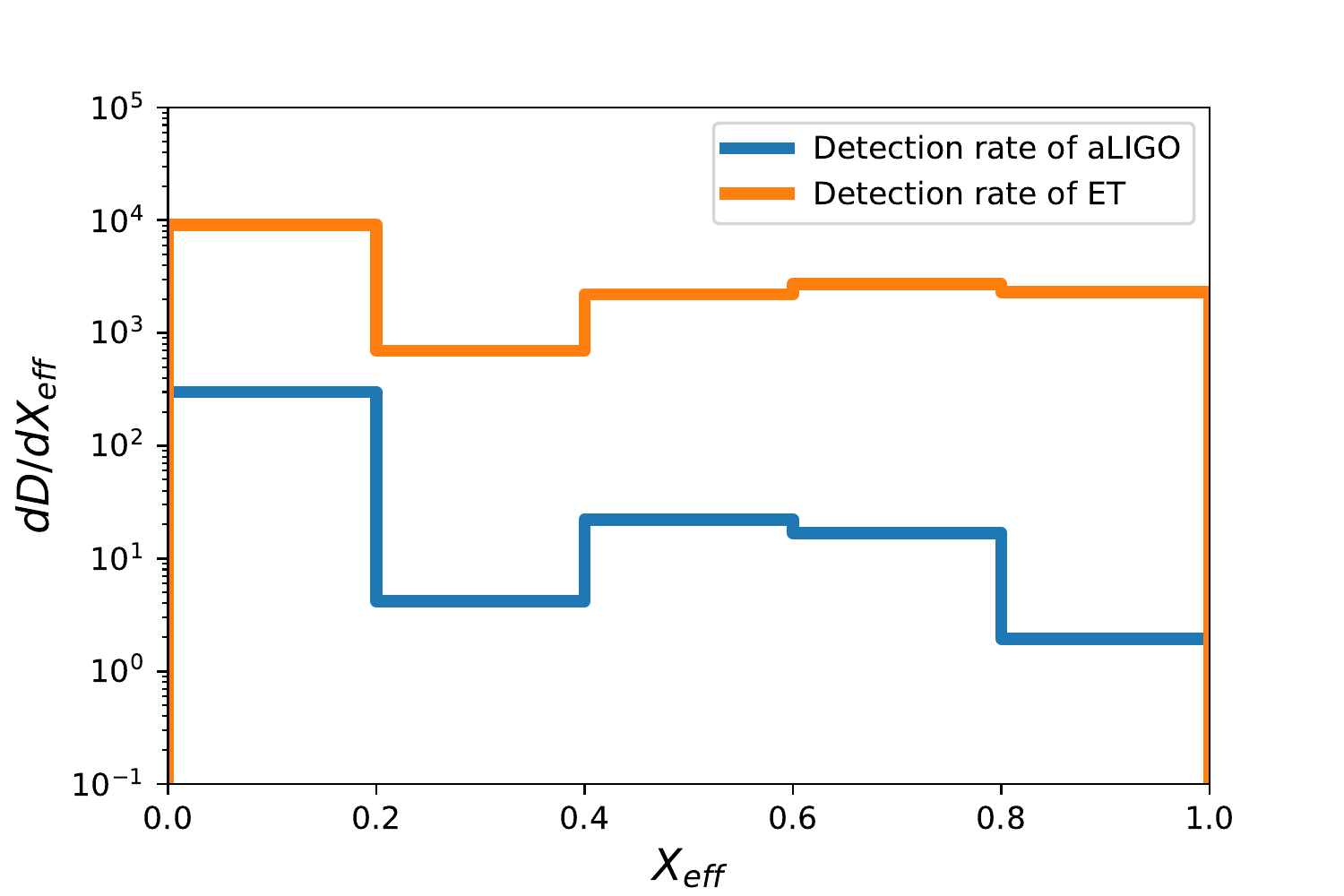}
        \subcaption{$\beta$=0.5 model}
        \label{MT05_Spin}
      \end{minipage}

      \begin{minipage}[t]{0.5\hsize}
        \centering
        \includegraphics[keepaspectratio, scale=0.5]{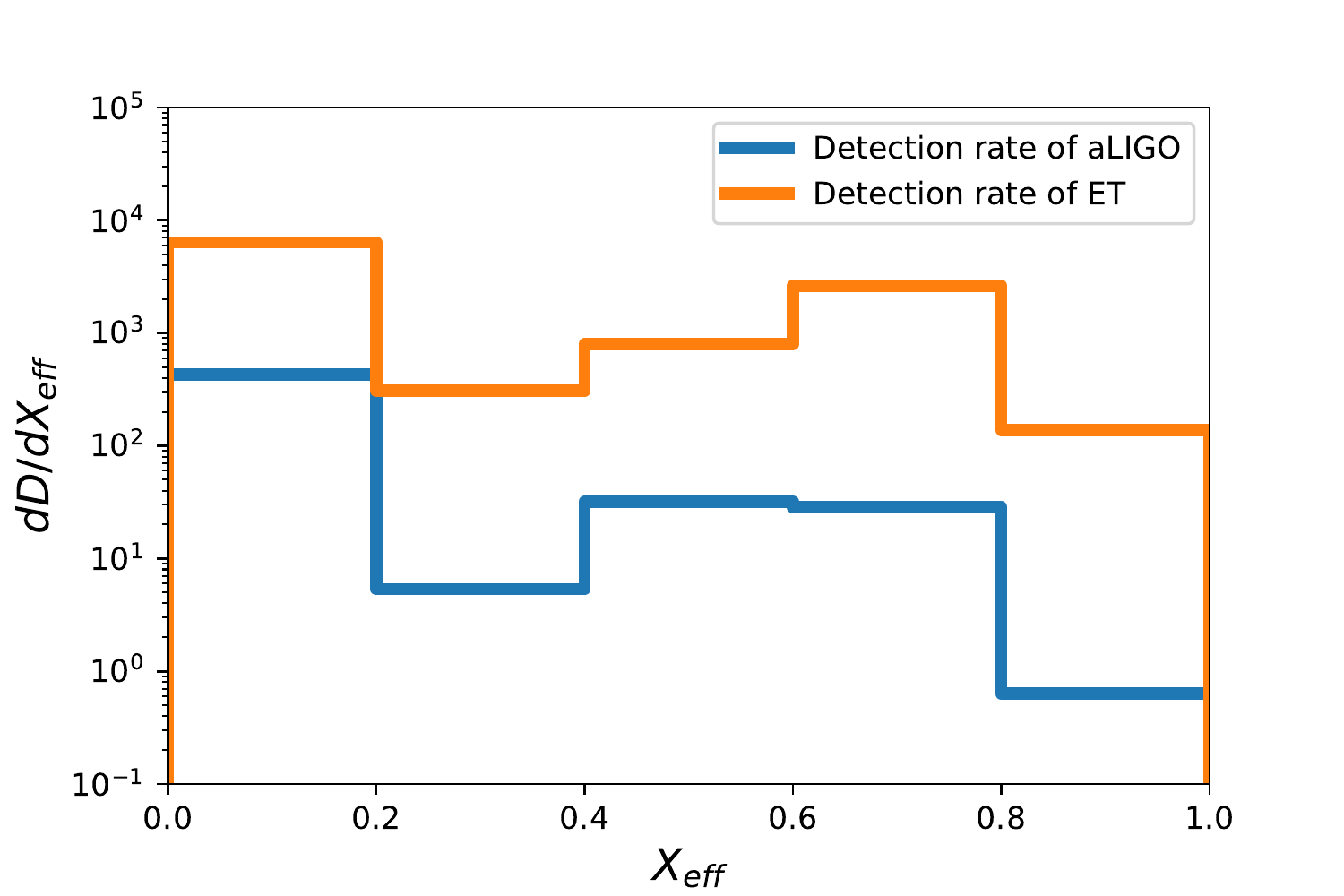}
        \subcaption{$\alpha\lambda$=0.1 model}
        \label{al01_Spin}
      \end{minipage}\\
         \begin{minipage}[t]{0.5\hsize}
        \centering
        \includegraphics[keepaspectratio, scale=0.5]{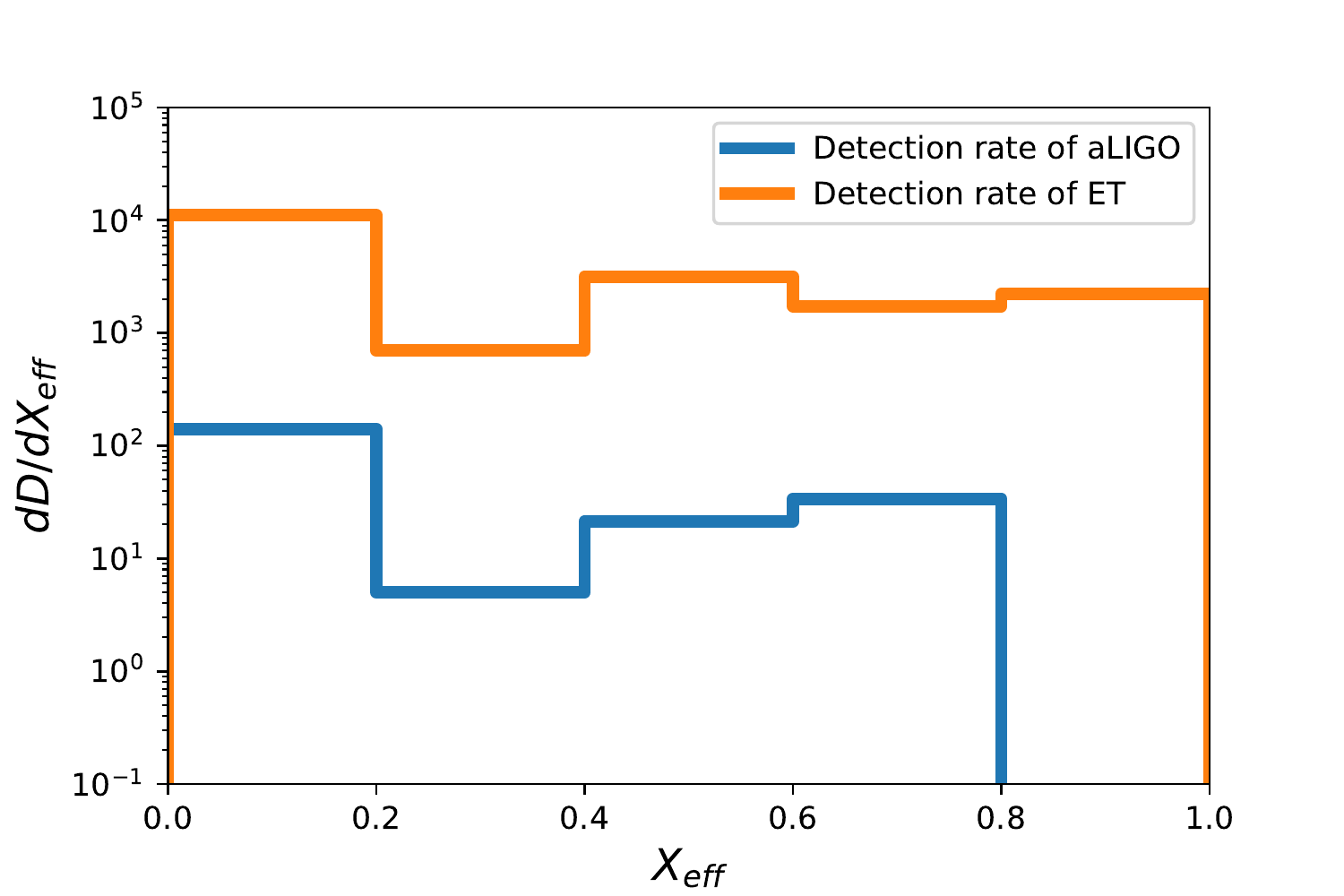}
        \subcaption{M100 model}
        \label{M100_Spin}
      \end{minipage}
      
           \begin{minipage}[t]{0.5\hsize}
        \centering
        \includegraphics[keepaspectratio, scale=0.5]{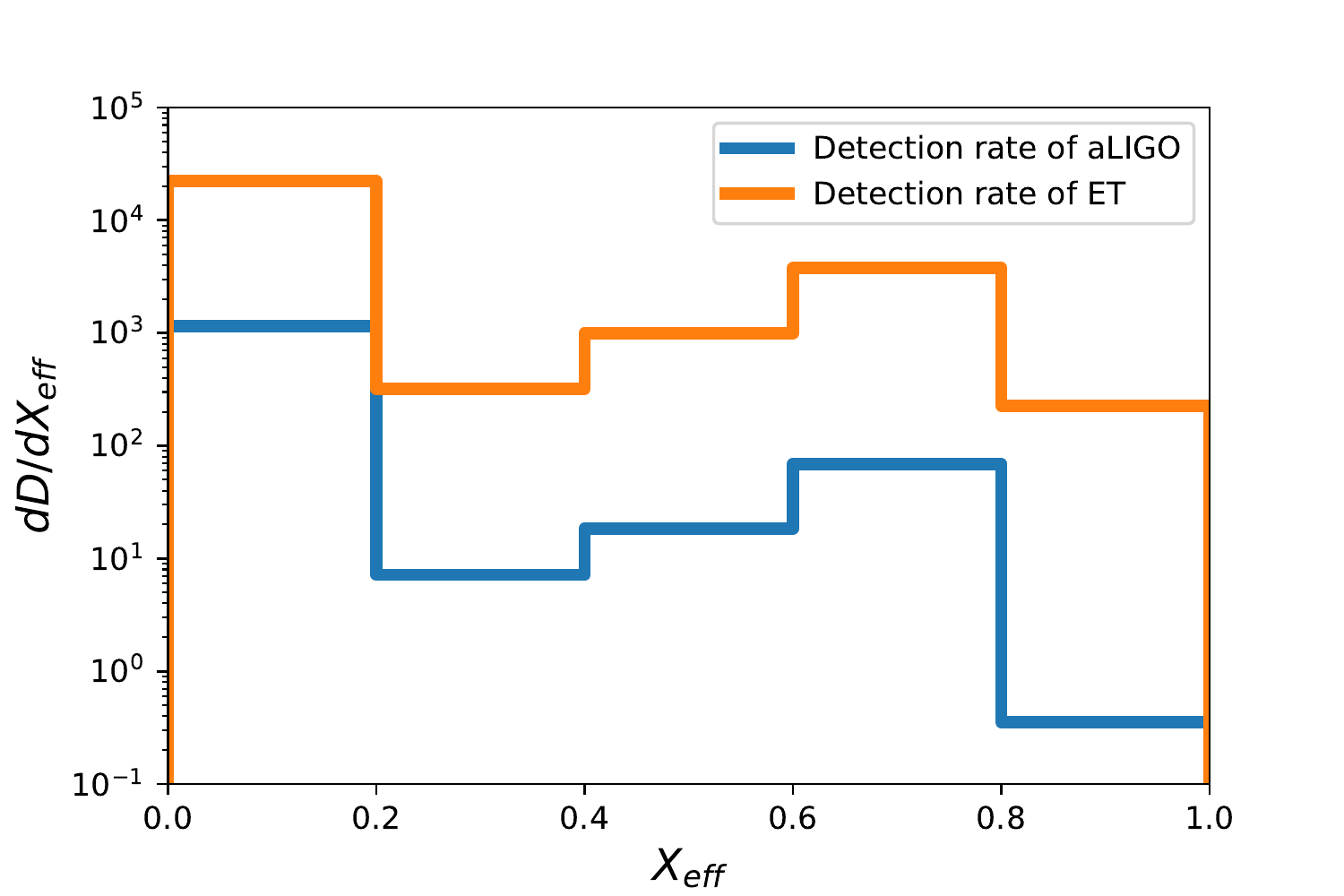}
        \subcaption{FS1 model}
        \label{FS1_Spin}
      \end{minipage}\\
      \begin{minipage}[t]{0.5\hsize}
        \centering
        \includegraphics[keepaspectratio, scale=0.5]{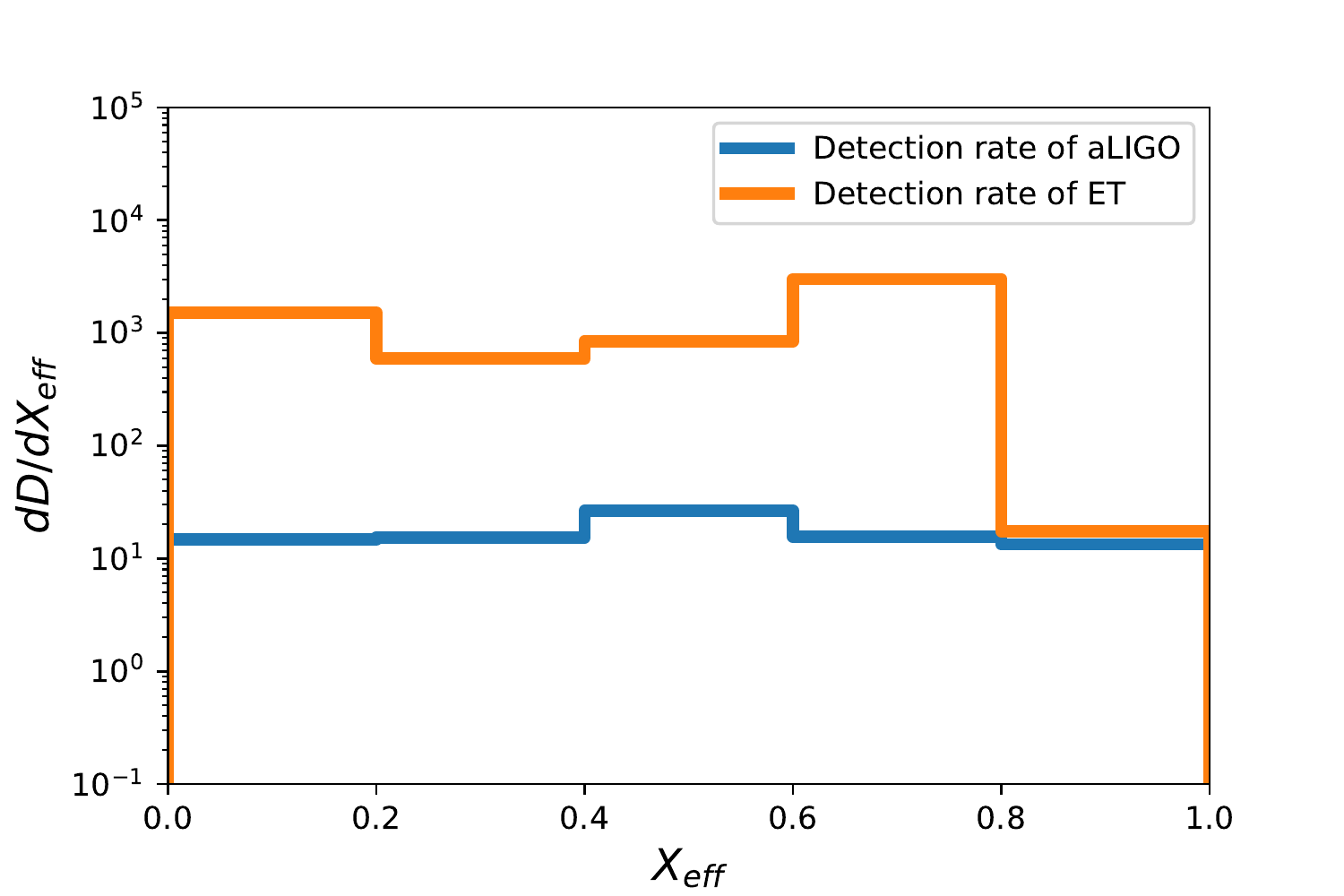}
        \subcaption{FS2 model}
        \label{FS2_Spin}
      \end{minipage}
      
    \end{tabular}
    \caption{Effective spin ($\chi_{\rm eff}$) distribution of detectable Pop III BBHs for aLIGO and ET.
        The line styles are the same as Fig.~\ref{Spin1}.}
        \label{Spin2}
  \end{figure*}

Figure \ref{Spin2} shows the effective spin distributions of detectable Pop III BBHs for the second-generation detectors and ET.

In the case of $\beta=0.5$, $\alpha\lambda=0.1$, M100, and FS1 models, the feature of detectable BBHs is almost same as that of the fiducial model.
On the other hand, in the case of FS2 model, the maximum mass of detectable BBHs is much smaller than that of the other models.
This is because of the same reason discussed in Sec. \ref{cmass}.

\end{document}